\documentclass[]{aa} 

\usepackage{natbib}
\bibpunct{(}{)}{;}{a}{}{,} 
\usepackage{epsfig}

\newcommand{\beq}{\begin{equation}} 
\newcommand{\eeq}{\end{equation}} 
\newcommand{\beqa}{\begin{eqnarray}} 
\newcommand{\eeqa}{\end{eqnarray}} 
\newcommand{\bea}{\begin{array}} 
\newcommand{\ea}{\end{array}} 

\newcommand{\dd}{{\rm d}}
\newcommand{\pl}{\partial}
 
\newcommand{\lag}{\langle} 
\newcommand{\rag}{\rangle} 

\newcommand{\ii}{{\rm i}}

\newcommand{\Om}{\Omega_{\rm m}} 
\newcommand{\rhob}{\overline{\rho}}

\newcommand{\vk}{{\bf k}}

\newcommand{\vq}{{\bf q}}

\newcommand{\vv}{{\bf v}}

\newcommand{\vx}{{\bf x}}
\newcommand{\vOmega}{{\bf \Omega}}
\newcommand{\vPsi}{{\bf \Psi}}

\newcommand{\tdelta}{{\tilde{\delta}}}

\newcommand{\trho}{{\tilde{\rho}}}
\newcommand{\tu}{{\tilde{u}}}
\newcommand{\tW}{{\tilde{W}}}

\newcommand{\cD}{{\cal{D}}}
\newcommand{\cF}{{\cal{F}}}

\newcommand{\cO}{{\cal{O}}}

\newcommand{\FoH}{{F_{1\rm H}}}
\newcommand{\FtH}{{F_{2\rm H}}}
\newcommand{\PoH}{{P_{1\rm H}}}
\newcommand{\PtH}{{P_{2\rm H}}}

\newcommand{\Ci}{{\rm Ci}}
\newcommand{\Si}{{\rm Si}}

\newcommand{\simlt}{\hspace{0.3em}\raisebox{0.4ex}{$<$}\hspace{-0.75em}\raisebox{-.7ex}{$\sim$}\hspace{0.3em}}
\newcommand{\simgt}{\hspace{0.3em}\raisebox{0.4ex}{$>$}\hspace{-0.75em}\raisebox{-.7ex}{$\sim$}\hspace{0.3em}}

\begin{document} 

\topmargin =0.3cm

\title{Combining perturbation theories with halo models}    
\author{P. Valageas \and T. Nishimichi}   
\institute{Institut de Physique Th\'eorique, CEA Saclay, 91191 Gif-sur-Yvette, 
France \and Institute for the Physics and Mathematics of the Universe, University of Tokyo, Kashiwa, Chiba 277-8568, Japan}
\date{Received / Accepted } 
 
\abstract
{}
{We investigate the building of unified models that can predict the matter-density
power spectrum and the two-point correlation function from very large to small
scales, being consistent with perturbation theory at low $k$ and with halo models
at high $k$.}
{We use a Lagrangian framework to re-interpret the halo model and to decompose
the power spectrum into ``2-halo'' and ``1-halo'' contributions, related to
``perturbative'' and ``non-perturbative'' terms. We describe a simple implementation
of this model and present a detailed comparison with numerical simulations,
from $k \sim 0.02$ up to $100 h$Mpc$^{-1}$, and from $x \sim 0.02$ up to
$150 h^{-1}$Mpc.}
{We show that the 1-halo contribution contains a counterterm that ensures a
$k^2$ tail at low $k$ and is important not to spoil the predictions on the scales
probed by baryon acoustic oscillations, $k \sim 0.02$ to $0.3 h$Mpc$^{-1}$.
On the other hand, we show that standard perturbation theory is inadequate
for the 2-halo contribution, because higher order terms grow too fast at high $k$,
so that resummation schemes must be used. Moreover, we explain why such
a model, based on the combination of perturbation theories and halo models,
remains consistent with standard perturbation theory up to the order of the
resummation scheme. We describe a simple implementation, based on a 1-loop
``direct steepest-descent'' resummation for the 2-halo contribution that allows
fast numerical computations, and we check that we obtain a good match to
simulations at low and high $k$. We also study the dependence of such
predictions on the details of the underlying model, such as the choice of the
perturbative resummation scheme or the properties of halo profiles.
Our simple implementation already fares better than standard 1-loop
perturbation theory on large scales and simple fits to the power spectrum
at high $k$, with a typical accuracy of $1\%$ on large scales and $10\%$ on small
scales. We obtain similar results for the two-point correlation function.
However, there remains room for improvement on the transition
scale between the 2-halo and 1-halo contributions, which may be the most difficult
regime to describe.}
{}


\maketitle

\section{Introduction} 
\label{Introduction}

In the standard cosmological scenario, the large-scale structures we observe in
the recent Universe (galaxies, clusters, filaments, voids, ..) have formed
through the amplification by gravitational instability of small primordial
perturbations generated in the primordial Universe by quantum fluctuations
\citep{Peebles1980}. Moreover, at the beginning of matter domination
($z \sim 3000$), the power increases on small scales for cold dark
matter (CDM) models \citep{Peebles1982}, which leads to a ``hierarchical scenario''
where smaller scales become nonlinear first. Then, increasingly large and massive
structures form in the course of time, as small structures merge and larger scales
turn nonlinear. Therefore, on large scales or at early times, it is possible to use
linear theory or, more generally, perturbation theory, while on small scales or at late
times one must use numerical simulations or phenomenological models.
A quantity of great interest to characterize the density fields built by these processes
is the density power spectrum $P(k)$, which is the Fourier transform of the
density two-point correlation function. On very large linear scales this is actually
sufficient to fully determine the statistical properties of the matter distribution
if the initial conditions are (almost) Gaussian, as in standard scenarios of
single-field inflation. On smaller nonlinear scales, higher order correlations
are generated by the nonlinearities of the dynamics, but $P(k)$ remains
a standard tool to compare observations with theoretical predictions
(since higher order statistics are increasingly noisy or difficult to predict).

When one focuses on large scales, several observational probes, such as weak lensing surveys
\citep{Massey2007,Munshi2008}, or galaxy redshift surveys provide rich cosmological information.
For example, the signature of baryon acoustic oscillations (BAO) 
\citep{Eisenstein1998,Eisenstein2005}, 
redshift-space distortions \citep{Hamilton1992,Percival2009}, 
non-Gaussianities in the primordial perturbations \citep{Liguori2010,Desjacques2010}, and mass of neutrinos 
\citep{Swanson2010,Saito2010} are sensitive to the clustering on linear to weakly nonlinear scales, 
where the departures from the linear regime are already noticeable but still modest. 
In order to meet the accuracy of future observations a theoretical
accuracy on the order of $1\%$ is required, which is beyond the reach of
simple phenomenological models or fits to numerical simulations
(if one wishes to obtain robust predictions for a large range of initial conditions
and cosmological parameters). However, this regime should be within the
range of validity of perturbation theories, which offer the advantage of
systematic and reliable predictions, without fitting parameters.
This has led to a renewed interest in perturbative approaches in recent years, as it
may be possible to  improve over the standard perturbation theory
\citep{Goroff1986,Bernardeau2002} by using resummation schemes that allow
partial resummations of higher order terms
\citep{Crocce2006a,Crocce2006b,Valageas2007a,Matarrese2007,Taruya2008,Valageas2008,Pietroni2008,Matsubara2008}.
In particular, this somewhat improves the accuracy of the predicted matter
power spectrum on the large scales associated with BAO, as compared with
standard perturbation theory truncated at the same order 
\citep{Crocce2008,Carlson2009,Taruya2009}.

On the other hand, many observational probes, such as weak lensing surveys
on smaller angular scales and galaxy surveys, are sensitive to highly nonlinear
scales. There, systematic analytical approaches that have been proposed so
far do not apply, in particular because of the importance of shell crossing effects
that are not adequately taken into account
(an exception is the formalism developed in \citet{Valageas2004}, which
does not make the single-stream approximation since it deals with the phase-space
distribution $f(\vx,\vv;t)$, but leads to heavy computations).
Then, one must resort to numerical simulations or to phenomenological models.
One such model is the Lagrangian mapping presented in 
\citet{Hamilton1991} (see \citet{Peacock1996} in Fourier space), that writes the
nonlinear two-point correlation
or power spectrum in terms of its linear counterpart on a different scale.
A second model is the ``halo model'' \citep{Scherrer1991,Cooray2002}, where
the matter distribution is described as a collection of halos.
This yields two contributions for the power spectrum or two-point correlation
function, a ``1-halo'' term associated with particle pairs that belong to the same
halo, and a ``2-halo'' term associated with particles that belong to two different
halos. Then, the 1-halo term, which dominates on small scales, is sensitive
to the density profile and mass distribution of the halos, whereas the
2-halo term, which dominates on large scales, is sensitive to the correlation
between halos.
In practice, one often replaces the 2-halo term by the linear two-point
correlation, or power spectrum, to reproduce linear theory on large scales.
Previous works \citep{Cooray2002,Smith2003,Smith2007,Giocoli2010}
have shown that halo models provide a good match to numerical simulations,
especially at high $k$, whereas Lagrangian mappings based on the stable
clustering ansatz \citep{Peebles1980} do not fare as well.

It is clear that it is of great interest to develop unified models, which combine
the benefits of systematic perturbation theories on large scales with the
reasonable match to simulations on small scales provided by halo models.
The goal of the present paper is precisely to investigate the building of such
a model, that describes all scales, from the linear to the highly nonlinear regime.
We can already note here two points that must be addressed to reach
this goal: i) the increasingly large growth at high $k$ of higher order terms
obtained within standard perturbation theory, and ii) the nonzero limit for
$k\rightarrow 0$ of the 1-halo term (as written in previous works), which
eventually becomes non-negligible (and even dominant) on large scales,
since the linear power spectrum typically scales as $P_L(k) \sim k$ at low $k$
for CDM scenarios. 
Thus, the standard implementations of both frameworks are badly behaved
beyond their regime of validity, where they should actually become 
negligible.

In this work we address these issues, and we perform a detailed comparison between
different possible implementations of such a unified model, as well as with
numerical simulations, as follows. We first consider the computation of the
density power spectrum from a Lagrangian point of view in
Sect.~\ref{Decomposition}. This provides a decomposition over 2-halo and
1-halo terms that are slightly different from the versions found in previous
works, and we emphasize their relationship with ``perturbative'' and
``non-perturbative'' terms.
In particular, we explain why the 1-halo contribution contains a
counterterm, missed in previous studies, that ensures that it becomes truly
negligible on large scales, solving the second point ii) above.
Next, we present a simple implementation of this model in
Sect.~\ref{implementation}, and we explain how the point i) above is solved
by using resummation schemes instead of the standard perturbation theory.
We describe in Sect.~\ref{N-body-simulations} the N-body simulations that we
use to evaluate the accuracy of our models, and we perform detailed comparisons
for the nonlinear density power spectrum in Sect.~\ref{Comparison}, from
$k \sim 0.02 h$Mpc$^{-1}$ up to $k \sim 100 h$Mpc$^{-1}$, and for
redshifts $z=0.35$ up to $z=3$.
Then, we investigate in Sect.~\ref{ingredients} the dependence of our results
on various ingredients of the model, such as the properties of virialized halos
or the choice of the perturbative resummation scheme. We also evaluate
the importance of the 1-halo counterterm on large scales.
Next, we consider the real-space two-point correlation function in
Sect.~\ref{Real-space}.
Finally, we estimate in Sect.~\ref{accuracy} the accuracy of such unified models,
for both the power spectrum and the two-point correlation, and we conclude in
Sect.~\ref{Conclusion}.

In this paper we neglect the impact of baryonic physics on the matter power
spectrum, or more precisely we do not explicitly address this issue.
Thus, we assume that it can be incorporated in an effective manner through
the choice of the halo density profile that enters the model.

\section{Decomposition of the density power spectrum from a
Lagrangian point of view}
\label{Decomposition}

We show in this section how the density power spectrum can be split into
``perturbative'' and ``non-perturbative'' terms, and how they are related
to the 2-halo and 1-halo terms, from a Lagrangian point of view.

\subsection{``Perturbative'' and ``non-perturbative'' terms}
\label{Perturbative-and-non-perturbative}

Let us recall that in a Lagrangian framework one considers the trajectories $\vx(\vq,t)$
of all particles, of initial Lagrangian coordinates $\vq$ and Eulerian coordinates
$\vx$ at time $t$. In particular, at any given time $t$, this defines a mapping,
$\vq\mapsto\vx$, from Lagrangian to Eulerian space, which fully determines the
Eulerian density field $\rho(\vx)$ through the conservation of matter,
\beq
\rho(\vx) \, \dd \vx = \rhob \, \dd\vq ,
\label{continuity}
\eeq
where $\rhob$ is the mean comoving matter density of the Universe and we work
in comoving coordinates.
Then, defining the density contrast as
\beq
\delta(\vx,t) = \frac{\rho(\vx,t)-\rhob}{\rhob} ,
\label{deltadef}
\eeq
and its Fourier transform as 
\beq
\tdelta(\vk) = \int\frac{\dd\vx}{(2\pi)^3} \, e^{-\ii\vk\cdot\vx} \, \delta(\vx) ,
\label{tdelta}
\eeq
we obtain from Eq.(\ref{continuity})
\beq
\tdelta(\vk) = \int\frac{\dd\vq}{(2\pi)^3} \, \left( e^{-\ii\vk\cdot\vx(\vq)} 
- e^{-\ii\vk\cdot\vq} \right) .
\label{tdelta-q}
\eeq
Next, defining the density power spectrum as
\beq
\lag \tdelta(\vk_1) \tdelta(\vk_2) \rag = \delta_D(\vk_1+\vk_2) P(k_1) ,
\label{Pkdef}
\eeq
we obtain from Eq.(\ref{tdelta-q}), using statistical homogeneity,
\beq
P(k) = \int\frac{\dd\vq}{(2\pi)^3} \, \lag e^{\ii \vk \cdot \Delta\vx} - 
e^{\ii\vk\cdot\vq} \rag ,
\label{Pkxq}
\eeq
where we introduced the Eulerian-space separation $\Delta\vx$, 
\beq
\Delta\vx = \vx(\vq) - \vx(0) .
\label{Delta-x}
\eeq
The expression (\ref{Pkxq}) is fully general since it is a simple consequence
of the matter conservation Eq.(\ref{continuity}), and of statistical homogeneity.
In particular, it holds for any dynamics, such as the one associated with the
Zeldovich approximation \citep{Zeldovich1970}. There, the mapping $\vx(\vq)$
is given by the linear displacement field $\vPsi_L$, as $\vx(\vq)=\vq+\vPsi_L(\vq)$,
so that for Gaussian initial conditions (i.e. $\vPsi_L(\vq)$ is Gaussian) one can
easily compute the average (\ref{Pkxq}) and derive an explicit analytic expression
for the density power spectrum \citep{Schneider1995,Taylor1996,Valageas2010a}.

The second term in Eq.(\ref{Pkxq}), $e^{\ii\vk\cdot\vq}$, merely gives
a Dirac factor, $\delta_D(\vk)$, and can often be discarded in exact or systematic
computations (e.g., perturbative expansions) if one consider $k\neq 0$.
Within perturbative expansions of $P(k)$ over powers of the linear power spectrum,
$P_L(k)$, it simply cancels out the zeroth-order term so that $P(k) = P_L(k)$ at
lowest order. However, as we shall see below, it will prove important
in the following analysis to keep track of this factor. Indeed, it will be split over
two parts, along with the first term $e^{\ii \vk \cdot \Delta\vx}$, and it is
clear that one should either discard it or keep it in a consistent manner in both parts,
contrary to what is implicitly done in usual versions of the halo model.

For the three-dimensional gravitational dynamics it is not possible to
derive an explicit analytical expression for the average  (\ref{Pkxq}).
Therefore, as in the Eulerian framework, one must resort to perturbative
expansions \citep{Bouchet1995,Matsubara2008}
or phenomenological models.
In this paper, our goal is precisely to use Eq.(\ref{Pkxq}) as a starting point
to decompose the density power spectrum, $P(k)$, over two parts, associated
with ``perturbative'' and ``non-perturbative'' contributions.

Here and in the following, we call ``perturbative'' those contributions that
can be obtained from perturbative expansions over powers of the linear
power spectrum $P_L(k)$ within the fluid approximation. 
In practice, these may be derived from the equations of motion, written in terms
of the density and velocity fields in the usual hydrodynamical description of the
system, by writing these nonlinear Eulerian fields as perturbative series over
powers of the linear growing mode \citep{Goroff1986,Bernardeau2002}.
Partial resummations of this standard perturbative series have been recently
introduced \citep{Crocce2006a,Crocce2006b,Valageas2007a,Matarrese2007,Taruya2008,Valageas2008,Pietroni2008} but remain within this hydrodynamical regime.
Of course, it is possible to develop perturbative expansions, and associated
resummations, in Lagrangian space \citep{Bouchet1995,Matsubara2008},
by looking for an expansion of the displacement field $\vPsi(\vq)$ over
powers of the linear displacement field.
Taken at face value, these schemes imply some shell crossing. For instance, at
lowest order one recovers the Zeldovich dynamics for the displacement field.
However, this does not mean that they take into account in a systematic
fashion the physics beyond shell crossing, which remains beyond the reach
of current approaches of this kind. Technically, the gravitational potential
is obtained from the Poisson equation, where the density contrast is obtained
from the conservation of matter (\ref{continuity}) as
$\delta(\vx) = |\det (\pl\vx/\pl\vq)|^{-1}-1$, and one expands the Jacobian
over powers of the displacement $\vPsi$. In doing so, one discards the absolute
value and misses the changes of sign that occur after shell crossing.
On the other hand, these schemes give the same final expansion over powers of
$P_L$ for the nonlinear power spectrum as the Eulerian schemes, up to the
order of truncation of the computation.
Therefore, we consider both perturbative approaches as ``perturbative'' schemes,
in the sense of expansions over powers of $P_L$ that neglect shell crossing.
By contrast, we call ``non-perturbative'' those contributions that cannot be
obtained through these approaches, either because they are beyond the reach of
perturbative expansions (such as factors of the form $e^{-1/P_L}$ that cannot
be uncovered by Taylor series over $P_L$) or because they arise from the behavior
of the system beyond shell crossing (and require going beyond the single-stream
approximation).

In \citet{Valageas2010a} such a decomposition between perturbative and
non-perturbative contributions to the density power spectrum was obtained
within the framework of the Zeldovich dynamics. More precisely, a ``sticky model''
that only differs from the Zeldovich dynamics beyond shell crossing was introduced.
Then, whereas the Zeldovich density power spectrum is exactly given by a
resummation of perturbative terms (i.e., the non-perturbative contribution
is zero), the ``sticky model'' power spectrum contains an additional
non-perturbative contribution, while its perturbative part is identical to
the one of the Zeldovich dynamics.
Inspired by this simpler example, we perform a similar decomposition of
Eq.(\ref{Pkxq}) over perturbative and non-perturbative contributions.
However, we no longer have a simple criterion to distinguish particle
pairs that have undergone shell crossing, because for the gravitational dynamics
even at the perturbative level (at third order) the displacement field develops
rotational terms \citep{Buchert1994,BernardeauVal2008}. Therefore, we use
a phenomenological halo model, in the sense that we assume that the
matter distribution can be described as a collection of spherical halos defined
by a given nonlinear density contrast $\delta$. As usual, we can write the
halo mass function as
\beq
n(M) \dd M = \frac{\rhob}{M} f(\nu) \frac{\dd\nu}{\nu} , \;\; \mbox{with} \;\;
\nu = \frac{\delta_L}{\sigma(M)} ,
\label{nM-def}
\eeq
in terms of a scaling function $f(\nu)$ and of the reduced variable $\nu$.
Here, $\sigma(M)$ is the rms linear density contrast at scale $M$, or
Lagrangian radius $q_M$, with
\beq
\sigma(M) = \sigma(q_M)  \;\; \mbox{with} \;\; M= \rhob \frac{4\pi}{3} q_M^3 ,
\label{Mq}
\eeq
and
\beq
\sigma^2(q) = 4\pi \int_0^{\infty} \dd k \, k^2 P_L(k) \tW(kq)^2 ,
\label{sigma2-def}
\eeq
where $\tW(kq)$ is the Fourier transform of the top-hat of radius $q$, defined as
\beq
\tW(kq) = \int_V\frac{\dd \vq}{V} \, e^{\ii \vk\cdot\vq}
= 3 \, \frac{\sin(kq)-kq\cos(kq)}{(kq)^3} .
\label{tWdef}
\eeq
In the second Eq.(\ref{nM-def}) the linear density contrast $\delta_L$ is related
to the nonlinear density threshold $\delta$ that defines the halos through the
spherical collapse dynamics \citep{Valageas2009d},
\beq
\delta = \cF(\delta_L) ,
\label{FdeltaL-def}
\eeq
and for Gaussian initial conditions the large-mass tail shows the exponential
falloff
\beq
\nu \rightarrow \infty : \;\; \ln[f(\nu)] \sim -\frac{\nu^2}{2} ,
\label{nu-tail}
\eeq
whence
\beq
M \rightarrow \infty : \;\; n(M) \sim e^{-\delta_L^2/(2\sigma^2(M))} .
\label{M-tail}
\eeq
This corresponds to the Press-Schechter falloff \citep{Press1974}, but with
a somewhat lower threshold $\delta_L$ given by Eq.(\ref{FdeltaL-def})
(the usual Press-Schechter threshold actually corresponds to
$\delta_L = \cF^{-1}(\infty)$, associated with full collapse to a point).
However, the relation (\ref{FdeltaL-def}) only holds for $\delta < \delta_{\rm vir}$,
which typically gives $\delta< 200$, as larger density contrasts are associated
with inner shells where shell crossing plays a key role and modifies
Eq.(\ref{FdeltaL-def}), associated with spherical dynamics at constant mass
\citep{Valageas2009d}.
On the other hand, the halo mass function satisfies the normalization
\beq
\int_0^{\infty} \frac{\dd \nu}{\nu} \, f(\nu) =1 ,
\label{fnu-norm}
\eeq
which ensures that all the mass is contained within such halos (for linear
power spectra such that $\sigma(q)$ grows to infinity on small scales).
This also ensures that there is no overcounting (as would be the case if one
used a mass function with a normalization greater than unity).

Then, in Lagrangian space, the probability $\dd F$ for one particle $\vq_1$ to belong
to a halo of mass in the range $[M,M+\dd M]$ reads as
\beq
\dd F = f(\nu) \frac{\dd\nu}{\nu} .
\label{dF-def}
\eeq
This is also the fraction of matter enclosed within such halos. Next, making the
approximation that each halo comes from an initial spherical region in
Lagrangian space, the probability for a second particle $\vq_2$, at distance
$q=|\vq_2-\vq_1|$, to belong to the same halo reads as
\beq
F_M(q) = \frac{\int_V \dd\vq_1 \int \dd\vOmega_q \, \theta(\vq_2\in V)}
{\frac{4\pi}{3} q_M^3 \, 4\pi} .
\label{F_M-def}
\eeq
Here we integrate over all positions $\vq_1$ within the spherical volume $V$
of radius $q_M$, and we integrate over all directions $\vOmega_q$ of the
Lagrangian vector $\vq=\vq_2-\vq_1$. The top-hat factor $\theta(\vq_2\in V)$ is
unity if $\vq_2$ is located within the volume $V$, and zero otherwise.
By isotropy the result only depends on the length $q$, and performing the
integrations yields
\beq
0 \leq q \leq 2 q_M : \;\; F_M(q) = \frac{(2 q_M-q)^2 (4 q_M+q)}{16 q_M^3} ,
\label{F_M-1}
\eeq
and $F_M(q)=0$ for $q>2 q_M$. Therefore, combining Eqs.(\ref{dF-def}) and
(\ref{F_M-1}), we obtain the probability that a pair of particles of Lagrangian
separation $q$ belongs to the same halo of mass $M$ as
\beq
q_M > \frac{q}{2} : \;\; \dd F_q(M) =  \frac{(2 q_M-q)^2 (4 q_M+q)}{16 q_M^3} \,
f(\nu) \frac{\dd\nu}{\nu} ,
\label{Fq-M-def}
\eeq
and $\dd F_q(M)=0$ for $q_M<q/2$. In particular, the probability that the pair
$\{\vq_1,\vq_2\}$ belongs to a single halo writes as
\beq
\FoH(q) = \int_{\nu_{q/2}}^{\infty} \frac{\dd\nu}{\nu} \, f(\nu) \,
\frac{(2 q_M-q)^2 (4 q_M+q)}{16 q_M^3} ,
\label{F1H-def}
\eeq
where $\nu_{q/2}$ is defined as in Eq.(\ref{nM-def}), for the Lagrangian radius
$q_M=q/2$. On the other hand, the probability $\FtH(q)$ that the Lagrangian pair 
does not belong to a single halo (whence the two points belong to two different
halos) reads as
\beq
\FtH(q) = 1-\FoH(q) .
\label{F2H-def}
\eeq

Then, we split the average in Eq.(\ref{Pkxq}) over two terms, $\PoH$ and $\PtH$,
associated with pairs $\{0,\vq\}$ that belong to a single halo or to two different
halos,
\beq
P(k) = \PoH(k) + \PtH(k) ,
\label{Pk-halos}
\eeq
with
\beq
\PoH(k) = \int\frac{\dd\vq}{(2\pi)^3} \, \FoH(q) \, \lag e^{\ii \vk \cdot \Delta\vx} - 
e^{\ii\vk\cdot\vq} \rag_{1\rm H}
\label{Pkxq-1H}
\eeq
and
\beq
\PtH(k) = \int\frac{\dd\vq}{(2\pi)^3} \, \FtH(q) \, \lag e^{\ii \vk \cdot \Delta\vx} - 
e^{\ii\vk\cdot\vq} \rag_{2\rm H} .
\label{Pkxq-2H}
\eeq
Here the averages $\lag ... \rag_{1\rm H}$ and $\lag ... \rag_{2\rm H}$
are the conditional averages, knowing that the pair of length $q$ belongs to
a single halo or to two halos. The decomposition (\ref{Pk-halos}) clearly corresponds
to the 1-halo and 2-halo terms of the usual halo model \citep{Cooray2002}.
Then, to make the connection with the distinction between perturbative and
non-perturbative terms, we note that at a perturbative level $\FoH$ is
identically zero and $\FtH$ unity,
\beqa
\FoH= 0 & \mbox{and} & \FtH= 1 , \nonumber \\
&& \mbox{at all orders of perturbation theory} . 
\label{F1H-F2H-pert}
\eeqa
Indeed, the large-mass tail (\ref{M-tail}) is actually a rare-event limit that holds
both in the large-mass limit, at fixed linear density power spectrum 
\citep{Valageas2002IV,Valageas2009d}, and in the quasi-linear limit at fixed mass,
where the amplitude of the linear density power spectrum goes to zero
\citep{Valageas2002II,Valageas2009d}. It is this second regime which corresponds
to usual perturbation theories, where as recalled above we look for expansions
over powers of the amplitude of the linear power spectrum.
Then, because of the exponential decay of the form $e^{-1/\sigma^2(M)}$ we
can see that the expansion over powers of $P_L$ of $\FoH(q)$ defined in
Eq.(\ref{F1H-def}), at fixed $q$, is identically zero.
From Eq.(\ref{F2H-def}) this also yields $\FtH=1$ at all orders of perturbation theory.
Therefore, we can see from Eqs.(\ref{Pkxq-1H})-(\ref{Pkxq-2H}) that the
1-halo contribution is a fully non-perturbative term, while the 2-halo
contribution is (almost) the perturbative term multiplied by the factor
$\FtH(q)$.

The factor $\FoH$ being non-perturbative
is not mainly related to shell crossing, but simply to the fact that it cannot be
recovered by a series expansion over powers of $P_L$. 
However, for halos that would be defined by a high density threshold, typically
$\delta > 200$, the exponential falloff (\ref{M-tail}) is modified by shell crossing
(i.e. the factor $\delta_L$ is no longer given by Eq.(\ref{FdeltaL-def}), see
\citet{Valageas2009d}) so that it would also be non-perturbative in this sense.
On the other hand, even for lower threshold $\delta$, the exact form of the factor
$\FoH(q)$, and in particular the low-mass tail of the halo mass function, depends
on the behavior of the system beyond shell crossing.

The decomposition (\ref{Pk-halos}), with the Lagrangian-based interpretation
(\ref{Pkxq-1H})-(\ref{Pkxq-2H}), has the advantage to automatically satisfy the
conservation of matter. Thus, thanks to Eq.(\ref{F2H-def}) we count all particle
pairs once. By contrast, in the Eulerian derivation of the halo model, where we
first write the density field $\rho(\vx)$ as a sum of halo profiles, we would need
to pay attention to possible overlaps between halos, which arise when we use 
a spherical approximation.
Thus, the splitting (\ref{Pk-halos}) is more easily expressed in this framework,
and one can independently focus on the modelization of the averages
$\lag e^{\ii \vk \cdot \Delta\vx} -  e^{\ii\vk\cdot\vq} \rag$. The latter also offers
a closer link to the dynamics, through the mapping $\vq\mapsto\vx$.
We shall not make much use of this relationship in the following, as we 
use simple approximations that allow us to recover the usual Eulerian expressions,
with the addition of simple prefactors and counterterms, but this may provide
a route to more accurate modeling.

\subsection{``2-halo'' contribution}
\label{2-halo}

We first consider the 2-halo contribution (\ref{Pkxq-2H}).
Since the conditional average involves the constraint that the pair $\{0,\vq\}$ does
not belong to the same halo, the mean of $e^{\ii \vk \cdot \Delta\vx}$ is ``biased''
as compared with a mean over all possible pairs. However, in order to simplify
the computation of this term, we note that at a perturbative level
$\FtH= 1$, as seen in (\ref{F1H-F2H-pert}), so that we can replace the average
$\lag ... \rag_{2\rm H}$ by the mean over all pairs, as given by perturbation
theory,
\beq
\PtH(k) \simeq \int\frac{\dd\vq}{(2\pi)^3} \, \FtH(q) \, \lag e^{\ii \vk \cdot \Delta\vx} - 
e^{\ii\vk\cdot\vq} \rag_{\rm pert} ,
\label{Pkxq-pert1}
\eeq
where the subscript ``pert'' denotes quantities obtained from standard perturbation
theory (or equivalently its various resummation schemes).
Thus Eq.(\ref{Pkxq-pert1}) is still exact at all orders of perturbation theory.
This expression is best suited for perturbation theories developed within the
Lagrangian framework. Unfortunately, as we shall discuss in Sect.~\ref{Lagrangian}
below, Lagrangian perturbation theories built so far are not as efficient as their
Eulerian counterparts, especially when we consider available resummation schemes.
Then, in order to make contact with Eulerian perturbation theories we further
approximate Eq.(\ref{Pkxq-pert1}) as
\beqa
\PtH(k) & \simeq & \FtH(1/k) \int\frac{\dd\vq}{(2\pi)^3} \,
\lag e^{\ii \vk \cdot \Delta\vx} - e^{\ii\vk\cdot\vq} \rag_{\rm pert} \nonumber \\
& = & \FtH(1/k) P_{\rm pert}(k) ,
\label{Pkxq-pert2}
\eeqa
where we have replaced the $q-$dependent factor $\FtH(q)$ by its value at
a typical scale $q\sim 1/k$.
Again, this is legitimate at a perturbative level, where $\FtH=1$, so that
Eq.(\ref{Pkxq-pert2}) remains exact at all orders of perturbation theory.
To obtain the second line we simply used the exact expression (\ref{Pkxq}),
which implies the same equality in terms of perturbative expansions.

Let us recall that from (\ref{F1H-F2H-pert}) the 1-halo contribution is
zero at all orders of perturbation theory, so that the full power spectrum $P(k)$
of Eq.(\ref{Pk-halos}) automatically agrees with perturbation theory at all orders,
whether we use Eq.(\ref{Pkxq-pert1}) or Eq.(\ref{Pkxq-pert2}).
Then, within these approximations the only effect of non-perturbative corrections
to the 2-halo term is to multiply the perturbative power spectrum
$P_{\rm pert}(k)$ by the prefactor $\FtH(1/k)$. 

Within the usual halo model  the 2-halo term reads as \citep{Cooray2002}
\beqa
P_{2\rm H}^{\rm h.m.}(k) & = & \int \frac{\dd\nu_1 \dd\nu_2} {\nu_1\nu_2}
f(\nu_1) f(\nu_2) \tu_{M_1}(k) \tu_{M_2}(k) P_{M_1M_2}(k) \nonumber \\
& \simeq & P_L(k) \label{P2H-H1} ,
\eeqa
where $\tu_M(k)$ is the normalized halo density profile, defined in Eq.(\ref{uM-k-def})
below, and $P_{M_1M_2}(k)$ is the halo power spectrum.
The second line (\ref{P2H-H1}) is obtained in the low-$k$ limit, so that
$\tu_M(k) \rightarrow 1$ and $P_{M_1M_2}(k) \simeq b(M_1) b(M_2)
P_L(k)$, with a halo bias $b(M)$ that is normalized to unity.
It is also possible to combine perturbation theory and nonlinear halo bias to make
the expression above consistent with standard perturbation theory while
building a model for the halo power spectrum itself, see \citet{Smith2007}.

Of course, in order to describe the weakly nonlinear regime one can as well
replace $P_L(k)$ by $P_{\rm pert}(k)$ in Eq.(\ref{P2H-H1}), which gives
an expression very similar to Eq.(\ref{Pkxq-pert2}).
Then, we can see that a first difference between the halo-model expression
(\ref{P2H-H1}) and Eq.(\ref{Pkxq-pert2}) is that we did not need to introduce any halo
bias to derive Eq.(\ref{Pkxq-pert2}). In fact, although the contribution (\ref{Pkxq-2H})
is associated with a 2-halo term, by making the simple approximation
(\ref{Pkxq-pert1}) we avoid any need to consider halo biasing, and as explained
above this does not spoil the agreement with perturbation theory.
The second difference is the prefactor $\FtH(1/k)$ in Eq.(\ref{Pkxq-pert2}).
As seen from Eq.(\ref{Pkxq}) and the splitting (\ref{Pk-halos}), this term is
required by self-consistency, to ensure that the two averages
$\FoH(q) \, \lag e^{\ii \vk \cdot \Delta\vx}\rag_{1\rm H}$ and
$\FtH(q) \, \lag e^{\ii \vk \cdot \Delta\vx}\rag_{2\rm H}$ sum up to
$\lag e^{\ii \vk \cdot \Delta\vx}\rag$, and in particular that they sum up to
unity for $k \rightarrow 0$.
Within the usual halo model, this factor is implicitly set to unity by taking the
large-scale limit in Eq.(\ref{P2H-H1}) and ignoring exclusion constraints on the
halos. This is also valid within perturbation theory, as seen
in (\ref{F1H-F2H-pert}). However, in order to describe the weakly nonlinear
regime, where the 1-halo term is nonzero, it is best to keep the prefactor
$\FtH(1/k)$ in Eq.(\ref{Pkxq-pert2}), to keep a consistent model and to avoid
any overcounting.

\subsection{``1-halo'' contribution}
\label{1-halo}

From Eqs.(\ref{F1H-def}) and (\ref{Pkxq-1H}) the 1-halo contribution to the
density power spectrum reads as
\beqa
\PoH(k) & = & \int\frac{\dd\vq}{(2\pi)^3} \int_{\nu_{q/2}}^{\infty} \frac{\dd\nu}{\nu}
\, f(\nu) \, \frac{(2 q_M-q)^2 (4 q_M+q)}{16 q_M^3} \nonumber \\ 
&& \times \lag e^{\ii \vk \cdot \Delta\vx} -  e^{\ii\vk\cdot\vq} \rag_M .
\label{P-1H-M}
\eeqa
In order to compute the average in Eq.(\ref{P-1H-M}), within a halo of mass $M$,
we make the approximation of fully virialized spherical halos.
Thus, we describe the halos as spherical objects, truncated at a radius $r_M$
such that the mean density contrast within this radius is the nonlinear
threshold $\delta$ used to define these objects, and the enclosed mass is equal
to $M$,
\beq
M= \rhob \frac{4\pi}{3} q_M^3 = (1+\delta)  \rhob \frac{4\pi}{3} r_M^3 .
\label{rM-def}
\eeq
We introduce as usual the normalized Fourier transform of this halo radial
profile,
\beq
\tu_M(k) = \frac{\int\dd\vx \, e^{-\ii\vk\cdot\vx} \rho_M(x)}{\int\dd\vx \, \rho_M(x)}
= \frac{1}{M} \int\dd\vx \, e^{-\ii\vk\cdot\vx} \rho_M(x) ,
\label{uM-k-def}
\eeq
where $\rho_M(x)$ is the halo density profile.
Next, using the approximation of fully virialized halos, that is, that the two Lagrangian
particles ``0'' and ``$\vq$'' have lost all memory of their initial locations and
are independently located at random within the halo, we write
\beqa
\lag e^{\ii \vk \cdot \Delta\vx} \rag_M & = & \frac{1}{M^2} \int\dd\vx_1 \dd\vx_2 \,
\rho_M(\vx_1) \rho_M(\vx_2) \, e^{\ii\vk\cdot(\vx_2-\vx_1)}  \label{mean-M1} \\
& = & \tu_M(k)^2 . \label{mean-M2} 
\eeqa
Substituting into Eq.(\ref{P-1H-M}) and exchanging the order of integration gives
\beqa
\PoH(k) & = & \int_0^{\infty} \frac{\dd\nu}{\nu} f(\nu) \int_0^{2 q_M} 
\frac{\dd\vq}{(2\pi)^3} \, \frac{(2 q_M-q)^2 (4 q_M+q)}{16 q_M^3} \nonumber \\
&& \times \left( \tu_M(k)^2 - e^{\ii\vk\cdot\vq} \right) ,
\label{uM-k-2}
\eeqa 
and the integration over $\vq$ yields
\beq
\PoH(k) = \int_0^{\infty} \frac{\dd\nu}{\nu} f(\nu) \frac{M}{\rhob (2\pi)^3}
\left(  \tu_M(k)^2 - \tW(k q_M)^2 \right) .
\label{Pk-1H}
\eeq
Therefore, we recover the usual 1-halo term of the halo model \citep{Cooray2002},
with the addition of the new counterterm $\tW(k q_M)^2$, where $\tW$
was defined in Eq.(\ref{tWdef}).

This counterterm originates from the second term in Eq.(\ref{Pkxq}), which
actually sums up to a Dirac factor $\delta_D(\vk)$. However, since we
eventually split the nonlinear power spectrum as in Eq.(\ref{Pk-halos}) and we
use different approximations for the 2-halo and 1-halo contributions, it is best
to keep track of this factor in both contributions. In particular,  as pointed out
in Sect.~\ref{Perturbative-and-non-perturbative}, one should avoid keeping it
in one contribution and disregarding it in the other one. Thus, it has been
explicitly used in Eq.(\ref{Pkxq-pert2}) to cancel out the zeroth-order term
$F_{2\rm H} \delta_D(\vk)$, so that one should keep it in the 1-halo term
(\ref{Pkxq-1H}). Within the usual halo model this term is usually missed because
the splitting between the 2-halo and 1-halo terms is not treated so
carefully. In particular, as noticed in Sect.~\ref{2-halo}, the usual halo model
implicitly takes $\FtH=1$ (which is valid in perturbation theory) while taking
$\FoH \neq 0$ by adding the 1-halo term, which is not fully consistent.

In fact, the counterterm of Eq.(\ref{Pk-1H}) can be recovered without going through
the previous steps, associated with a Lagrangian point of view, by the following
simple argument. Let us consider a uniform universe, where there are no
density fluctuations and the density is everywhere equal to $\rhob$.
Then, within the spirit of the usual (Eulerian) halo model, and making the same
approximations
(e.g., neglecting geometrical constraints, associated with exclusion constraints, and
departures from spherical symmetry), we can consider that all the matter
is within halos of constant density $\rhob$ with an arbitrary distribution of
halo radii. That is, we can split this uniform system over an arbitrary set of
cells, which we can call halos. Then, within such a halo model the usual 
1-halo term for the full density correlation, $P^{\rho\rho}_{1\rm H}(k)$
(defined as $\lag\trho(\vk_1)\trho(\vk_2)\rag =\delta_D(\vk_1+\vk_2)
P^{\rho\rho}(k_1)$, as in Eq.(\ref{Pkdef})), reads as the counterterm of
Eq.(\ref{Pk-1H}), with a positive sign and multiplied by $\rhob^2$.
Indeed, the density profile of such halos is simply the top-hat of radius $q_M$,
since $\delta=0$ whence $r_M=q_M$, so that the normalized Fourier
transform of the halo radial profile defined in Eq.(\ref{uM-k-def}) is equal to
the Fourier transform (\ref{tWdef}), $\tu_M(k)= \tW(k q_M)$.
This means that within a halo model, for any halo mass function, the 1-halo
contribution to the
power spectrum associated with an uniform-density medium reads as 
the counterterm of Eq.(\ref{Pk-1H}). Since the density-contrast power spectrum
can also be defined as $\rhob^2 P(k) = P^{\rho\rho}(k)-P^{\rhob\,\rhob}(k)$
(i.e. we consider the connected two-point correlation), we must subtract
this uniform-density term as in Eq.(\ref{Pk-1H}), as the first term 
$\tu_M(k)^2$ corresponds to the total density power spectrum, $P^{\rho\rho}(k)$.

In agreement with this result, we can note by an integration over $\vk$ of the
second term of Eq.(\ref{uM-k-2}) that the second term of Eq.(\ref{Pk-1H}) is indeed
a well-normalized approximation to $\delta_D(\vk)$, where the Dirac peak
is broadened over a size $\sim 1/q_M$ associated with the typical halo scale
(thus, as could be expected, this becomes an increasingly good approximation
to $\delta_D(\vk)$ as the halo size goes to infinity and the associated power is
repelled to $k \rightarrow 0$).
Clearly, the fact that this does not yield exactly a Dirac factor, whence that it does not
vanish for nonzero $k$, is due to the fact that it is only a partial contribution to the
full power spectrum, since by construction within such halo models there is always
a 2-halo term, and it is only the sum of both contributions that contains such
a Dirac factor. (This also agrees with the fact that we recover a Dirac factor
in the limit $q_M \rightarrow \infty$, since at fixed scale $1/k$ the 2-halo
term clearly disappears in this limit).

In practice, the counterterm of Eq.(\ref{Pk-1H}) is small on most scales of interest,
especially in the nonlinear regime. However, it plays an important role on
very large scales. Indeed, since $\tu_M(0)=1$ we can see that the first term
of Eq.(\ref{Pk-1H}), associated with the usual halo model, goes to a nonzero
constant as $k\rightarrow 0$. Since for CDM-like initial power spectra
$P_L(k) \propto k^{n_s}$ with $n_s \simeq 1$ as $k\rightarrow 0$, this implies
that the usual 1-halo term dominates on very large scale, which is not correct.
In fact, from the conservation of matter a small-scale redistribution of matter
generically yields a $P(k)\propto k^2$ tail at low $k$, while taking into account
momentum conservation one obtains a $k^4$ tail \citep{Peebles1974}.
To solve this problem the use of compensated halo profiles (i.e. with
$\tu_M(0)=0$) was investigated in \citet{Cooray2002}. However, they noticed
that this recipe fails because it also spoils the 2-halo term, as seen
from the first line in Eq.(\ref{P2H-H1}).
Here we can see that this problem is automatically solved within our approach
by the counterterm $\tW(kq_M)^2$ in Eq.(\ref{Pk-1H}). In particular we recover
at low $k$ the expected slope $\PoH(k) \propto k^2$, associated with small-scale
rearrangements (indeed, the 1-halo term only describes the redistribution within
small clumps, and is not sensitive to the long-range correlations between
clumps, that is described by the 2-halo term).

Looking at the steps of the derivation of Eq.(\ref{Pk-1H}) we can also see why we
fail to recover the $k^4$ tail due to momentum conservation. Indeed, in the
approximation (\ref{mean-M2}), where we have assumed full virialization,
we have erased all memory of the initial positions and velocities of the
two Lagrangian particles $\{0,\vq\}$, which clearly implies that we have
disregarded the constraints associated with momentum conservation.
Therefore, in order to obtain a $k^4$ tail one should improve this approximation,
by taking care at some level of the particle velocities.
Of course, a simpler remedy would be to modify the counterterm
$\tW(kq_M)^2$ in Eq.(\ref{Pk-1H}), by using a window $\tW_2(kq_M)^2$
that cancels both the terms $k^0$ and $k^2$ of $\tu_M(k)^2$.
However, since the CDM linear power spectrum roughly behaves as
$P_L(k) \propto k$ at low $k$ the $k^2$ tail is sufficient to make the 1-halo term
subdominant on large scales, as we shall check in Sect.~\ref{Comparison} below.
Therefore, we shall keep the simple approximation (\ref{mean-M2}) in
the following, and Eq.(\ref{Pk-1H}) for the 1-halo term,
as it appears to be sufficient to reach a good agreement with
numerical simulations.

\subsection{Extension to the galaxy power spectrum}
\label{galaxy}

Although in this article we focus on the dark matter power spectrum, we briefly
discuss in this section how the previous results can be used for the galaxy power
spectrum, which is also a quantity of great practical interest.
This necessarily involves additional ingredients, as we must define the relationship
between the dark matter density field and the galaxies. As usual
\citep{Seljak2000,Cooray2002}, we keep the splitting (\ref{Pk-halos}) over 1-halo
and 2-halo terms. Then, we simply write the galaxy 2-halo contribution as
\beq
P^{\rm gal}_{2\rm H}(k) = \lag b\rag^2 \; \PtH(k) ,
\label{Pgal-2H}
\eeq
where $\lag b \rag$ is the mean bias of the galaxy population that we consider.
This can be computed using one of the bias models that have been proposed in
previous works \citep{MW1996,Sheth1999,SMT2001,Valageas2009d,Valageas2010c}
or fits to numerical simulations \citep{Tinker2010}.
Next, following the Eulerian interpretation of the counterterm of Eq.(\ref{Pk-1H})
discussed in the previous section, as arising from the difference
$\lag n^{\rm gal}n^{\rm gal}\rag - \lag n^{\rm gal} \rag \lag n^{\rm gal} \rag$,
we write the 1-halo term as
\beqa
P^{\rm gal}_{1\rm H}(k) & = & \int_0^{\infty} \frac{\dd\nu}{\nu} f(\nu) \, 
\frac{M}{\rhob (2\pi)^3} \, \frac{\rhob^2}{\overline{n}^2} \, \frac{\lag N(N-1)\rag}{M^2}
\nonumber \\
&& \times \left(  \tu_M(k)^p - \tW(k q_M)^p \right) ,
\label{Pgal-1H}
\eeqa
where $\overline{n}$ is the mean galaxy number density, $\lag N ( N - 1)\rag$ is the
mean squared number of galaxies (minus a factor $N$) in halos of mass $M$, and
$p=1$ or $2$ depending on whether the former quantity is smaller or greater than
unity \citep{Seljak2000,Cooray2002}. Thus, we have simply subtracted
to the usual expression the counterterm 
associated with galaxies set at random within the Lagrangian radius $q_M$
(i.e., corresponding to a uniform density universe).
Again, the quantity $\lag N (N-1)\rag$ must be computed from
a model defined for the population of galaxies that we consider.

The derivation of Eq.(\ref{Pgal-1H}) is more phenomenological than the derivation
of Eq.(\ref{Pk-1H}) from the Lagrangian point of view (\ref{P-1H-M}). However, since
one always needs to introduce some phenomenological model to relate galaxies to
the matter distribution (here through the two quantities $\lag b \rag(M)$ and
$\lag N(N-1)\rag(M)$), this should be sufficient.

The contribution (\ref{Pgal-1H}) to the galaxy power spectrum vanishes as
$k^2$ in the large-scale limit, which again solves the unphysical nonzero limit
obtained in previous implementations.
This behavior remains valid even though galaxies are discrete objects, since this
property only implies a constant shot-noise asymptote for $P^{\rm gal}(k)$ in the
high-$k$ limit. This shot-noise has actually been subtracted in Eq.(\ref{Pgal-1H}),
using the quantity $\lag N(N-1)\rag$ instead of $\lag N^2\rag$, to follow the
usual convention \citep{Peebles1980,Seljak2000,Cooray2002}.
At low $k$, we are dominated by the 2-halo contribution and in case of
constant large-scale bias we recover the slope $P^{\rm gal}(k) \propto k^{n_s}$
of the primordial matter power spectrum.\footnote{A simple illustration of a density
field made of discrete objects that obeys such behaviors is provided by the
adhesion model \citep{Gurbatov1989,Vergassola1994,BernardeauVal2010b}.
This yields a power spectrum with a universal $k^0$ slope at high $k$, because
of the pointlike density peaks, while the slope at low $k$ depends on the initial
conditions \citep{Valageas2010d}, and for the simple cases $n_s=0$ and $n_s=-2$
in 1D one can obtain its exact expression \citep{Valageas2009c,Valageas2009a}.}

\section{A simple implementation}
\label{implementation}

We describe in more detail in this section how we implement Eqs.(\ref{Pkxq-pert2})
and (\ref{Pk-1H}) to compute the nonlinear density power spectrum.

\subsection{``2-halo'' contribution}
\label{2-halo-1}

We first consider the 2-halo contribution (\ref{Pkxq-pert2}), and more
precisely the perturbative part $P_{\rm pert}(k)$, as we shall discuss
the non-perturbative prefactor $\FtH(1/k)$ in Sect.~\ref{1-halo-1} below.

Since we cannot compute and resum all terms of the perturbative expansion
of $P(k)$, contrary to the simpler Zeldovich dynamics, we must use for
$P_{\rm pert}(k)$ an approximation associated with a truncation at a finite order.
This leads to several possible choices, since a priori we can use the standard
perturbation theory \citep{Goroff1986,Bernardeau2002} as well as any of the
various resummation schemes that have been introduced in the recent years
\citep{Crocce2006a,Crocce2006b,Valageas2007a,Matarrese2007,Valageas2008,Taruya2008,Pietroni2008}.
Indeed, all such methods give expressions for $P_{\rm pert}(k)$ that agree up
to the order of the computation, and only differ by higher order terms.

However, it turns out that standard perturbation theory cannot suit our purposes
because higher order terms grow increasingly fast at high $k$
\citep{Bernardeau2002}. As we shall check in Sect.~\ref {Eulerian} below,
at one-loop order it already gives a steep contribution that remains non-negligible
at high $k$, because the prefactor $\FtH(1/k)$ does not go to zero very fast
(because for CDM power spectra $\sigma(q)$ only grows logarithmically at small
$q$ if $n_s=1$, and even remains finite if $n_s<1$).
This implies that to get a well-behaved 2-halo term we need to use
other schemes, that provide a perturbative part $P_{\rm pert}(k)$ that
remains small at high $k$ as compared with the 1-halo term.
Therefore, we must use one of the resummation schemes that have been
recently developed and show a well-behaved high-$k$ behavior.

It is interesting to point out that, although such approaches were devised
in order to improve the accuracy of perturbation theory on large scale
and to increase its range of validity (by improving its convergence through
partial resummations of an infinite number of higher order terms),
our study shows that a second important benefit of these schemes is to
provide well-behaved expressions for $P_{\rm pert}(k)$.
Here the point is not that the high-$k$ behavior is accurately reproduced
(since anyway perturbation theory does not apply in this range) but that
it remains within reasonable bounds (typically close to or smaller
than the linear power) so as to be subdominant.
Thus, by giving systematic expansions of $P_{\rm pert}(k)$, which agree
with perturbation theory up to the order of the computation while
remaining well-behaved at high $k$, these schemes allow the construction of
unified models such as those studied in this article, based on Eq.(\ref{Pk-halos}), 
that show a correct behavior at both low and high $k$.
(In the same spirit, we have checked in Sect.~\ref{1-halo} that the 1-halo
term is well-behaved at low $k$.)
This provides a second important motivation for such resummation schemes,
as compared with standard perturbation theory.

\begin{figure}[htb]
\begin{center}
\epsfxsize=8.5 cm \epsfysize=4.1 cm {\epsfbox{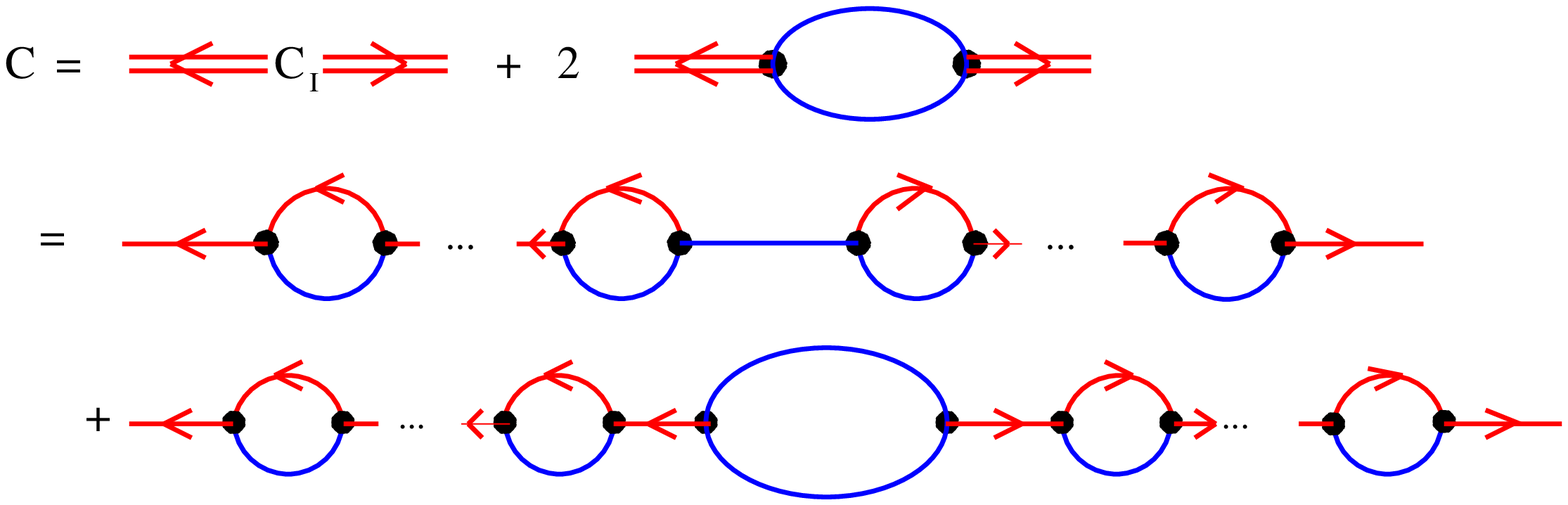}}
\end{center}
\caption{The resummation performed by the ``direct steepest-descent'' method
at one-loop order, for the two-point correlation $C$. The blue single lines
are the linear correlation $C_L$ and the red single lines are the linear response
(propagator) $R_L$. The double red lines in the first equality are the nonlinear
response $R$ (at that order), which contains an infinite series of bubble diagrams,
and explicit substitution gives the series of diagrams shown in the second
equality below.}
\label{fig_Csd}
\end{figure}

In this article we focus on the ``direct steepest-descent method'' introduced
in \citet{Valageas2007a}. In terms of diagrams it means that the two-point
correlation $C$ is given at one-loop order by the resummations shown in
Fig.~\ref{fig_Csd}, where single lines are the linear correlation and response
functions $C_L$ and $R_L$, while the double lines are the nonlinear
correlation and response functions $C$ and $R$ obtained at that order,
see also \citet{Valageas2008}.
At this one-loop order the standard perturbation theory consists in keeping
only the three diagrams with zero or one bubble among the infinite series
shown in the second equality in Fig.~\ref{fig_Csd}.
We recall the details of the computation of the (perturbative) density power spectrum
$P_{\rm pert}(k)$ within this approach in App.~\ref{direct-steepest-descent}.

This direct steepest-descent method is not necessarily the most accurate
resummation scheme. In particular, it yields a response function that does not
decay at high $k$ or late times, but shows increasingly fast oscillations with an
amplitude that follows the linear response function. This is not realistic,
since one expects a Gaussian-like decay for Eulerian response functions,
as can be seen from theoretical arguments and numerical simulations
\citep{Crocce2006a,Crocce2006b,Valageas2007b,BernardeauVal2010a}.
However, the fast oscillations still provide an effective damping in a weak
sense (that is when the response function is integrated over).
The reason why we consider this direct steepest-descent method here is that
it provides a simple and efficient method, which satisfies our requirements
and proves to be reasonably accurate, as we shall check in Sect.~\ref{Comparison}
below.
Indeed, while by construction it agrees with standard perturbation theory
at one-loop order, it is well-behaved at high $k$ as it remains close to the linear
power spectrum on all scales (when we truncate the computation at one-loop
order, as in this article), see \citet{Valageas2007a}.

It is also particularly efficient because the integrals involved in the computation
of the power spectrum factorize (because the bubble in the upper right diagram
of Fig.~\ref{fig_Csd} involves a product of linear two-point correlation functions,
whose time-dependence can be factorized).
This factorization can be clearly seen in Eq.(\ref{C-fact}), as the two-time correlation
$C(k;\eta_1,\eta_2)$ can be written as the sum of three products, each of them
being of the form $\varphi(\eta_1) \times \varphi(\eta_2)$. Therefore, there is
no need to compute two-dimensional time integrals, over both $\eta_1$ and
$\eta_2$.
A second property is that the integro-differential equations satisfied by the
nonlinear response $R(k;\eta_1,\eta_2)$ (which is needed at intermediate
stages to obtain the power spectrum) can be reduced to (third-order) ordinary
differential equations, as seen in Eqs.(\ref{R01diff})-(\ref{R02diff}).
This avoids the need to compute integrals over past times at each time step
over $\eta_1$.
These two properties allow a fast numerical computation, which can be
a useful feature for practical purposes.

\subsection{``1-halo'' contribution}
\label{1-halo-1}

We now turn to the 1-halo contribution (\ref{Pk-1H}), which requires a model
for the halo mass function and for the halo density profiles.
In this article we consider the scaling function $f(\nu)$ given by
\beq
f(\nu) = 0.502 \left[ (0.6\,\nu)^{2.5}+(0.62\,\nu)^{0.5} \right] \, e^{-\nu^2/2} .
\label{f-fit}
\eeq
It satisfies the normalization constraint (\ref{fnu-norm}) and it has already been
shown to provide a good match to numerical simulations for halos defined by
a density contrast $\delta=200$ \citep{Valageas2009d}.
Thus, substituting Eq.(\ref{f-fit}) into Eq.(\ref{nM-def}) we obtain the mass function
of these halos, with a linear threshold $\delta_L = \cF^{-1}(200)$ in the
second Eq.(\ref{nM-def}). At redshift $z=0$ this gives $\delta_L \simeq 1.59$
for a $\Lambda$CDM cosmology, and it increases slightly to $\delta_L \simeq 1.6$
at high $z$, see \citet{Valageas2009d} and the fit of Eq.(12) therein.
Next, the halo mass function also gives the probabilities $\FoH(q)$ and
$\FtH(q)$ defined in Eqs.(\ref{F1H-def})-(\ref{F2H-def}).
This yields in turn the prefactor $\FtH(1/k)$ that enters the 2-halo term
(\ref{Pkxq-pert2}).

For the halo density profile we use the usual NFW profile \citep{NFW1997},
\beq
\rho_M(x) = \frac{\rho_s}{x/r_s (1+x/r_s)^2} ,
\label{NFW-rho}
\eeq
which we truncate at the radius $r_{200}$, associated with the overall density
contrast $\delta=200$.
The scale radius $r_s$ is defined in terms of the outer radius $r_{200}$
through the concentration $c(M_{200})$,
\beq
c(M_{200}) = \frac{r_{200}}{r_s} ,
\label{cM-def}
\eeq
while the characteristic density is given by
\beq
\rho_s = \rhob \, \frac{201}{3} \, 
\frac{c^3}{\ln(1+c)-c/(1+c)} .
\label{rhos-def}
\eeq
Therefore, it remains to specify the concentration parameter as a function of the
halo mass, which we have labeled $M_{200}$ above to remind that it corresponds
to the mass defined by the nonlinear density contrast $\delta=200$.
Although we shall also consider fits to numerical simulations that have already
been proposed in the literature, we shall find in Sect.~\ref {Comparison}
below that a better match to numerical simulations for the density power spectrum
(especially in the high-$k$ tail) is obtained with the following prescription,
\beq
c(M_{200}) = 10.04 \, \left( \frac{M_{200}}
{2\times 10^{12} h^{-1} M_{\odot}}\right)^{-0.1} \, (1+z)^{-0.8} ,
\label{cM-1}
\eeq
which is intermediate between the behaviors found in \citet{Dolag2004} and
\citet{Duffy2008}. However, this should not be trusted beyond the range
studied in this paper ($z\leq 3$).

Since we obtain Eq.(\ref{cM-1}) from a comparison with the density
power spectrum measured in simulations, it can be understood as an
``effective'' concentration parameter that also includes the mean effect
of halo substructures. Although it is possible to build more sophisticated halo
models that explicitly describe substructures \citep{Giocoli2010}, we do
not investigate such refinements here. In the same spirit, it could be possible
to include the mean effect of baryons on the total matter profile of the halos
through such an ``effective'' concentration parameter (or another choice 
for the shape of the halo profile than the NFW formula (\ref{NFW-rho})).
However, this is also beyond the scope of the present study, and we leave
such investigations to future works.
In any case, it is clear that any modeling of halo profiles (either at an ``effective''
level or through detailed explicit models) can be incorporated in the framework
we describe in this paper. In particular, this could improve the accuracy at high
$k$ as simulations probe smaller scales and provide tighter (and more robust)
constraints on halo properties.

For practical purposes it is convenient to use halo profiles that have an explicit
expression for the Fourier transform $\tu_M(k)$ defined in Eq.(\ref{uM-k-def}),
especially if one is interested in high wavenumbers $k$ where the integrand
shows fast oscillations. Let us recall that for the NFW profile (\ref{NFW-rho})
one has \citep{Scoccimarro2001}
\beqa
\tu_M(k) & = &  \left [ \ln(1+c)-\frac{c}{1+c} \right]^{-1} 
\biggl \{ \frac{-\sin(ckr_s)}{(1+c)kr_s} \nonumber \\
&& + \cos(kr_s) \bigl [ \Ci[(1+c)kr_s] - \Ci(kr_s) \bigl ]  \nonumber \\
&& + \sin(kr_s) \bigl [ \Si[(1+c)kr_s] - \Si(kr_s) \bigl ] \biggl \} , 
\eeqa
where $\Ci(x)= -\int_x^{\infty} \cos t \, \dd t/t$ and
$\Si(x)= \int_0^x \sin t \, \dd t/t$ are the Cosine and Sine integrals
\citep{Abramowitz}.

\section{N-body simulations}
\label{N-body-simulations}

In this section we describe the details of our N-body simulations, and we show
the results of some tests of their possible systematic errors.

\begin{figure*}
\begin{center}
\epsfxsize=6.05 cm \epsfysize=6 cm {\epsfbox{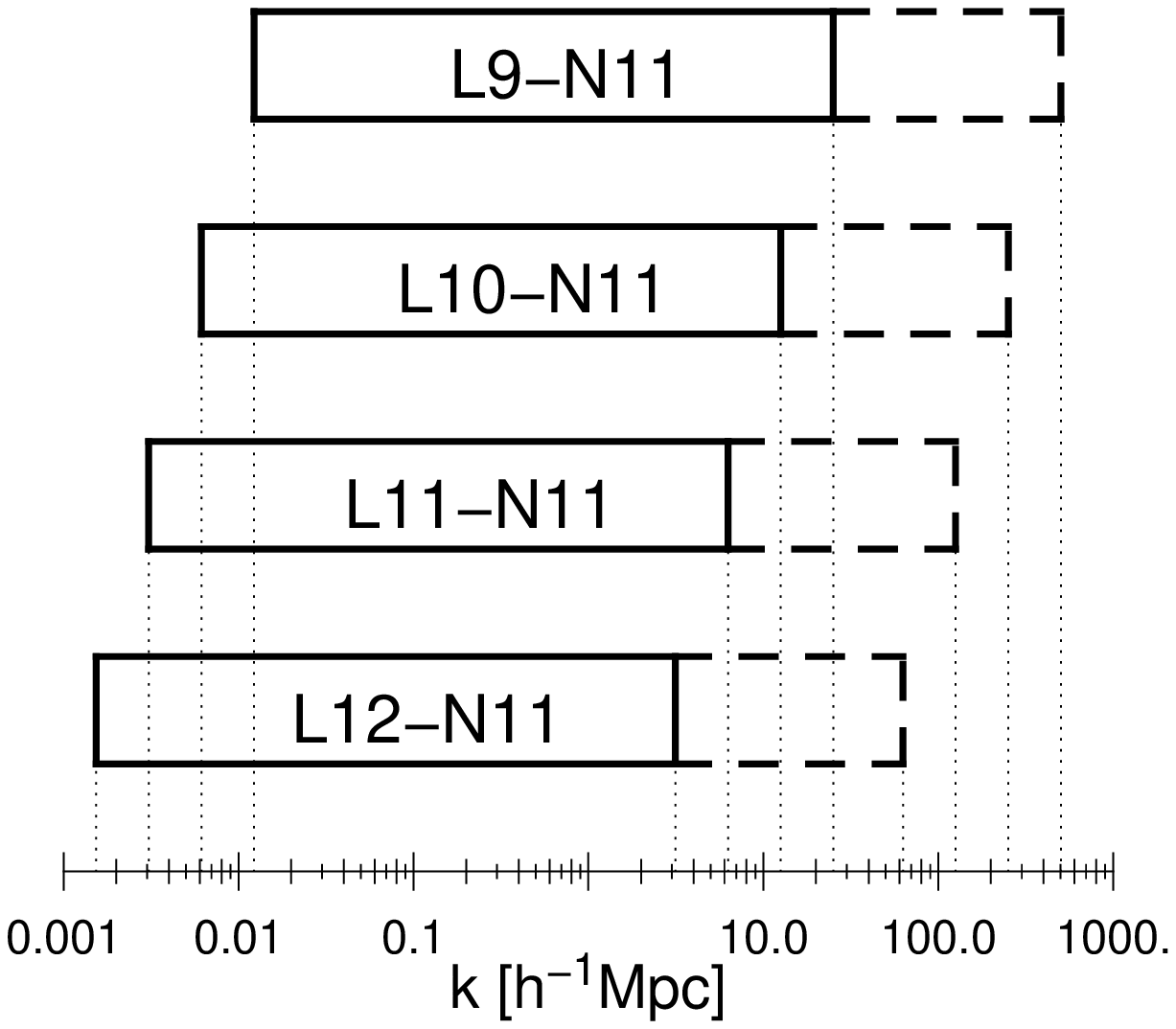}}
\epsfxsize=6.05 cm \epsfysize=6 cm {\epsfbox{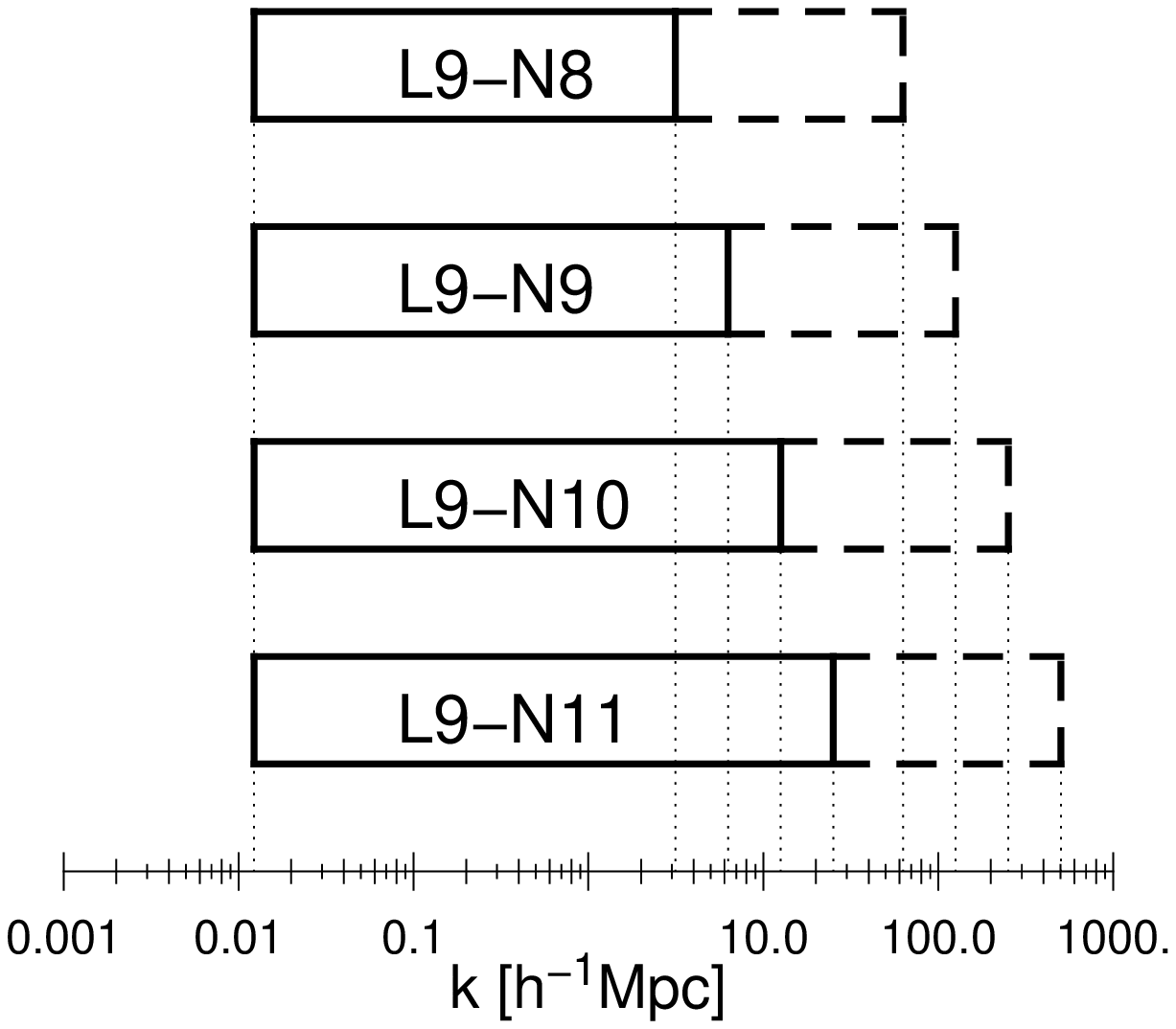}}
\epsfxsize=6.05 cm \epsfysize=6 cm {\epsfbox{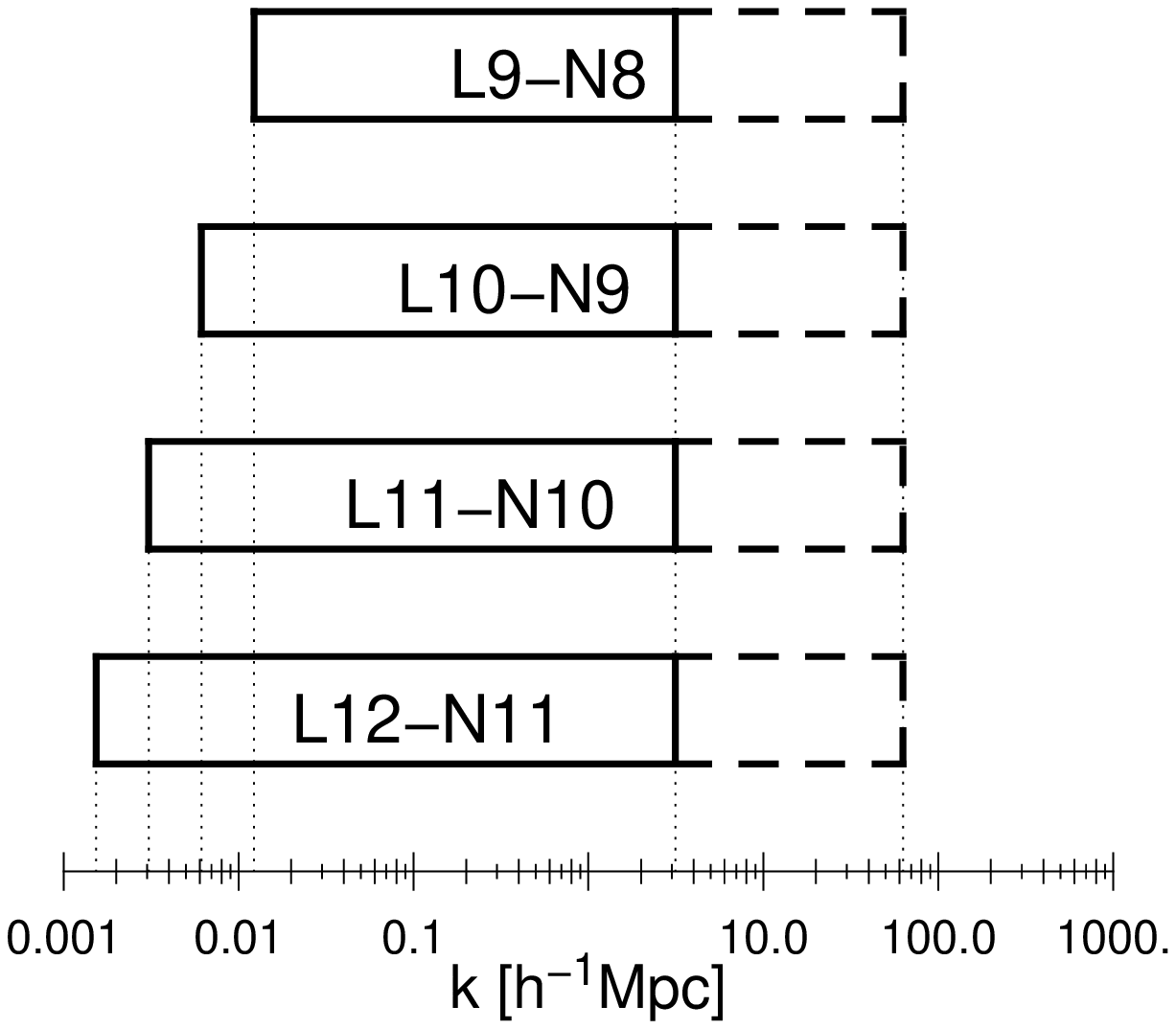}}
\caption{Wavenumber ranges covered by our N-body simulations, from the
fundamental mode (left edge) to the mean inter-particle scale (solid right edge)
and to the softening scale (dashed right edge).
{\it Left}: the four main simulations, {\it middle}: simulations used in Test 1, {\it right}: simulations used in Test 2.}
\label{fig:cover}
\end{center}
\end{figure*}

\subsection{Set up}
\label{subsec:setup}

We adopt a flat $\Lambda$CDM model with cosmological parameters  
$(\Omega_{\rm m},\Omega_{\rm b},h,\sigma_8,n_{\rm s})
= (0.279, 0.046035, 0.701, 0.817, 0.96)$,
which is consistent with 5 year observation of WMAP \citep{Komatsu2009}. 
We use a publicly available code, {\tt CAMB} \citep{Lewis2000}, to compute the
linear power spectrum.
All the initial conditions for the simulations are created by adding displacements
to particles put on a regular lattice by second-order Lagrangian perturbation theory
\citep{Scoccimarro1998a,Crocce2006c} at $z=99$, 
and evolved by a publicly available Tree Particle-Mesh code, {\tt GADGET2}
\citep{Springel2005}.
We employ $N=2048^3$ particles in four periodic boxes with different box sizes 
($L_{\rm box}=512, 1024, 2048$ and $4096\,h^{-1}$Mpc). 
We call these runs as {\tt L9-N11}, {\tt L10-N11}, {\tt L11-N11} and {\tt L12-N11},
respectively.
The softening lengths for the tree force are set to be $5\%$ of the mean
inter-particle distance.

In addition to the four main simulations, we run five smaller simulations 
({\tt L9-N8}, {\tt L9-N9}, {\tt L9-N10}, {\tt L10-N9} and {\tt L11-N10}) for convergence
tests. The box sizes and number of particles of these simulations are summarized
in Tab.~\ref{tab:list}.

\begin{table}[h!]
\caption{List of N-body simulations presented in this paper. The box size,
$L_{\rm box}$, is shown in units of $h^{-1}$Mpc.}
\begin{center}
\begin{tabular}{|c||c|c|c|c|c|}
\hline
name & $L_{\rm box}$ & $N^{1/3}$ & Main & Test1 & Test2 \\
\hline\hline
L9-N11 & 512 & 2048 & $\surd$ & $\surd$ &\\
\hline
L10-N11 & 1024 & 2048 & $\surd$ & &\\
\hline
L11-N11 & 2048 & 2048 & $\surd$ & &\\
\hline
L12-N11 & 4096 & 2048 & $\surd$ & & $\surd$\\
\hline
L9-N8 & 512 & 256 & & $\surd$ & $\surd$ \\
\hline
L9-N9 & 512 & 512 & & $\surd$ & \\
\hline
L9-N10 & 512 & 1024 & & $\surd$ & \\ 
\hline
L10-N9 & 1024 & 512 & & & $\surd$\\
\hline
L11-N10 & 2048 & 1024 & & & $\surd$\\
\hline
\end{tabular}
\end{center}
\label{tab:list}
\end{table}

\subsection{Measuring power spectrum and two-point correlation function}
\label{subsec:measure}

\begin{figure*}
\begin{center}
\epsfxsize=6.05 cm \epsfysize=6 cm {\epsfbox{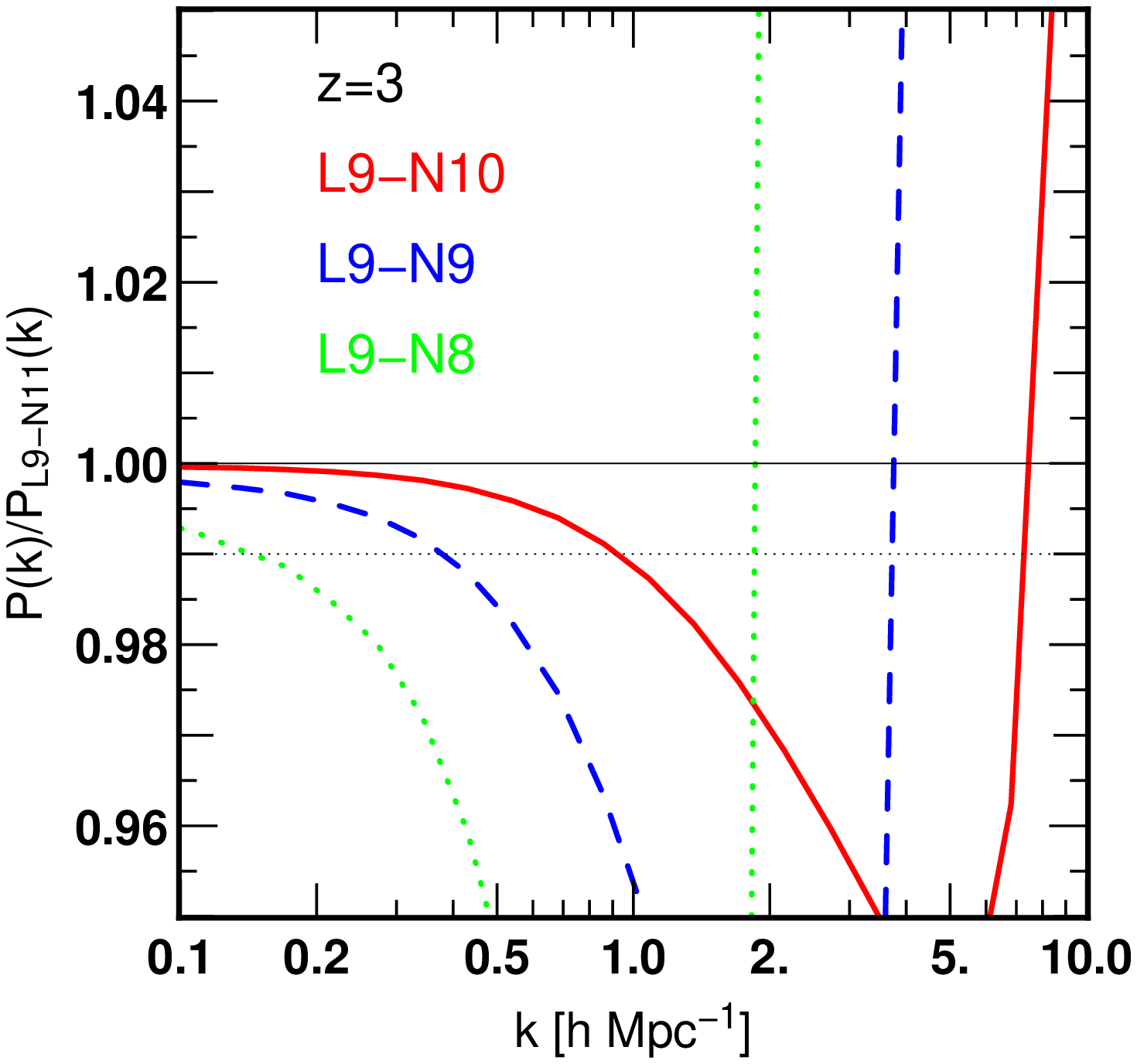}}
\epsfxsize=6.05 cm \epsfysize=6 cm {\epsfbox{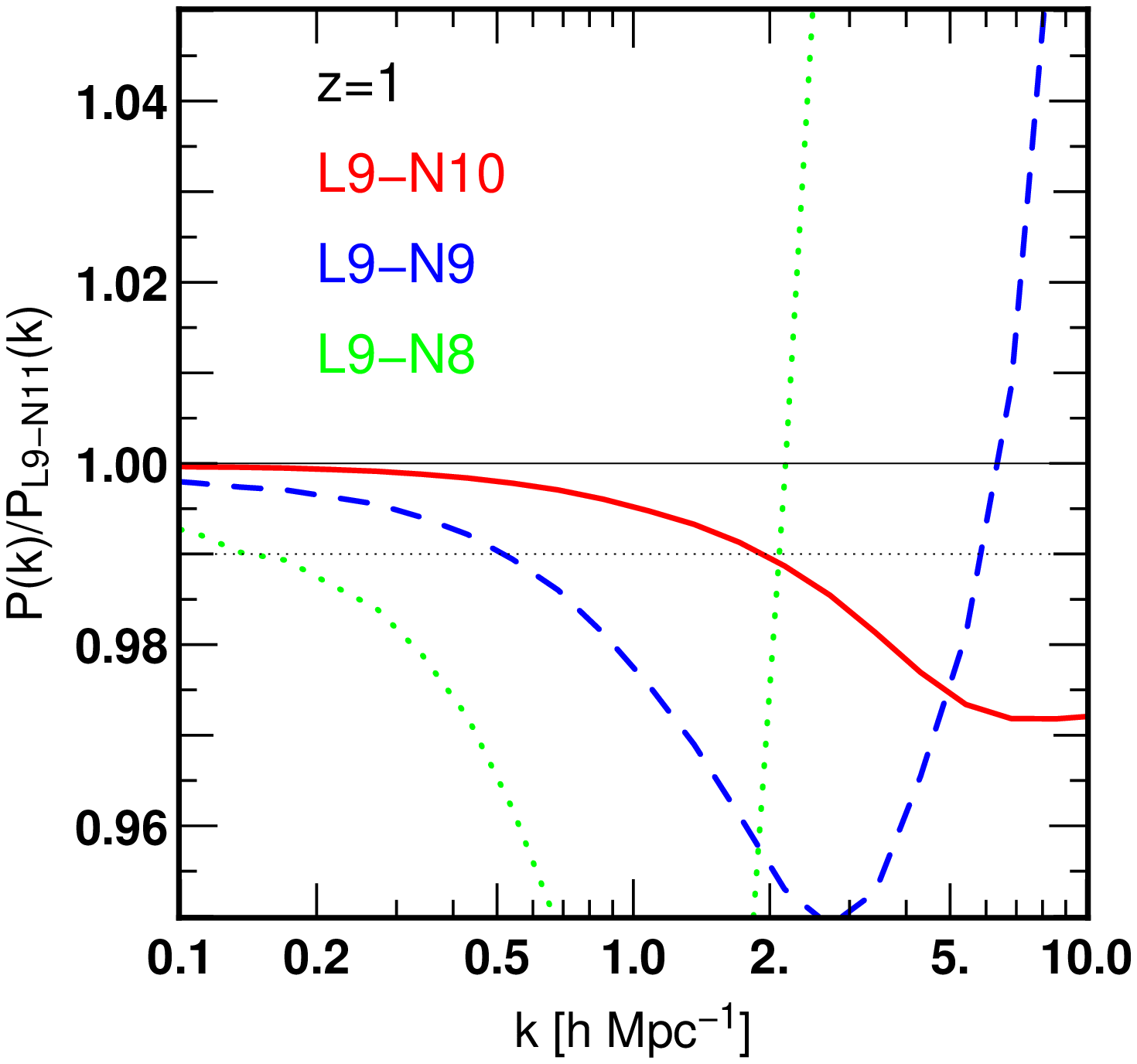}}
\epsfxsize=6.05 cm \epsfysize=6 cm {\epsfbox{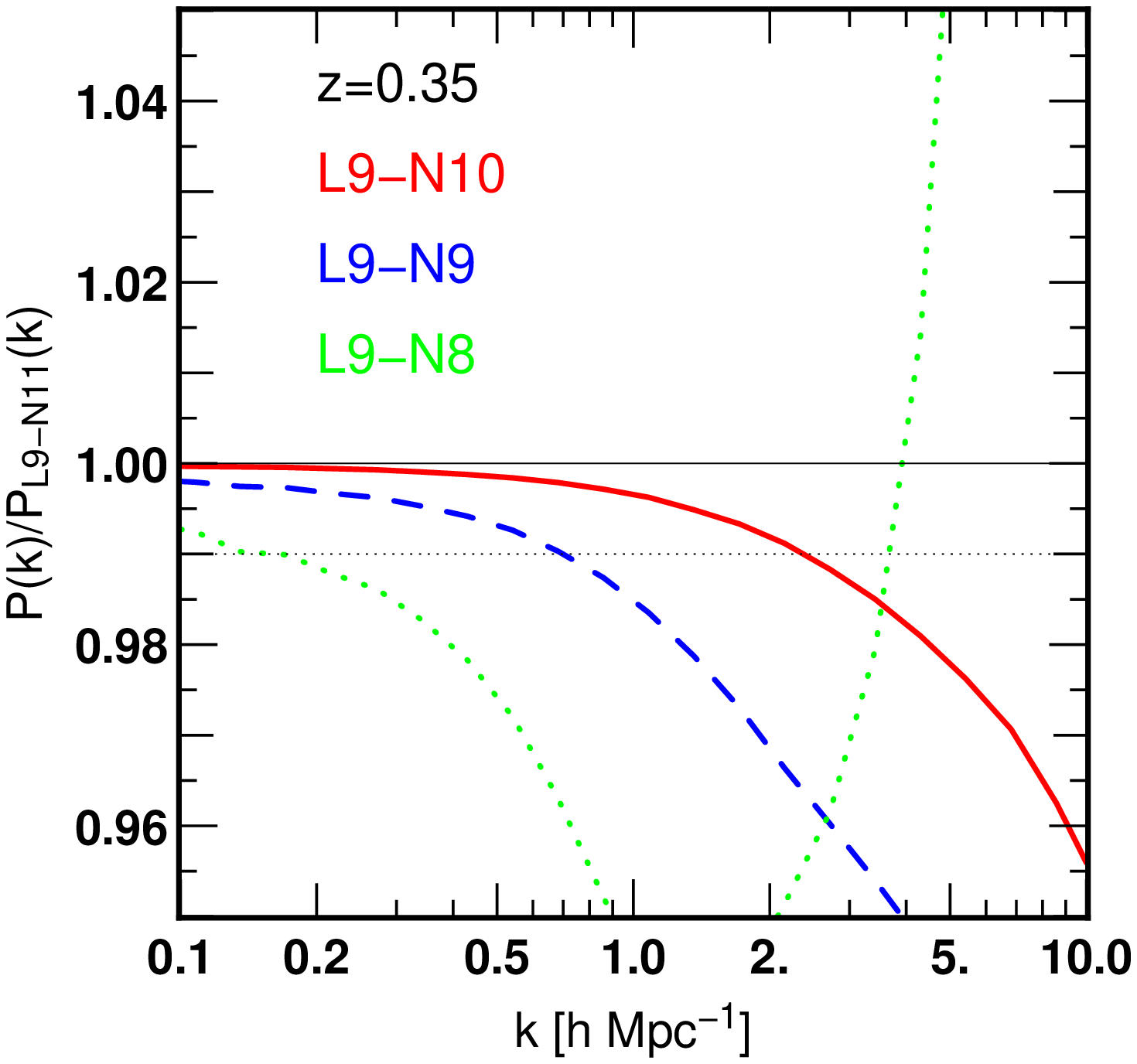}}
\caption{The ratio of the power spectrum to the highest resolution run,
{\tt L9-N11}. The three lines correspond to {\tt L9-N10}, {\tt L9-N9} and {\tt L9-N8}
from top to bottom. {\it Left}: $z=3$, {\it middle}: $z=1$, {\it right}: $z=0.35$.}
\label{fig:test1}
\end{center}
\end{figure*}

\begin{figure*}
\begin{center}
\epsfxsize=6.05 cm \epsfysize=6 cm {\epsfbox{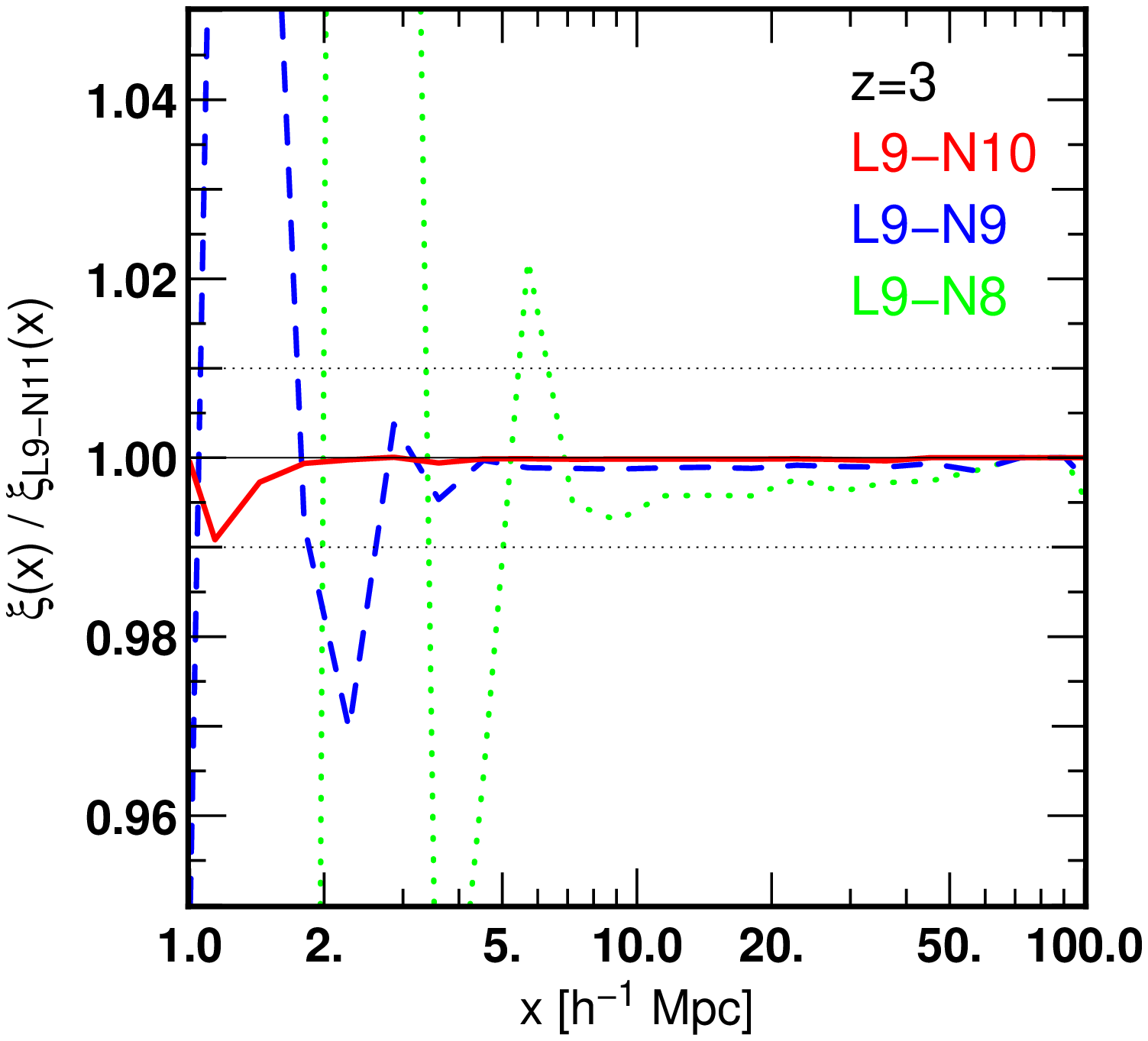}}
\epsfxsize=6.05 cm \epsfysize=6 cm {\epsfbox{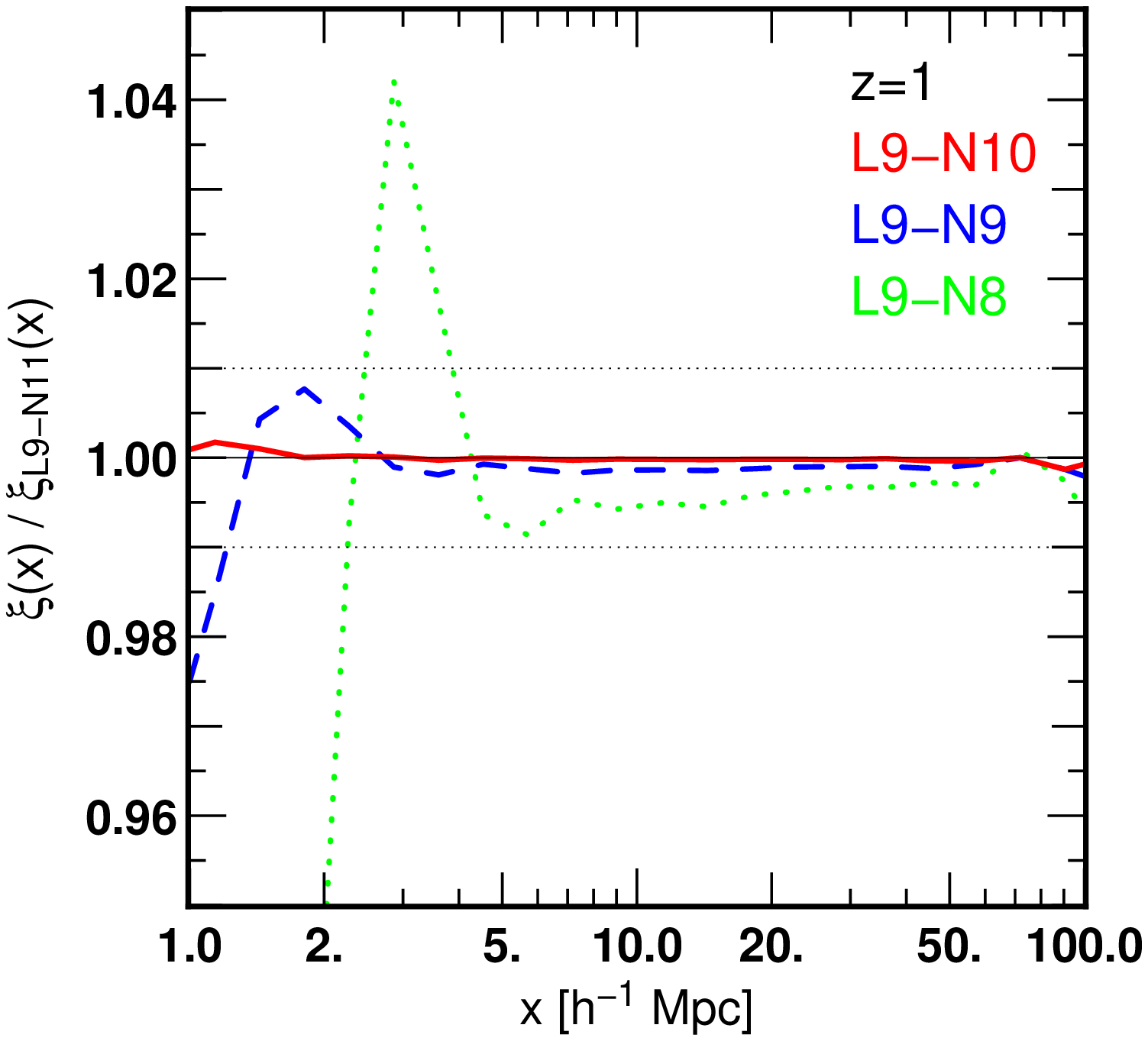}}
\epsfxsize=6.05 cm \epsfysize=6 cm {\epsfbox{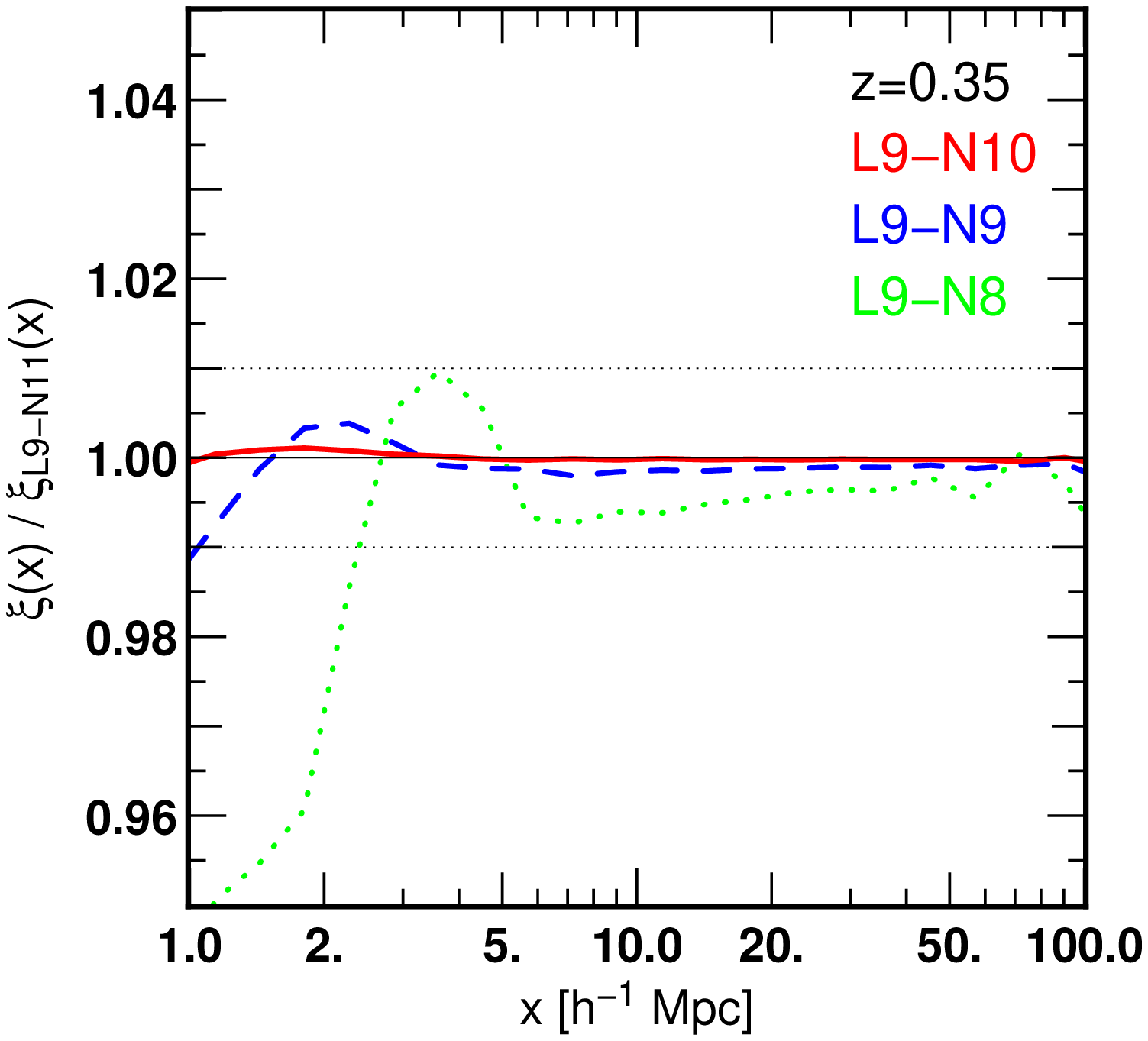}}
\caption{Same as Fig.~\ref{fig:test1}, but for two-point correlation function.}
\label{fig:test1-xi}
\end{center}
\end{figure*}

\begin{figure*}
\begin{center}
\epsfxsize=6.05 cm \epsfysize=6 cm {\epsfbox{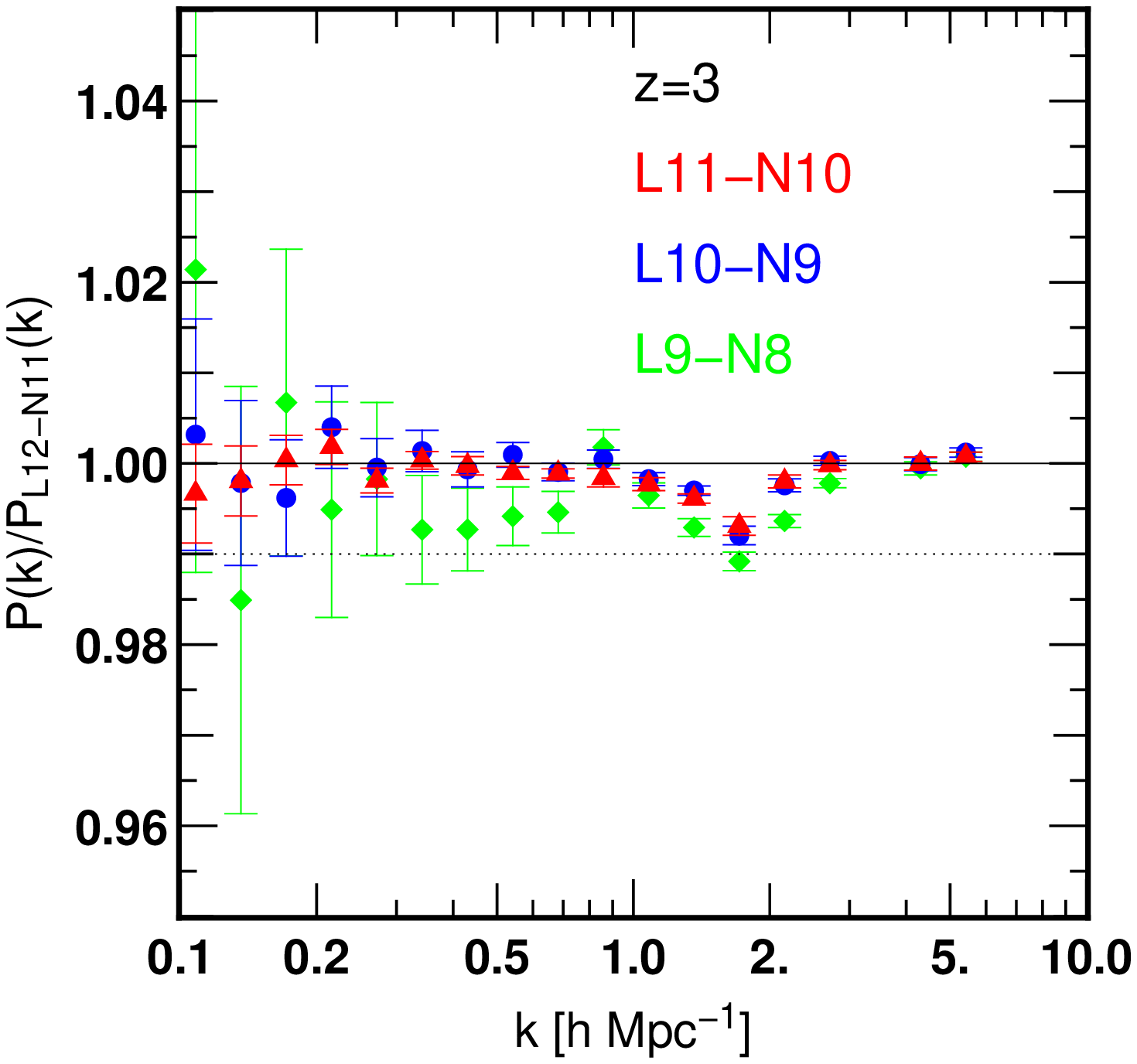}}
\epsfxsize=6.05 cm \epsfysize=6 cm {\epsfbox{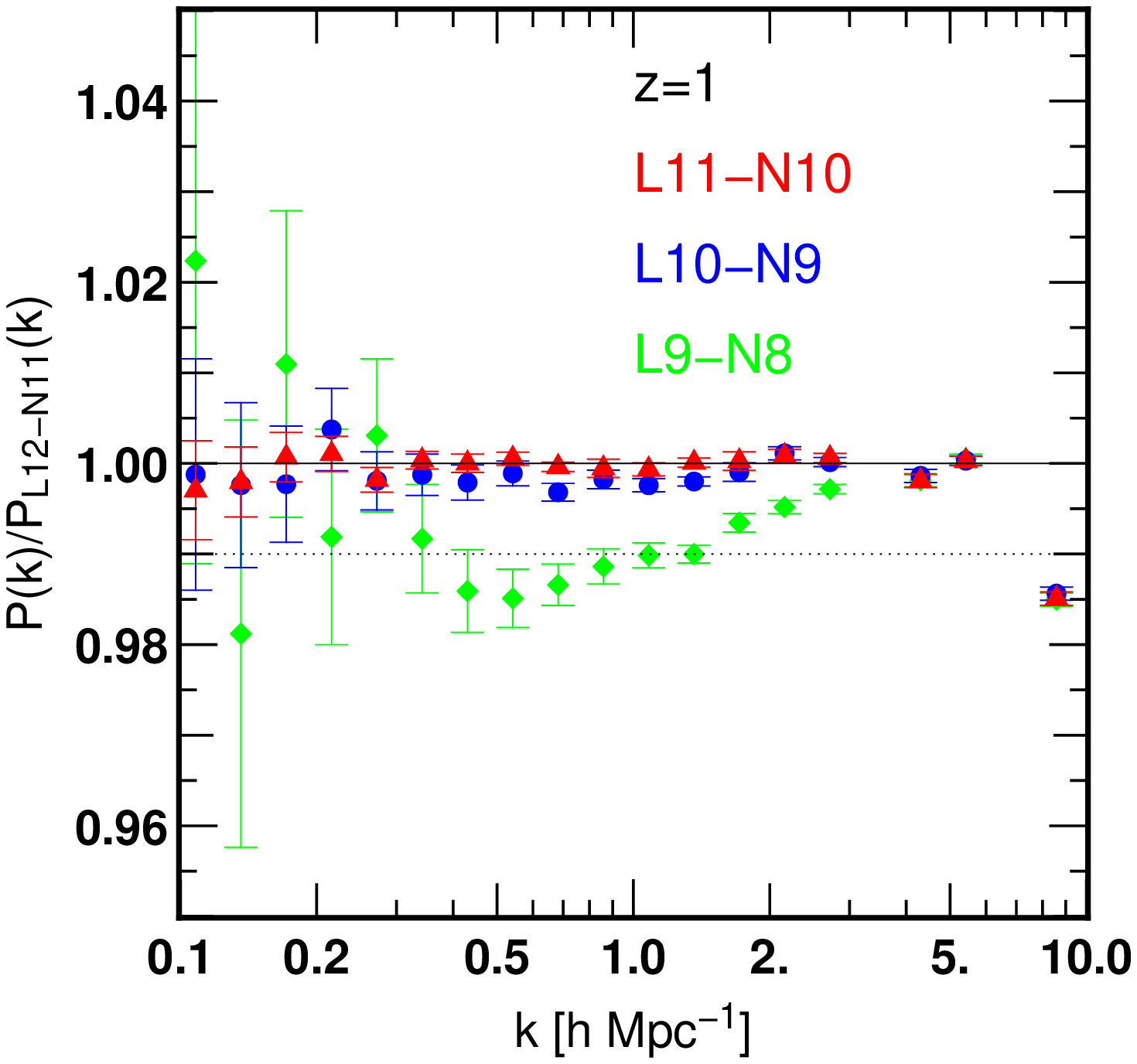}}
\epsfxsize=6.05 cm \epsfysize=6 cm {\epsfbox{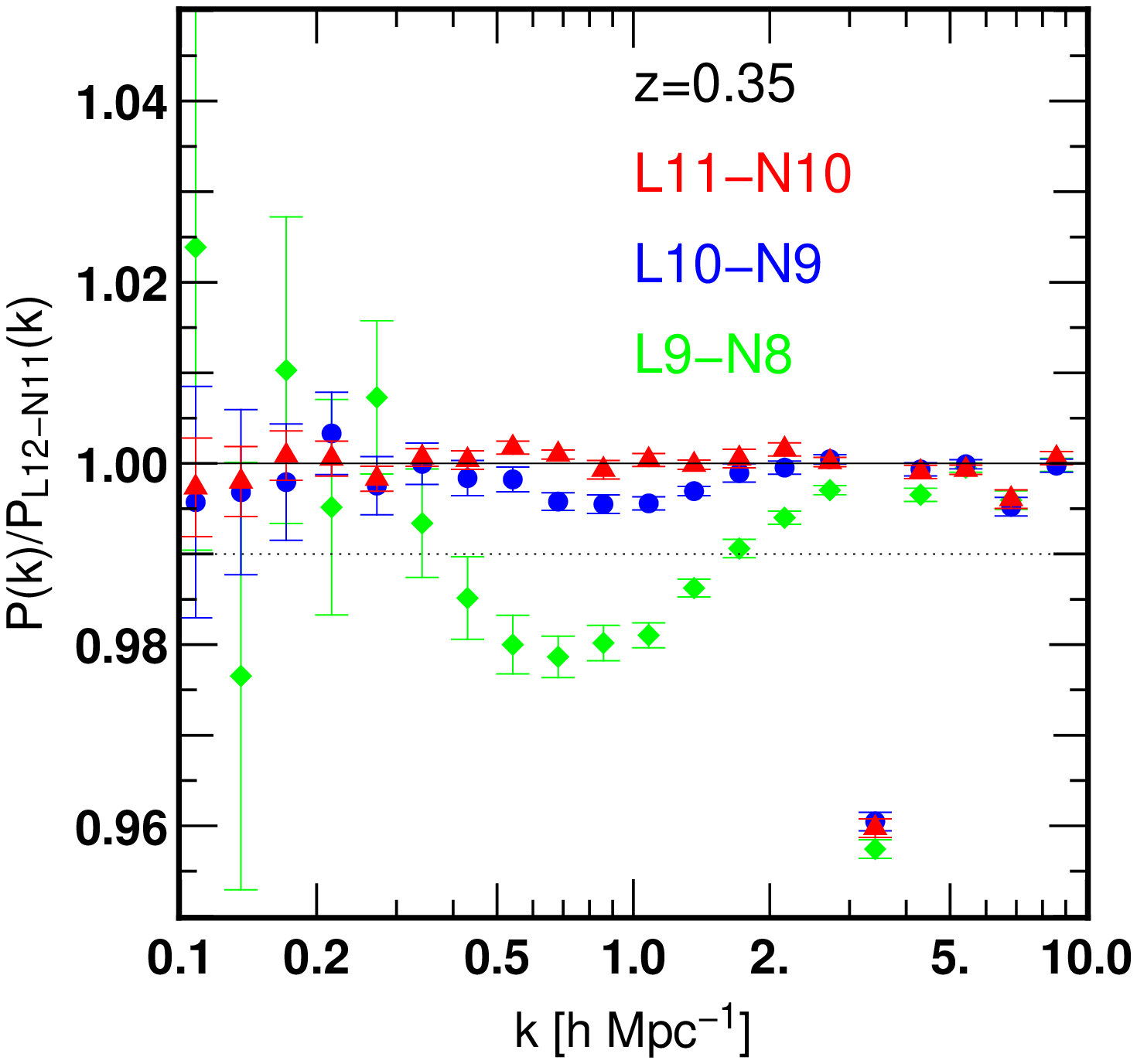}}
\caption{The ratio of the power spectrum to the largest volume run, {\tt L12-N11}. 
The three symbols correspond to {\tt L11-N10} (triangles), {\tt L10-N9} (circles)
and {\tt L9-N8} (diamonds). {\it Left}: $z=3$, {\it middle}: $z=1$, {\it right}:
$z=0.35$.}
\label{fig:test2}
\end{center}
\end{figure*}

\begin{figure*}
\begin{center}
\epsfxsize=6.05 cm \epsfysize=6 cm {\epsfbox{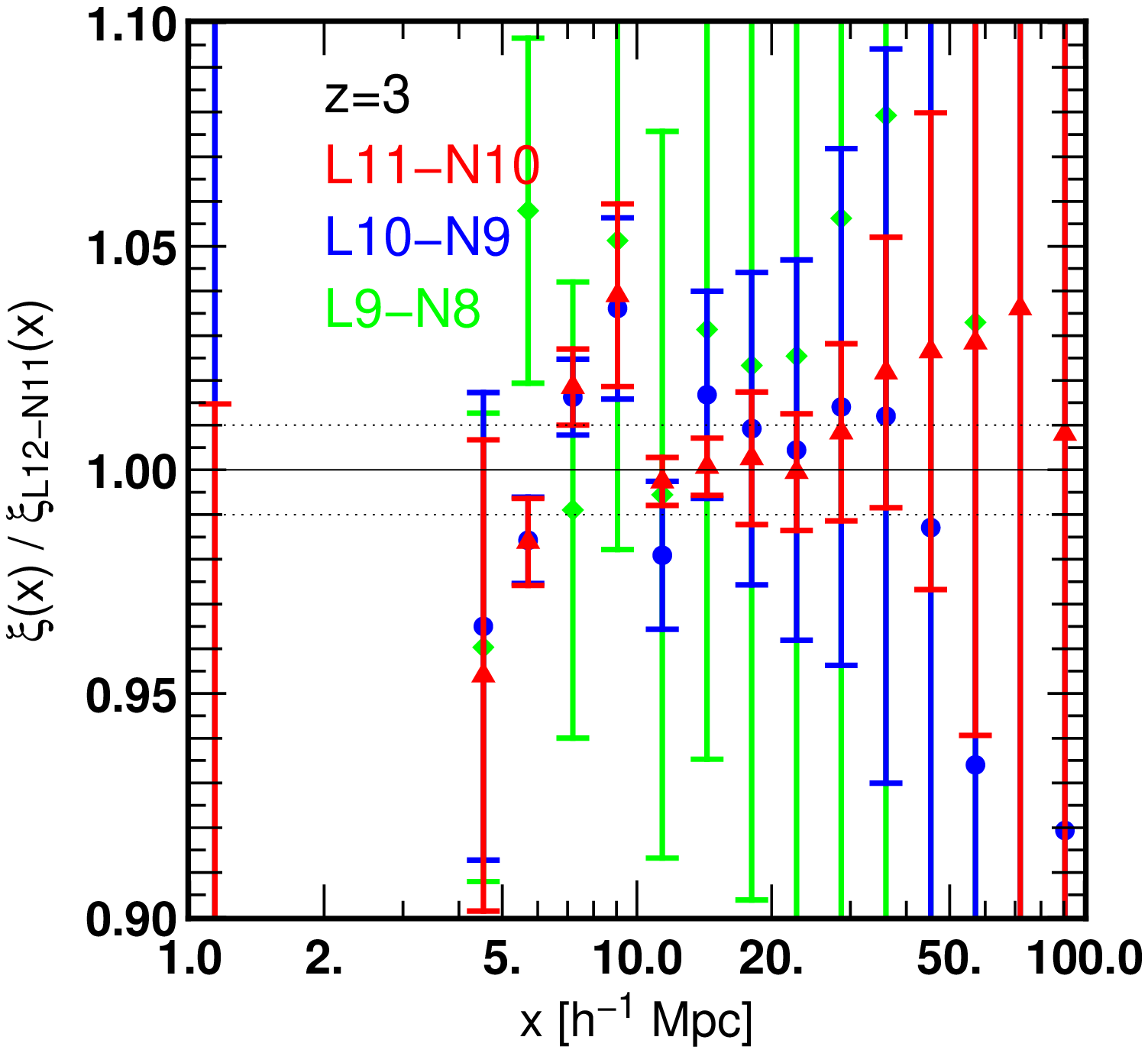}}
\epsfxsize=6.05 cm \epsfysize=6 cm {\epsfbox{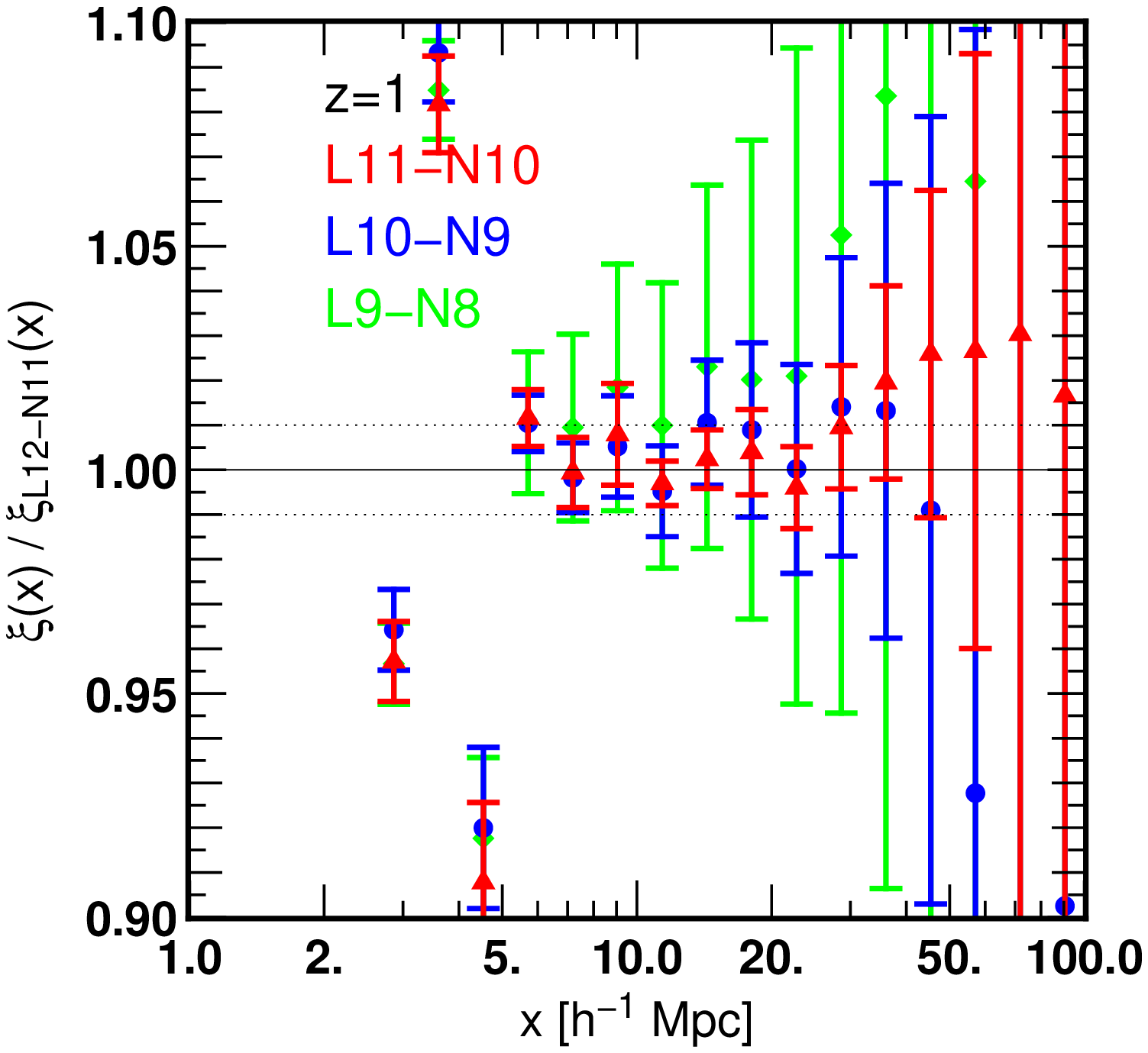}}
\epsfxsize=6.05 cm \epsfysize=6 cm {\epsfbox{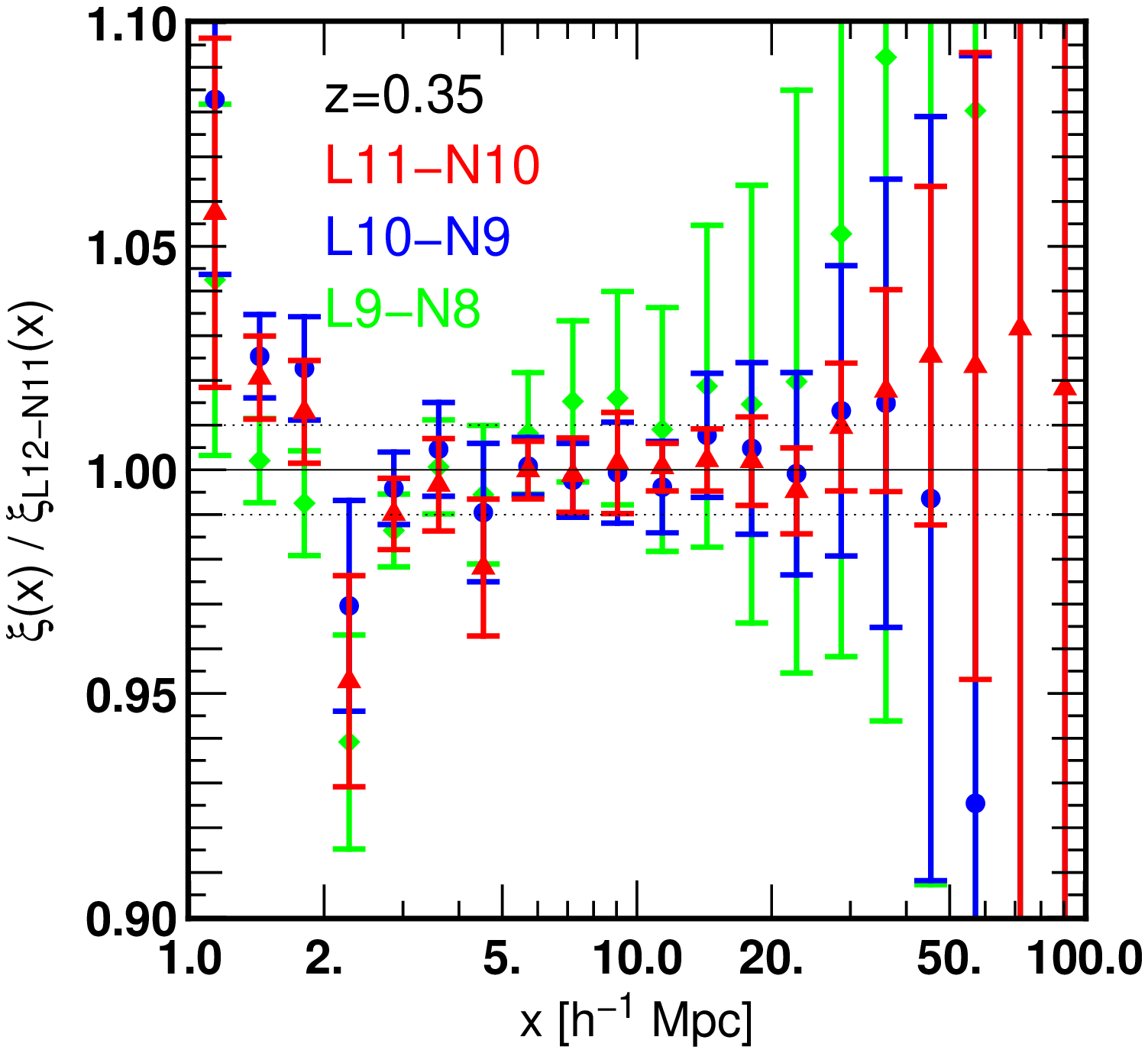}}
\caption{Same as Fig.~\ref{fig:test2}, but for the two-point correlation function.}
\label{fig:test2-xi}
\end{center}
\end{figure*}

\begin{figure}[htb]
\begin{center}
\includegraphics[width=6cm]{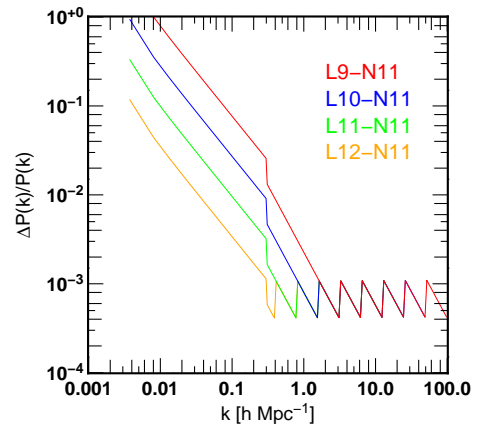}
\caption{$1$-$\sigma$ statistical error of the power spectrum (fractional)
estimated from Eq.~(\ref{eq:error}) for L9-N11, L10-N11, L11-N11 and L12-N11
(from top to bottom).}
\label{fig:statistical}
\end{center}
\end{figure}

We basically follow the ordinary FFT method to measure both the power spectrum
and the two-point correlation function. 
Before assigning particles to the mesh, however, we fold the particle distribution
into a smaller box to avoid the systematic error caused by the assignment on a
finite number of grid points, which is important on small scales near the grid
interval \citep[see][for a similar discussion]{Colombi2009}. Namely, we perform 
the following operation to the particle position, $\bf x$:
\beq
{\bf x} \to {\bf x} \,\%\, (L_{\rm box}/2^n),
\label{folding}
\eeq
where the operation, $a\,\%\,b$, gives the reminder of the division of $a$ by $b$,
and we vary the integer $n$ from $0$ to $6$.
After that, we assign particles on the $2048^3$ grid points of the folded small
box with side length of $L_{\rm box}/2^n$ using the CIC algorithm
\citep{Hockney1981}.
We then Fourier transform the density contrast and take the average of
$|\tilde\delta({\bf k})|^2$ within $k$-bins to obtain the power spectrum. The
binning is chosen to be logarithmic on small scale with 20 bins per
decade ($k>0.3h\,{\rm Mpc}^{-1}$), but we adopt a linear binning with the interval 
$\Delta k=0.005h\,{\rm Mpc}^{-1}$ on large scale ($k<0.3h\,{\rm Mpc}^{-1}$) to see
the feature of BAOs more clearly. While the folding procedure makes the
systematic effect caused by assignment smaller, it reduces the number of available 
modes. We choose the value of $n$ for the folding depending on the wavenumber
so that the number of modes is large enough.
We plot in Fig.~\ref{fig:statistical} the resulting statistical error ($1$-$\sigma$ level)
estimated from
\beq
\frac{\Delta P(k)}{P(k)} = \frac{1}{\sqrt{N_{\rm mode}(k)}},
\label{eq:error}
\eeq
where $N_{\rm mode}(k)$ denotes the number of independent mode in that bin.
This is exact when $\tilde{\delta}({\bf k})$ is Gaussian
\citep{Feldman1994}, and this can be shown to remain a good estimate in the
current situation (see e.g., \cite{Takahashi2009}).
One may notice a dip at $k\sim0.3h$Mpc$^{-1}$. This corresponds to the change
of binning from linear to logarithmic. Another feature is a characteristic zig-zag
pattern on smaller scale. This pattern corresponds to the change in the value of
$n$ used in the folding. Even after reducing the number of available modes by 
folding, however, we keep enough Fourier modes to achieve accuracy of $\sim$
sub-percent level on small scale ($k \simgt 1h$Mpc$^{-1}$).

The two-point correlation function is measured in a similar manner. We simply
perform an inverse FFT to the three dimensional power spectrum calculated on
grid points. We estimate the statistical error using the formula:
\beq
\left|\Delta\xi(x)\right|^2 = 64\pi^4\int\frac{{\rm d}k}{V} \, k^2 \,  
\left\{j_0(kx)P(k)\right\}^2,
\label{eq:xi_error}
\eeq
where the quantity $j_0$ is the spherical Bessel function of the first kind. 
We use the power spectrum measured from the same simulation in the integrand
for consistency, see \citet{Taruya2009} for more details.

\subsection{Convergence tests}
\label{subsec:convergence}
One may naively think that one can model the power spectrum by combining
simulations with different box sizes. However, it is not trivial to determine which
simulation gives the best measurement on a given scale. The simulation in a larger
volume lacks the power on small scale, while that in a smaller box does so on large 
scale.

A characteristic wavenumber for the finite box size is given by $2\pi/L_{\rm box}$,
the fundamental mode.
For the finiteness of mass/force resolution, one may define two important
wavenumbers.
First, the initial density fluctuation cannot be input on scales smaller than the
mean inter-particle distance ($2\pi/L_{\rm box}\times N^{1/3}$ in wavenumber).
Second, the force calculation is not accurate below the softening scale. This
wavenumber is given by $2\pi/L_{\rm box}\times N^{1/3}\times20$ for
our simulations (i.e., $5\%$ of the mean inter-particle distance). One can basically
trust the simulation results from the fundamental mode to the inter-particle
distance, or at most to the softening scale. These wavenumbers are shown in the
left panel of Fig.~\ref{fig:cover} for the four main simulations. 

After nonlinear evolution, however, the initial small systematic error on one scale
may propagate to different scales due to the coupling between different modes.
This makes the situation complicated. To separate the two systematic effects
caused by the finiteness of mass/force resolution and simulated volume, we
perform two tests using additional simulations in this subsection.
We will show that both finiteness effects lead to a lack of power on certain scales.
We especially focus on the scales where these effects are more than $1\%$,
and avoid to use the data points on these scales in later discussions.

\subsubsection{Test 1: finite mass/force resolution}

We first examine the effect of the finiteness of the mass/force resolution. We use
{\tt L9-N8}, {\tt L9-N9}, {\tt L9-N10} and {\tt L9-N11} for this test
(see Tab.~{\ref{tab:list}}). These simulations have the same volume but different
number of particles, so that we can see the systematic effect purely from the
finiteness of the resolution (see the middle panel of Fig.~\ref{fig:cover}).
We set the same random linear density field in creating the initial conditions of
the four simulations to make the comparison clearer.
Since the three additional simulations, {\tt L9-N8}, {\tt L9-N9} and {\tt L9-N10},
have the same resolution as {\tt L12-N11}, {\tt L11-N11} and {\tt L10-N11},
respectively, we can assess the impact of the systematic effect for the "main"
simulations by this test.

The results are shown in Fig.~\ref{fig:test1}. We plot the ratio of the power
spectrum measured from the three lower resolution runs to that from the highest
resolution one (i.e., {\tt L9-N11}).
One can clearly see that all curves blow up at large wavenumbers. This corresponds
to the scale where the shot noise contribution becomes important. In addition to
this feature, the curves show a lack of power on intermediate scales, which is more
important for lower resolution runs at higher redshifts. We can interpret this as
follows. At higher redshifts, the lack of power on around mean inter-particle scale
survives. But the power generated by the pure nonlinear growth gradually
dominates over the initial power and the difference between the four runs become
smaller at lower redshifts.

We also test the effect of finite resolution on the two-point correlation function.
Similar trends can be seen in Fig.~\ref{fig:test1-xi} for the two-point correlation
function.

\subsubsection{Test 2: finite box size}

We next discuss the effect of the finiteness of the simulated volume.
We use {\tt L9-N8}, {\tt L10-N9}, {\tt L11-N10} and {\tt L12-N11} for this test
(see Tab.~\ref{tab:list}).
These runs have the same mass/force resolution but different box size (see
the right panel of Fig.~\ref{fig:cover}). Again, since the three smaller simulations
have the same volume as the main runs, {\tt L9-N11}, {\tt L10-N11} and
{\tt L11-N11}, we can test the possible systematics of these main simulations.

We show the results in Fig.~\ref{fig:test2}.
In contrast to the test for the finite resolution, the data points seem noisy.
This is because we cannot set the same random realization of the linear density
field for simulations with different box sizes.
When one sees the result of {\tt L9-N8}, the systematic effect is prominent at
$0.3 \sim 2h$Mpc$^{-1}$, and is growing with time. This is contrary to what we
see in the finite resolution test, which as we showed was decaying. This growing
nature of the systematics suggests that the effect comes from the mode coupling
of large-scale modes. The systematic effect is not important (i.e., below $1\%$
level) for the rest of the simulations.

We test the convergence of the two-point correlation function which is shown in
Fig.~\ref{fig:test2-xi}.
Unfortunately, however, we cannot determine a strict range of the trustable scales
due to large statistical error. Since the two-point correlation function is a weighted
sum of the power spectrum over all scales, a large error bar at a certain
wavenumber may propagate to the two-point correlation function over a wide
range of scales.
Furthermore, the two-point correlation function is not positive definite unlike the
power spectrum.
Thus, the size of its {\it fractional} error can be very large on scales where the
two-point correlation function is close to zero.
Finally, we cannot control the size of error bars by changing the width of the bins,
since it does not explicitly depend on this width [see Eq.~(\ref{eq:xi_error})].

The ratio of the two-point correlation function on scales above $\simgt5h^{-1}$Mpc
seems consistent with unity at all the three redshifts.
As seen in Fig.~\ref{fig:test2-xi}, the finite resolution effect is important below
$\simlt5h^{-1}$Mpc for {\tt L9-N8}. 
This prevents us from a clear test for the finite box size since the four simulations
used in this test have the same resolution as {\tt L9-N8}.
We conclude here that we do not find any clear evidence of systematic error on the
two-point correlation function arising from the finiteness of volume,
and we will determine which simulation to use following only the resolution test.
We leave more complete tests for the finite volume to future works.

\section{Comparison with numerical simulations}
\label{Comparison}

\begin{figure}
\begin{center}
\epsfxsize=7 cm \epsfysize=5.5 cm {\epsfbox{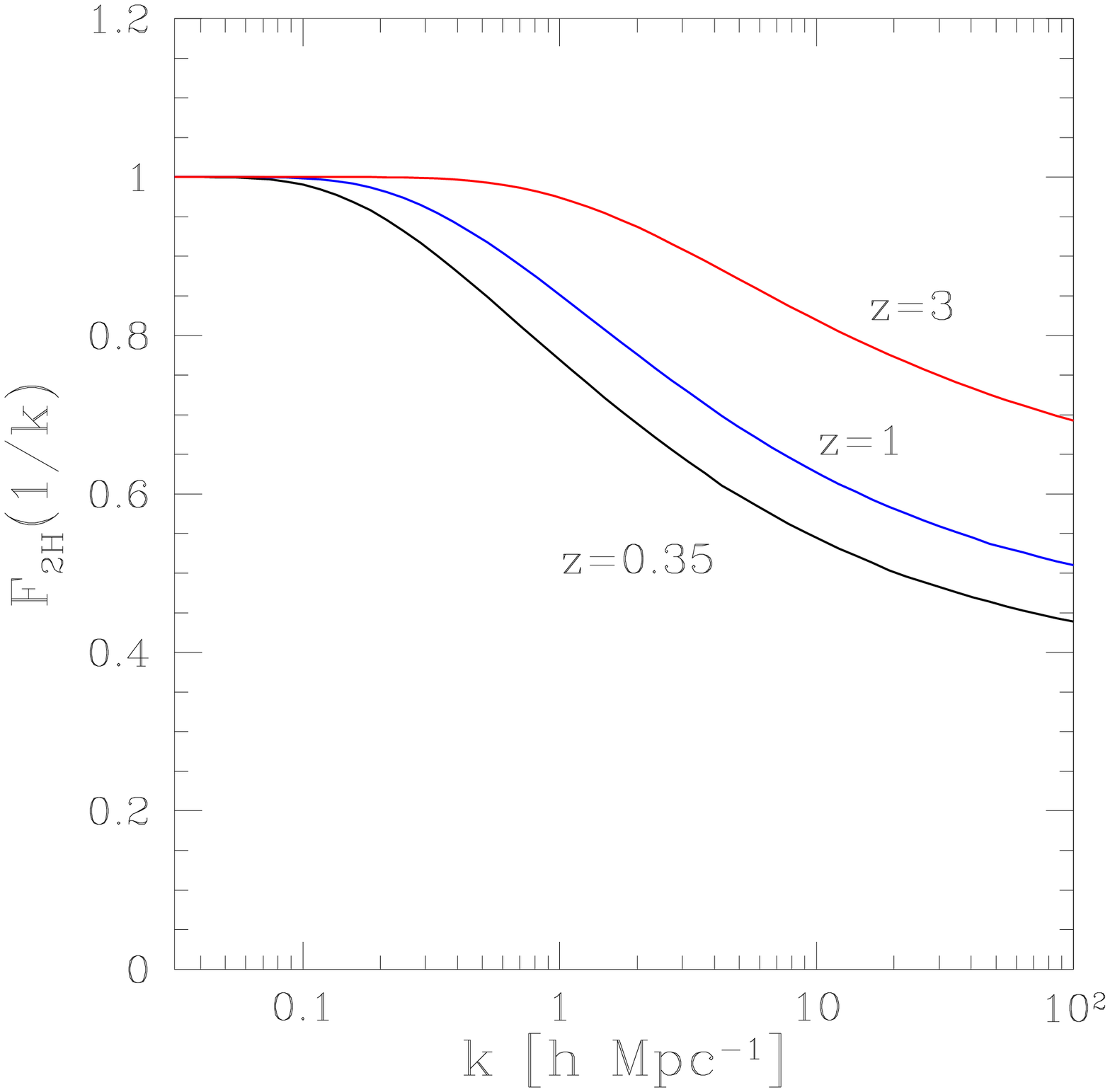}}
\end{center}
\caption{The probability $\FtH(1/k)$ that a Lagrangian pair of distance $q=1/k$
belongs to different halos, from Eq.(\ref{F2H-def}). We show our results for
$z=0.35$, $1$, and $3$.}
\label{fig_F2H}
\end{figure}

We first show in Fig.~\ref{fig_F2H} the quantity $\FtH(1/k)$, which is defined by
Eq.(\ref{F2H-def}) and enters the 2-halo term (\ref{Pkxq-pert2}), at redshifts
$z=0.35$, $1$, and $3$.
At fixed comoving scale, $q=1/k$, it decreases at lower redshift, since the probability
$\FoH$ for a Lagrangian pair of distance $q$ to belong to the same collapsed
halo grows with time, as larger scales turn nonlinear, and $\FtH=1-\FoH$.
We can note that it decreases very slowly at high $k$.
In fact, since we consider an initial power spectrum with a primordial
slope $n_s=0.96$ that is smaller than unity, the rms linear density
contrast $\sigma(q)$ of Eq.(\ref{sigma2-def}) does not diverge to infinity
as $q\rightarrow 0$ but goes to a finite asymptote. This implies that
the probability $\FtH(1/k)$
does not go to zero at high $k$ (i.e. on small scales), but reaches the
nonzero asymptote $\FtH(0)=\int_0^{\delta_L/\sigma(0)}\dd\nu f(\nu)/\nu$.
Although this value is unlikely to be exact, because of the approximations
involved in the derivation of $\FtH(q)$, it is possible that for such power
spectra without significant power on small scales there always remains some
fraction of matter which has never experienced shell crossing and remains
described by the single-stream fluid approximation.
Even though this is an interesting point from a theoretical perspective,
regarding the dynamical properties of the system, this is beyond the subject
of the present work and it plays no role for our purposes, since at high $k$
the power spectrum is clearly dominated by the 1-halo term, as we shall
check in Fig.~\ref{fig_lDk-200} below.
On the other hand, we can see that if we require a high accuracy, the deviation
of $\FtH$ from unity at $k\sim 0.3h$Mpc$^{-1}$ for $z=1$ cannot be completely
neglected and has some effect on the 2-halo term (\ref{Pkxq-pert2}).

\begin{figure*}
\begin{center}
\epsfxsize=6.1 cm \epsfysize=6 cm {\epsfbox{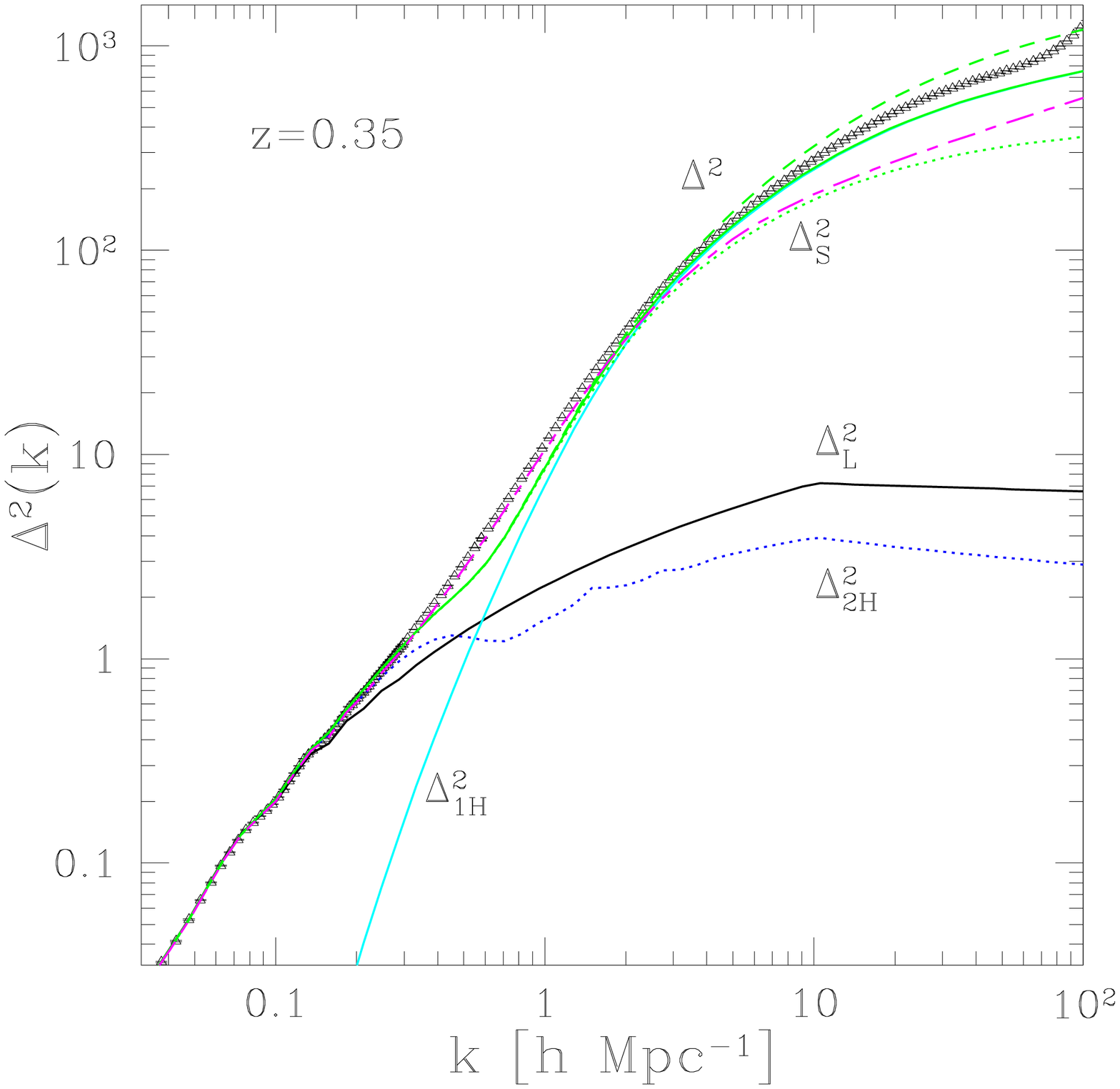}}
\epsfxsize=6.05 cm \epsfysize=6 cm {\epsfbox{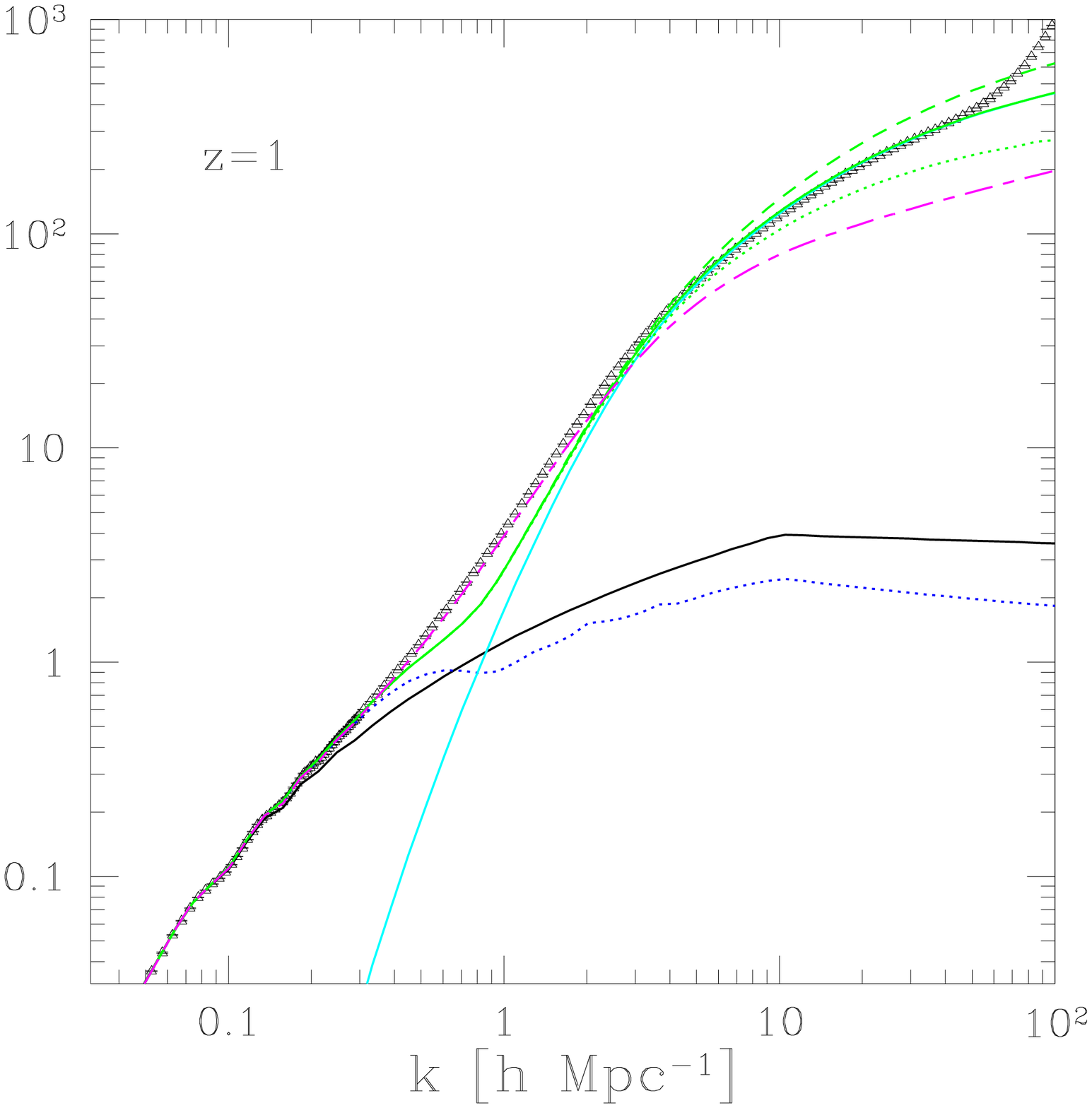}}
\epsfxsize=6.05 cm \epsfysize=6 cm {\epsfbox{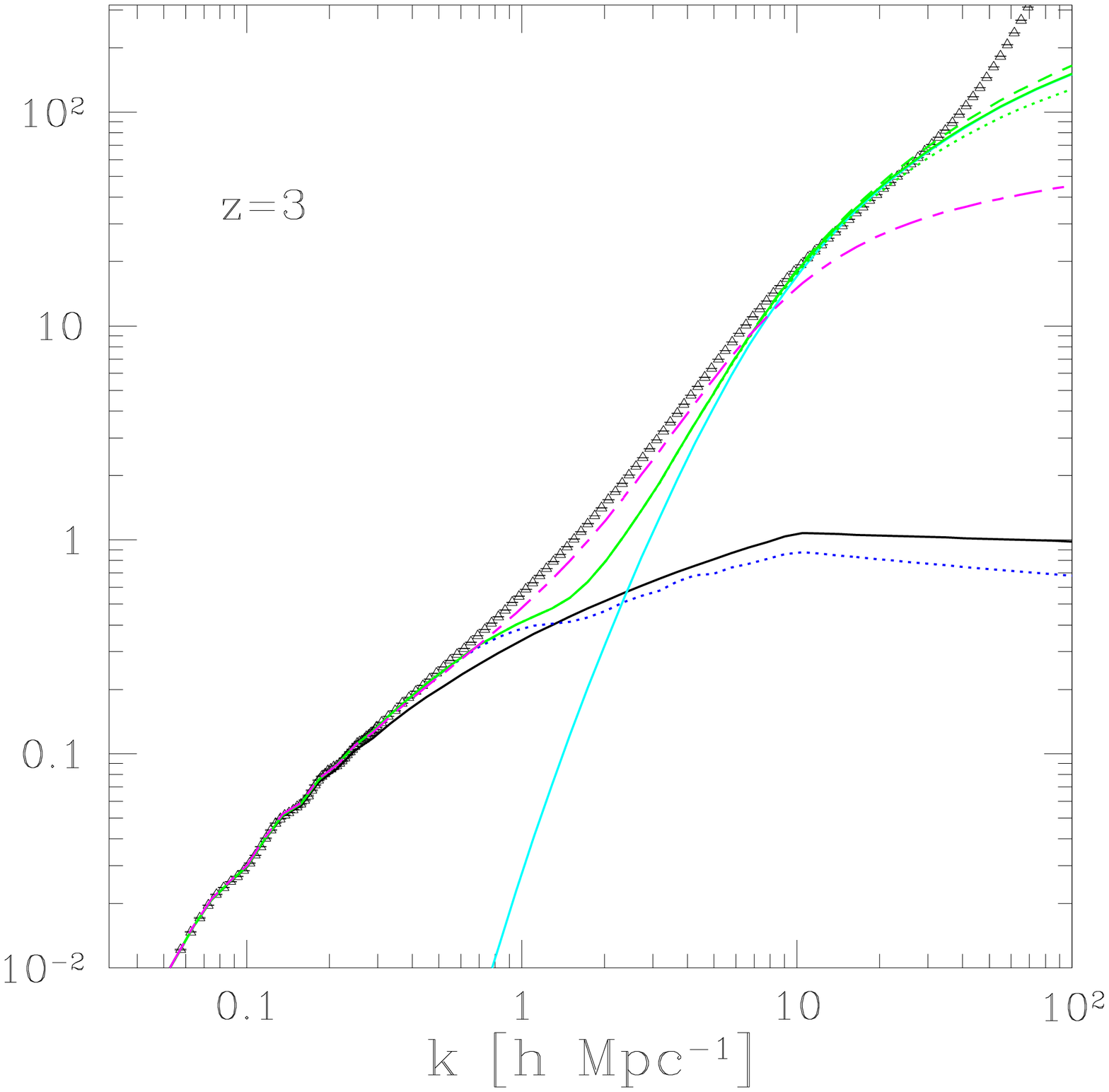}}
\end{center}
\caption{The power per logarithmic interval of $k$, as defined in
Eq.(\ref{Delta2-def}), at redshifts $z=0.35$, $1$, and $3$. The symbols are
the results from the numerical simulations described in
Sect.~\ref{N-body-simulations}.
The black solid line is the linear power spectrum, $\Delta^2_L$, and the blue
dotted line that is somewhat below at high $k$ is the 2-halo contribution
(\ref{Pkxq-pert2}), $\Delta^2_{2\rm H}$. The steep solid line $\Delta^2_{1\rm H}$
is the 1-halo contribution (\ref{Pk-1H}), for the halo model described in
Sect.~\ref{1-halo-1}. The green solid line is the full nonlinear power spectrum,
$\Delta^2=\Delta^2_{2\rm H}+\Delta^2_{1\rm H}$. The dashed (resp. dotted) green
line, that is slightly larger (resp. smaller) at high $k$, is the result obtained
by using the concentration relation for $c(M)$ given by \citet{Dolag2004}
(resp. \citet{Duffy2008}) instead of Eq.(\ref{cM-1}). The magenta dot-dashed line
$\Delta^2_S$, that is somewhat below these new N-body results at high $k$,
is the fit to older simulations from \citet{Smith2003}.}
\label{fig_lDk-200}
\end{figure*}

We now compare our model for the density power spectrum, that combines
a systematic perturbative approach with a phenomenological halo model,
with results obtained from numerical simulations.
We show in Fig.~\ref{fig_lDk-200} the power per logarithmic interval of $k$,
defined as
\beq
\Delta^2(k) = 4\pi k^3 P(k) ,
\label{Delta2-def}
\eeq
for the three redshifts $z=0.35$, $1$, and $3$.
The sudden rise of the N-body results at very high $k$ (which is more apparent
in the right panel at $z=3$) is due to the finite resolution and shot noise,
see Sect.~\ref{N-body-simulations}. This marks the highest wavenumber 
where we can compare the simulations with our theoretical models.

As discussed in Sect.~\ref{2-halo-1}, we can check that the 2-halo term
$\Delta^2_{2\rm H}$ remains well-behaved at high $k$ and is subdominant
with respect to the 1-halo term $\Delta^2_{1\rm H}$. It falls somewhat
below the linear power $\Delta^2_L$, contrary to the one-loop resummation
of $\Delta^2_{\rm pert}$ given by the direct steepest-descent scheme, which
becomes very close to $\Delta^2_L$ as seen in \citet{Valageas2007a},
because of the prefactor $\FtH(1/k)$ in Eq.(\ref{Pkxq-pert2}).

In agreement with the discussion of Sect.~\ref{1-halo}, we can also check that
the 1-halo term shows a fast decline at low $k$. At high $k$ we can see
that it is possible to reach a good agreement with the numerical simulations
by using an appropriate prescription for the concentration parameter $c(M)$,
such as the one given in Eq.(\ref{cM-1}) and represented by the green solid
line. The formula given in Eq.(\ref{cM-1}) is obtained by looking for values
of the free parameters (the normalization and the two exponents) that
provide a reasonable match with the power spectrum measured in the
simulations at high $k$ (looking among a few values close to the fits
already proposed in the literature).
Nevertheless, it is interesting to note that using fits for $c(M)$ proposed in
previous analysis of the halo profiles formed in numerical simulations, one
obtains predictions that are either larger \citep{Dolag2004} or smaller
\citep{Duffy2008} than the one associated with Eq.(\ref{cM-1}).
This shows that the nonlinear power spectrum is fully consistent in this
range with a simple halo model, such as (\ref{Pk-1H}), and with the properties
of halos seen in numerical simulations.
This also gives an estimate of the dependence of the power spectrum on the
prescription used for $c(M)$. In agreement with previous works
\citep{Huffenberger2003,Giocoli2010},
we recover the fact that the nonlinear power $\Delta^2(k)$ is larger and steeper
at high $k$ for concentration relations $c(M)$ that have a larger normalization
and a steeper dependence on $M$. A nice feature is that the nonlinear power
is largely independent of the details of the halo model up to $\Delta^2 < 100$,
so that models such as the one studied in this article remain quite predictive.
Even at higher $k$, we can see that up to $\Delta^2<10^3$, or
$k<100 h$Mpc$^{-1}$, using any of these prescriptions for $c(M)$ provides
a reasonable estimate of the power spectrum and even fares better than the
direct fit to $P(k)$ that was obtained from older simulations \citep{Smith2003}.
This suggests that models based on phenomenological ingredients such as
the halo model may prove more robust than direct fits to numerical results.
However, our model for $P(k)$ should not be trusted beyond the domain where
it has been checked, that is $k\leq 100 h$Mpc$^{-1}$ and $z\leq 3$.

In order to obtain predictions at much higher redshifts and wavenumbers,
one should use a prescription for $c(M)$ that is based on
some physical arguments rather than simple fits such as Eq.(\ref{cM-1}).
This would probably lead to a loss of accuracy in the range tested in
Fig.~\ref{fig_lDk-200}, as compared with the use of Eq.(\ref{cM-1}), but this is
likely to be more robust as we extrapolate to other regimes.
However, we shall not investigate the building of physical models for
$c(M)$ in this article, as this is a topic by itself.

\begin{figure*}
\begin{center}
\epsfxsize=6.1 cm \epsfysize=5 cm {\epsfbox{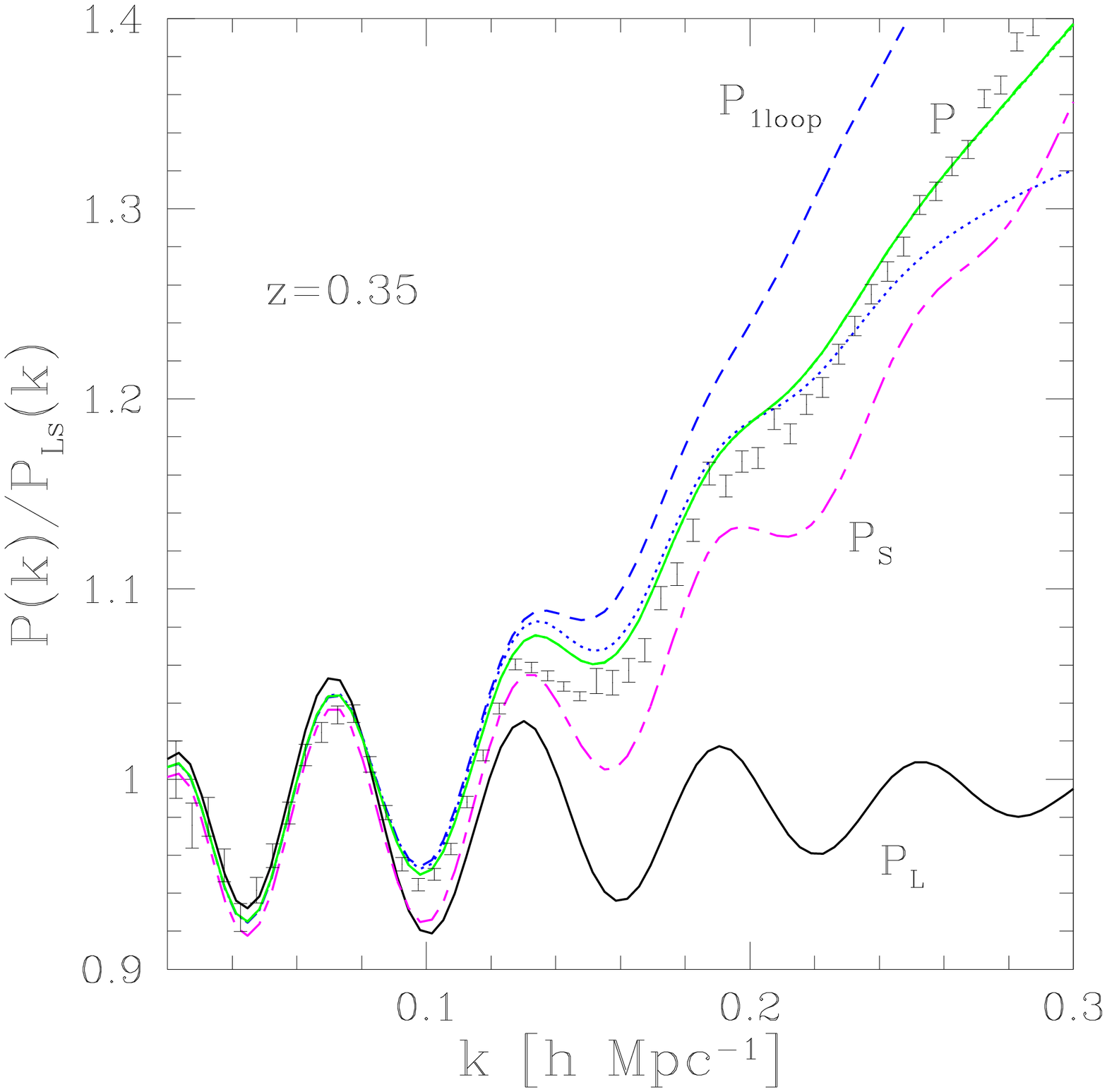}}
\epsfxsize=6.05 cm \epsfysize=5 cm {\epsfbox{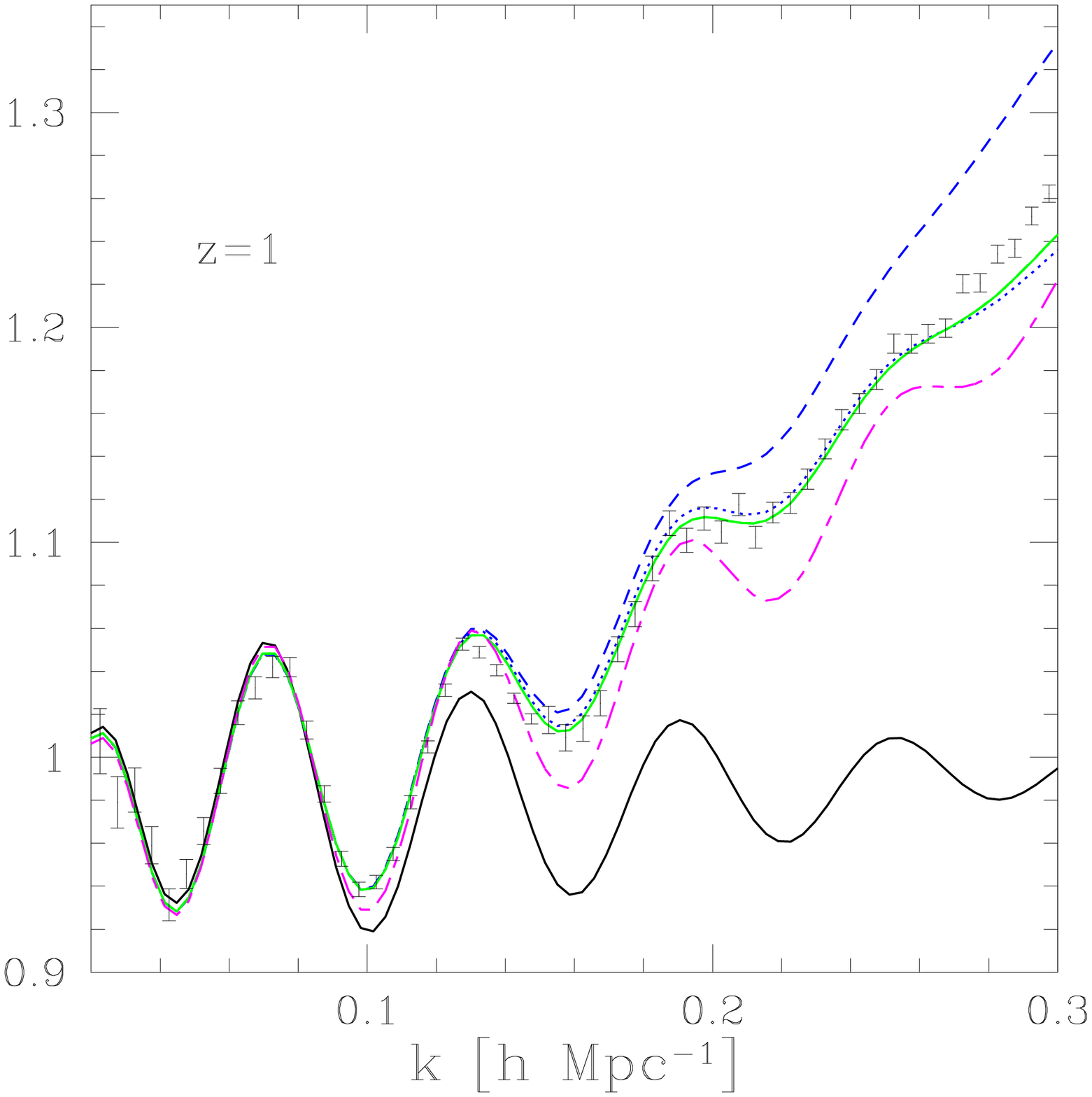}}
\epsfxsize=6.05 cm \epsfysize=5 cm {\epsfbox{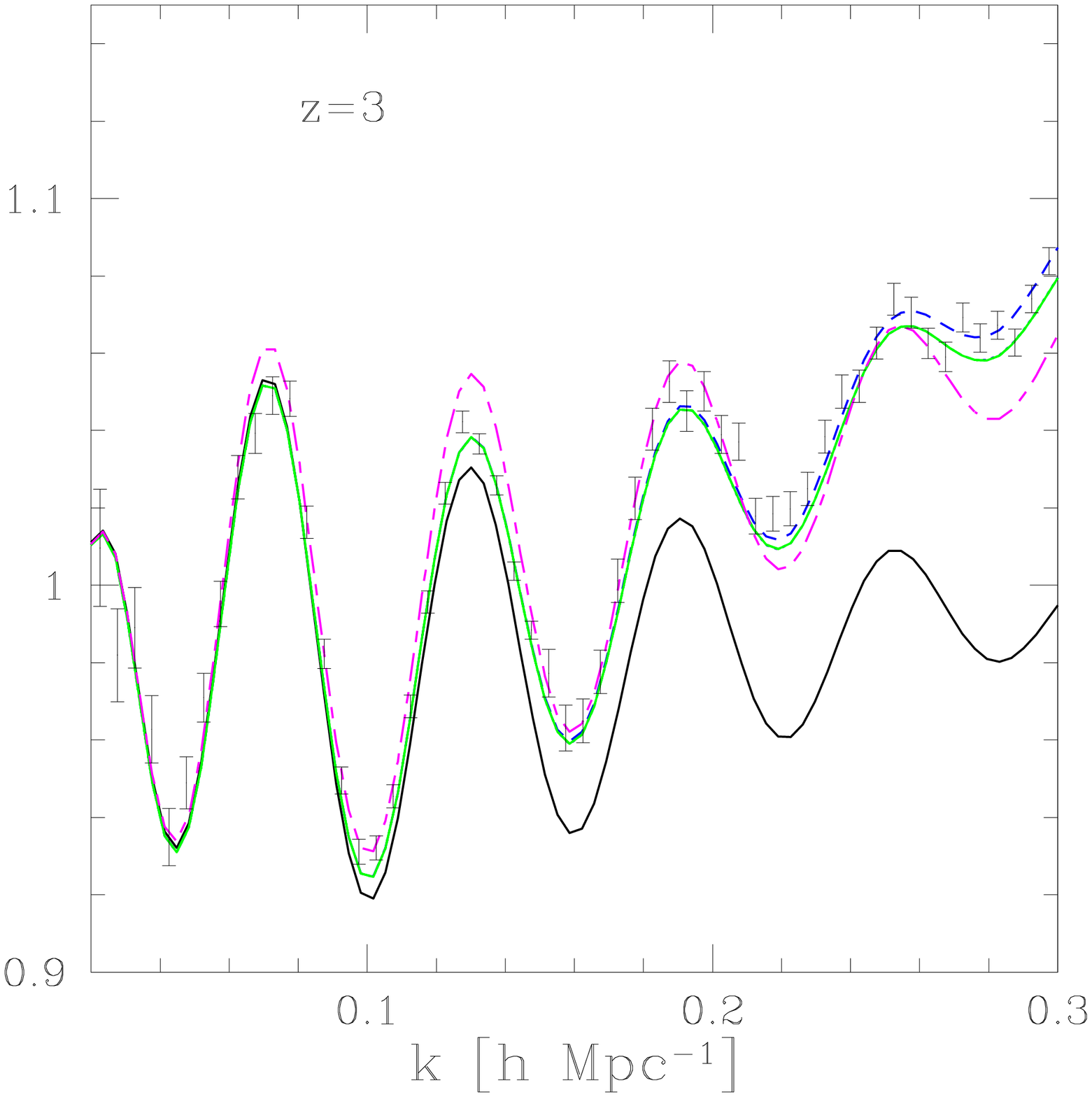}}
\end{center}
\caption{The ratio of the nonlinear power spectrum $P(k)$ to a smooth linear
power spectrum $P_{Ls}$ without acoustic baryonic oscillations, from
\citet{Eisenstein1999}.
The points with error bars are the results from N-body simulations.
The black solid line $P_L$ is the linear power spectrum, and the upper blue
dashed line $P_{1\rm loop}$ is the prediction of standard perturbation theory
up to one-loop order.
The green solid line $P$ is the full nonlinear power spectrum, from
Eq.(\ref{Pk-halos}), whereas the lower blue dotted line $P_{\rm pert}$ is the
perturbative part. The magenta dot-dashed line $P_S$ is the fit to simulations
from \citet{Smith2003} (which was not specifically designed to reproduce the baryon
oscillations).}
\label{fig_Pk-200}
\end{figure*}

We can see in Fig.~\ref{fig_lDk-200} that the full nonlinear power spectrum
(\ref{Pk-halos}) obtained by our approach, combining a perturbative expansion
with a phenomenological halo model, is able to reproduce the results measured
in numerical simulations, up to a reasonable accuracy.
Let us point out that the perturbative part,
$P_{\rm pert}(k)$, which dominates on large scales, contains no free parameter,
since it is given by a systematic perturbation theory. The 1-halo contribution
that dominates on small scales clearly contains some parameters, through the
choice of the halo mass function and halo density profile. However, the mass
function and the shape of the halo profile are already set by other measures
from numerical simulations, so that the main free parameter is the concentration
$c(M)$, which is not as well constrained. However, as seen in Fig.~\ref{fig_lDk-200},
this uncertainty only affects the very high-$k$ tail. Moreover, even in this region
the resulting model is competitive with direct fits to the power spectrum measured
in older simulations. Therefore, models such as the ones developed in this article
should prove useful to obtain reliable predictions for $P(k)$.
Another advantage of such models, as compared with simpler fitting formulas, is
that they should be quite accurate on large scales since they are consistent with
perturbation theory, up to the order of truncation of the computation.

Thus, we plot in Fig.~\ref{fig_Pk-200} the ratio of the power spectrum to
a ``no-wiggle'' linear power spectrum $P_{Ls}(k)$ which does not include
baryonic acoustic oscillations, in order to see more clearly the low-$k$ behavior.
We can check that in this range our model, based on the direct
steepest-descent resummation, shows a good match with the numerical
simulations. 
As could be expected, it fares better than a direct fit from
\citet{Smith2003} which was not designed to model this range with a high
accuracy (but still remains surprisingly good). Furthermore, the use of the
direct steepest-descent resummation proves to provide significantly more
accurate results than the standard perturbation theory, truncated at one-loop order,
given by the well-known expressions
\beq
P_{1\rm loop}(k) = P_L(k) + P_{22}(k) + P_{31}(k) ,
\label{P1loop-def}
\eeq
where formulas for the terms $P_{22}$ and $P_{31}$ may be found for instance
in \citet{Bernardeau2002}.
Let us recall here that the term $P_{\rm pert}(k)$ given by this
direct steepest-descent scheme contains no free parameter, nor any interpolation
procedure, and is consistent with standard perturbation theory up to one-loop
order (i.e. the difference between $P_{1\rm loop}$ and $P_{\rm pert}$ is due
to the partial resummation of higher order terms).
We can see in the left panel, at $z=0.35$, that around
$k \sim 0.14 h$Mpc$^{-1}$ the curve $P_{\rm pert}$ is slightly above the
full nonlinear power spectrum $P$. This is due to the prefactor $\FtH(1/k)$ in
Eq.(\ref{Pkxq-pert2}), which is slightly below unity. However, this is only a very
small effect on these scales. On the other hand, at higher $k$ (e.g.,
$k > 0.26 h$Mpc$^{-1}$ at $z=0.35$) the nonlinear power spectrum rises above
$P_{\rm pert}$ and keeps growing, while $P_{\rm pert}$ remains close to
$P_L$ at high $k$ as seen in Fig.~\ref{fig_lDk-200}. This is due to the 
1-halo contribution, which starts being non-negligible.
However, on these scales the dependence on the details of the halo
model is extremely weak. Indeed, the three green curves plotted in
Fig.~\ref{fig_lDk-200}, associated with the prescription (\ref{cM-1}) for the
concentration $c(M)$ and the two fits given by \citet{Dolag2004} and
\citet{Duffy2008}, are also plotted in Fig.~\ref{fig_Pk-200}. However, they
almost exactly fall on top of each other and cannot be distinguished in the
figure.

Thus, Figs.~\ref{fig_lDk-200} and \ref{fig_Pk-200} show that by combining
perturbation theories and halo models it is possible to obtain a good model
for the nonlinear density power spectrum, both on quasi-linear and highly
nonlinear scales. However, we can see in Fig.~\ref{fig_lDk-200} that in the
intermediate regime, where $\Delta^2 \sim 5$, our predictions fall below
the N-body results. This is also apparent in the high-$k$ parts of
Fig.~\ref{fig_Pk-200}, where the full nonlinear prediction $P(k)$ starts to grow
more slowly than the power measured in the numerical simulations.
This regime corresponds to the transition between the 2-halo and 1-halo
contributions (see Fig.~\ref{fig_lDk-200}) and as such it is at the limit of validity
of the approximations used for both terms.

On the perturbative side, that is the 2-halo term, the discrepancy can be due
to the truncation at one-loop order of the perturbative term. Indeed, we can expect
that by going to higher orders we can extend the range of validity of the
the perturbative term $P_{\rm pert}$ and push the downturn shown by the
blue dotted curve in Fig.~\ref{fig_Pk-200} to higher $k$.
Within such resummation schemes this means that we include all diagrams
up to $n$ loops, and partial resummations for higher order terms.
We can note that the discrepancy looks somewhat more severe at $z=3$
in Fig.~\ref{fig_lDk-200}. More precisely, the range of $k$ where there is
a noticeable mismatch before the 1-halo term becomes dominant
(larger than $P_L$) is somewhat more extended than at $z=0.35$.
This agrees with the results of \citet{Valageas2010a}, where it was found
(within the Zeldovich framework) that the scope of perturbation theory
is somewhat greater at higher redshift for CDM power spectra, in the sense
that the range where higher order perturbative terms are important
(i.e. larger than the non-perturbative correction associated with shell crossing
effects) is wider and that the perturbative expansion makes sense up to higher
orders.
However, we shall not try to go beyond one-loop order in this paper and we
leave such a task to future works.

On the non-perturbative side, it is clear that by definition halo models
are only phenomenological models and do not provide a systematic tool.
Even if we admit that it makes sense to ``recognize'' halos in the density
field (even though this is a descriptive tool rather than a direct ingredient of
the equations of motion), it is clear that realistic halos are not spherical
nor fully virialized. In particular, apart from the inaccuracies introduced by
the spherical approximations, outer halo radii are not fully virialized and
particles do not instantly loose all memory of their initial conditions, so that
the average in Eq.(\ref{mean-M2}) is only approximate. As we have discussed
in Sect.~\ref{1-halo}, such approximations also prevent us from exactly satisfying
momentum conservation. Since these effects mostly apply to the outer parts of
the halos they can be expected to have an impact on the moderately low-$k$
part of the power spectrum, that is the transition region.
Another effect that is likely to play a role is the truncation of the halo
profiles at the radius $r_{200}$. This may be investigated in a rather
straightforward manner by studying the
change in $P(k)$ as we modify this density threshold, and we shall consider
in Sect.~\ref{delta=50-1} below the results obtained for $\delta=50$.

On a more fundamental level, since the splitting between 2-halo and
1-halo contributions is necessarily approximate (it cannot be derived
in a systematic fashion from the equations of motion and always involves
some approximations, at least within analytical frameworks) it is not
surprising that the match is not perfect, especially in the transition range.

Another approximation used in this work is the
assumption of smooth halos, defined by the regular profile (\ref{NFW-rho}).
As noticed in Sect.~\ref{1-halo-1}, it is possible to extend the halo model
by including substructures \citep{Giocoli2010}, however this only has an effect
on the very high-$k$ tail of the power spectrum (as could be expected since these
are small-scale modifications).
We do not investigate such modifications in this work, as the agreement
we obtain with numerical simulations is already reasonably good on these
scales and we prefer not to introduce additional parameters.
These effects are also degenerate with the mean concentration relation $c(M)$,
as seen in Fig.~\ref{fig_lDk-200} by the comparison between the dashed and
dotted green curves, associated with different prescriptions from
\citet{Dolag2004} and \citet{Duffy2008}. In particular, as pointed out in
Sect.~\ref{1-halo-1}, the prescription (\ref{cM-1})
that we have obtained by requiring a good match at high $k$ can also be 
seen as an effective model, that describes the combined effects of both the
concentration of the mean radial profile and the presence of small
substructures. At some stage, if one wants to build a model where the
halo properties are independently obtained from the measure of
individual halos in N-body simulations, so that $c(M)$ can no longer
be tuned, it may nevertheless prove necessary to explicitly include such effects.

\section{Dependence on various ingredients of the model}
\label{ingredients}

We now investigate the impact of various ingredients of the model
on the predictions obtained for the density power spectrum.

\subsection{Halos defined by $\delta=50$}
\label{delta=50-1}

As we have discussed in the previous section, the lack of power on intermediate
scales seen in Fig.~\ref{fig_lDk-200}, around the transition between the 2-halo
and 1-halo terms, may be due in part to the truncation of the halo profiles at
the radius $r_{200}$. Indeed, realistic halos extend to larger radii, which should
generate some extra power on large scales. In order to take into account such
outer radii, while keeping the mass $M$ that enters the definition of the halo
mass function as the halo mass within the truncation radius (so as to avoid
overcounting and violations of mass normalizations), we consider defining halos
by a smaller nonlinear density contrast, such as $\delta=50$.
This lower threshold means that we include outer shells (as compared with the
case $\delta=200$), while halos still remain well separated from the background. 

For the halo mass function we still use the scaling function (\ref{f-fit}), but the
linear density contrast is now $\delta_L=\cF^{-1}(50)$, which gives for instance
$\delta_L \simeq 1.50$ at $z=1$. This automatically satisfies the normalization
(\ref{fnu-norm}) as well as the large-mass tail (\ref{M-tail}).
It was shown in \citet{Valageas2009d} that this recipe provides a reasonable
match to numerical simulations (but not as good as for $\delta=200$) for halos
defined by a density threshold $\delta=100$, and we shall assume that it
still provides a reasonable model for $\delta=50$. This should be sufficient for
our purposes, as the 1-halo term needed for the density power spectrum only
depends on integrals over the halo mass function and we have seen that both the
normalization and the large-mass tail are correctly reproduced.

\begin{figure}[htb]
\begin{center}
\epsfxsize=9 cm \epsfysize=8 cm {\epsfbox{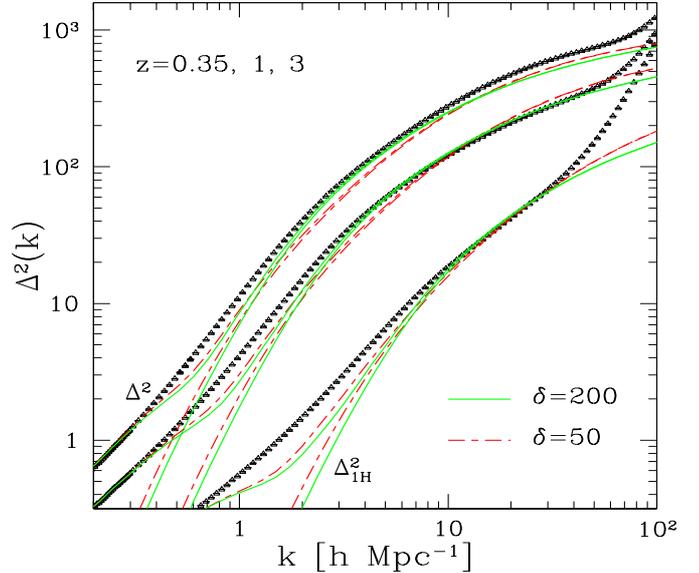}}
\end{center}
\caption{The power per logarithmic interval of $k$, as in Fig.~\ref{fig_lDk-200},
but for halos that are defined either by a density contrast $\delta=200$
(solid curves, identical to those of Fig.~\ref{fig_lDk-200}), or by $\delta=50$
(dot-dashed curves). Both the 1-halo contribution $\Delta_{1\rm H}$ and the
full nonlinear power spectrum $\Delta^2$ are shown.}
\label{fig_lDk-50}
\end{figure}

We still use the NFW profile of Eq.(\ref{NFW-rho}), but we now need to
obtain the scale radius $r_s$ in terms of $r_{50}$. To do so, we now define
the concentration parameter as
\beq
c(M_{50}) = \frac{r_{50}}{r_s} ,
\label{cM50-def}
\eeq
and the halo characteristic density now reads as
\beq
\rho_s = \rhob \, \frac{51}{3} \, 
\frac{c^3}{\ln(1+c)-c/(1+c)} .
\label{rhos50-def}
\eeq
We again look for a prescription for $c(M_{50})$ of the form of Eq.(\ref{cM-1}),
and we find in Fig.~\ref{fig_lDk-50} below that a reasonable agreement with
the numerical simulations can be achieved by using
\beq
c(M_{50}) = 13 \, \left( \frac{M_{50}}
{2\times 10^{12} h^{-1} M_{\odot}}\right)^{-0.1} \, (1+z)^{-0.7} .
\label{cM-50}
\eeq
This is mostly set by the behavior of the power spectrum at high $k$,
where $\Delta^2>100$, as seen in Fig.~\ref{fig_lDk-200} from the comparison
between various models for $c(M)$ for halos defined by $\delta=200$.
Thus, the properties of the density power spectrum on large scales,
where $\Delta^2<10$, are almost independent of the choice of parameters
in Eq.(\ref{cM-50}) (restricted to a reasonable range).

\begin{figure}[htb]
\begin{center}
\epsfxsize=9 cm \epsfysize=8 cm {\epsfbox{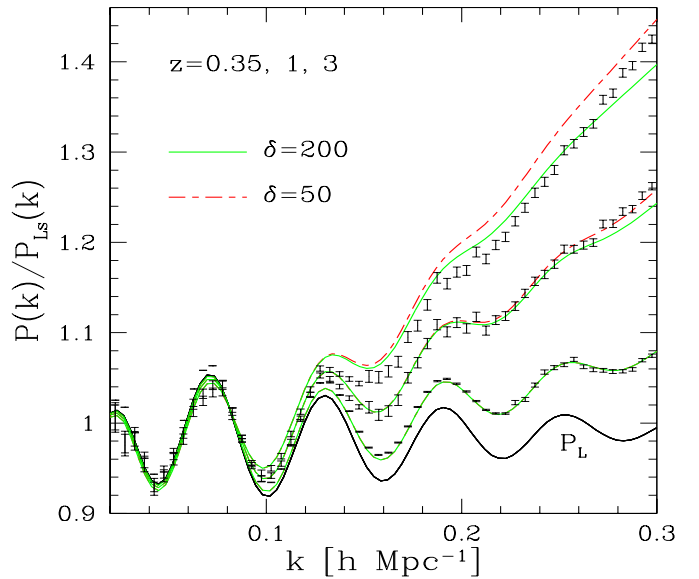}}
\end{center}
\caption{The ratio of the nonlinear power spectrum $P(k)$ to a smooth linear
power spectrum $P_{Ls}$ without acoustic baryonic oscillations, 
as in Fig.~\ref{fig_Pk-50}, for halos defined by $\delta=200$ (green solid curves,
identical to those of Fig.~\ref{fig_Pk-200}), or by $\delta=50$ 
(red dot-dashed curves).}
\label{fig_Pk-50}
\end{figure}

We compare in Fig.~\ref{fig_lDk-50} the power spectra obtained on nonlinear
scales by using halos defined either by $\delta=200$ or $\delta=50$
(while the perturbative term $P_{\rm pert}(k)$ is not modified).
As expected, truncating halo profiles at a larger radius yields some extra power
at large $k$ and improves somewhat the agreement with numerical simulations
in the intermediate range, where $\Delta^2 \sim 5$.
However, the change is quite small and the match to the N-body results is not perfect
yet. In fact, at redshift $z=0.35$, we obtain a smaller power spectrum around
$\Delta^2 \sim 100$ by using halos defined by $\delta=50$. However, one may
modify this by using a different concentration relation $c(M_{50})$
at that redshift.
In any case, these results suggest that the truncation at $r_{200}$
is not the unique reason for the discrepancy between the model and N-body
simulations in this transitory range, and as explained
in Sect.~\ref{Comparison} another possible source of inaccuracy is the partial
resummation of high-order terms in the perturbative term $P_{\rm pert}$.

Next, we compare in Fig.~\ref{fig_Pk-50} the power spectra obtained on quasi-linear
scales. We can see that on these very large scales, $k < 0.3 h$Mpc$^{-1}$,
the modification of the 1-halo term
has a very weak effect. However, it improves somewhat the agreement with the
simulations at low redshift, $z= 1$ (at $k\sim 0.3 h$Mpc$^{-1}$, as the gap below
the N-body results in the intermediate range is repelled to slightly higher $k$),
although it leads to a slight overestimate for
$P(k)$ at $z=0.35$. On the other hand, it appears to make almost
no change over these scales at higher redshift, $z= 3$.
In agreement with the discussion in Sect.~\ref{Comparison} and with
\citet{Valageas2010a}, this expresses the fact that for CDM power spectra
the scope of perturbative expansions is somewhat broader at higher $z$,
in the sense that the range of $k$ where higher order terms play a role
is wider. This corresponds to the interval $[k_{1\rm loop},k_{\rm s.c.}]$
where terms beyond linear order are already significant, as compared with the
linear power $P_L$, but still larger than the non-perturbative correction
associated for instance with shell crossing effects.

That our results for the density power spectrum only show a very weak
dependence on the details of the halo definition (by $\delta=200$ or $50$)
can actually be considered as a reassuring sign, rather than a problem
in the attempt to improve the agreement with simulations. Indeed,
if the splitting (\ref{Pk-halos}) and all subsequent computations were exact,
halos would only appear as an intermediate tool and would disappear in
the final result, since the matter density field is independent of how we
split it over halos at later stages. Therefore, the final result for $P(k)$ should
not depend on the details of the definition of the halos. In practice, this
cannot be the case because the splitting (\ref{Pk-halos}) itself is only approximate
as long as we consider spherical halos, and the subsequent treatment of both
2-halo and 1-halo contributions involves some approximations, as in
Eqs.(\ref{Pkxq-pert2}) and (\ref{mean-M1}). This leads to some artificial dependence
on the details of the halo model, but as shown in Figs.~\ref{fig_lDk-50} and
\ref{fig_Pk-50} this effect is rather small and argues for the robustness of the
model.

Therefore, Figs.~\ref{fig_lDk-50} and \ref{fig_Pk-50} suggest that there is still
room for systematic improvement, by including higher order terms in the
perturbative contribution $P_{\rm pert}$.
However, this may come at the price of heavier and slower computations,
which we do not investigate here.
On the other hand, they show that predictions on large scales are largely
independent of the details of the underlying halo model.

\subsection{Impact of the ``1-halo'' counterterm}
\label{counterterm}

\begin{figure}[htb]
\begin{center}
\epsfxsize=9 cm \epsfysize=8 cm {\epsfbox{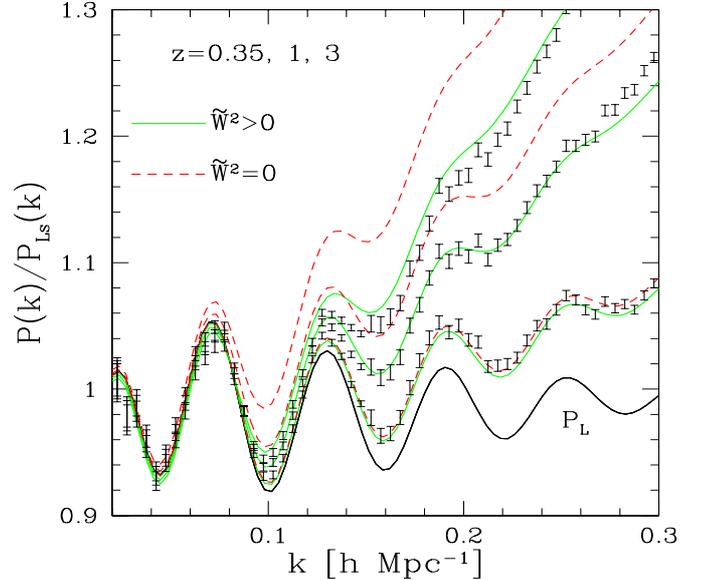}}
\end{center}
\caption{The ratio of the nonlinear power spectrum $P(k)$ to a ``no-wiggle''
linear power spectrum $P_{Ls}(k)$, as in Fig.~\ref{fig_Pk-200}, at redshifts
$z=0.35$, $1$, and $3$. The green solid lines correspond to our fiducial model,
with halos defined by a density contrast $\delta=200$, and are identical
to the green solid lines shown in Fig.~\ref{fig_Pk-200}, while the red dashed lines
are obtained by setting to zero the 1-halo counterterm $\tW(kq_M)^2$
of Eq.(\ref{Pk-1H}).}
\label{fig_Pk-noWth}
\end{figure}

We have seen in Sects.~\ref{Comparison} and \ref{delta=50-1}
that quantitative details of the 1-halo term
can have some effect on the high-$k$ tail (the concentration $c(M)$ of the halo
profiles) or on the intermediate range where $\Delta^2(k) \sim 5$
(the truncation radius of the halos).
We investigate here the role of a more fundamental ingredient of the 1-halo 
contribution, namely the counterterm $\tW(kq_M)^2$ in Eq.(\ref{Pk-1H}).
Thus, focusing on the case of halos defined by the nonlinear density
contrast $\delta=200$, we compare in Fig.~\ref{fig_Pk-noWth} the predictions
of our fiducial model with those
that are obtained by setting to zero the 1-halo counterterm $\tW(kq_M)^2$
of Eq.(\ref{Pk-1H}). Thus, we can see
that, even though the 1-halo counterterm plays a negligible role at high $k$,
it cannot be discarded at low $k$ if we require a high accuracy.
As explained in Sect.~\ref{1-halo}, this is also related to the fact that this counterterm
is required to obtain the $k^2$ tail at low $k$ expected for the 1-halo
contribution (if we do not insist on momentum conservation). Discarding this
counterterm leads to a nonzero asymptote at low $k$ for $P_{1\rm H}(k)$,
which eventually dominates over the linear power spectrum, which roughly decreases
as $P_L(k) \sim k$ for CDM cosmologies. On the scales probed in
Fig.~\ref{fig_Pk-noWth}, $k \sim 0.2 h$Mpc$^{-1}$, we have not entered yet the
very low $k$ regime where this would give rise to a spurious divergence.
However, we can clearly see the spurious extra power that would be generated
by the lack of this counterterm. 
Note that on these scales there are no fitting parameters
that could be tuned to compensate for this extra power. Indeed, the details of the
halo model, such as the concentration $c(M)$ (see Fig.~\ref{fig_Pk-200}),
or the truncation radius of the profiles (see Fig.~\ref{fig_Pk-50}), only have
a weak impact in this domain and could not balance this extra power.
On the other hand, the 2-halo term is also highly constrained, as the
blue dashed and dotted curves in Fig.~\ref{fig_Pk-200}, associated with standard
1-loop perturbation theory and steepest-descent resummation, give an estimate
of the possible scatter between the predictions that can be obtained from
various perturbative schemes.
This is most clearly seen around the peak at $k \sim 0.13 h$Mpc$^{-1}$ for
$z=0.35$ (as could be expected the extra power associated with such an incorrect
modelization of the 1-halo term grows at lower $z$ as larger halos form).
By contrast, the reasonably good agreement with simulations obtained by our
complete model (solid lines) suggests that it does not miss important
physical ingredients (as discussed in Sect.~\ref{Comparison} difficulties
mostly arise at higher $k$, associated with the transition between the 
2-halo and 1-halo contributions).

\subsection{Benefit of higher order perturbative terms}
\label{Benefit}

\begin{figure}[htb]
\begin{center}
\epsfxsize=9 cm \epsfysize=8 cm {\epsfbox{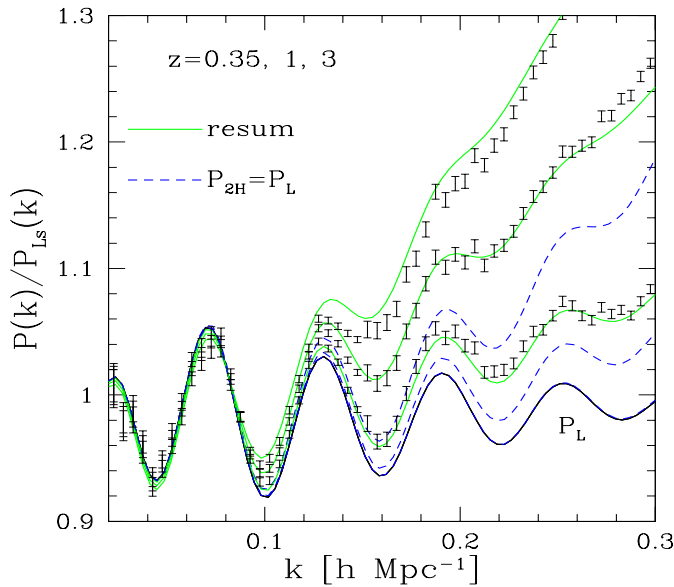}}
\end{center}
\caption{The ratio of the nonlinear power spectrum $P(k)$ to a ``no-wiggle''
linear power spectrum $P_{Ls}(k)$, as in Fig.~\ref{fig_Pk-200}, at redshifts
$z=0.35$, $1$, and $3$. The green solid lines ``resum'' correspond to our model
as in Fig.~\ref{fig_Pk-200}, whereas the blue dashed lines correspond to the
approximation $\PtH=P_L$ (it cannot be distinguished from the linear
power $P_L$ on these scales at $z=3$).}
\label{fig_Pk-PL}
\end{figure}

\begin{figure}[htb]
\begin{center}
\epsfxsize=9 cm \epsfysize=8 cm {\epsfbox{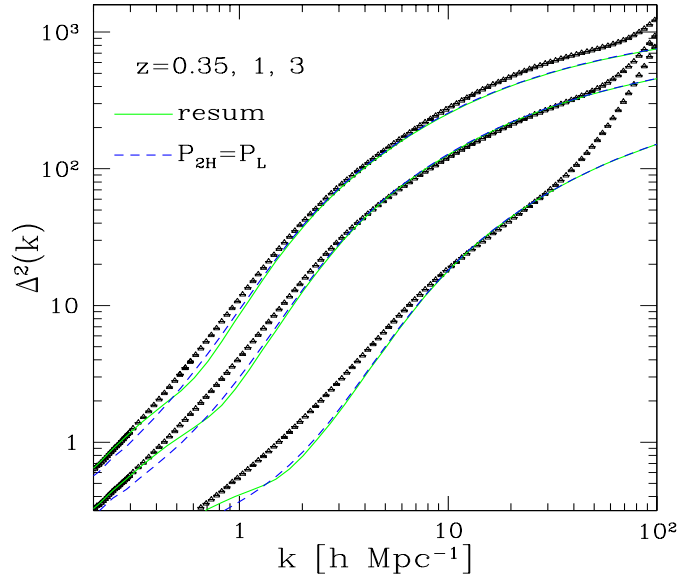}}
\end{center}
\caption{The power per logarithmic interval of $k$, as in Fig.~\ref{fig_lDk-200},
at redshifts $z=0.35$, $1$, and $3$. We plot the curves obtained with our model
(green solid lines) and with the approximation $\PtH=P_L$ (blue dashed lines),
as in Fig.~\ref{fig_Pk-PL}.}
\label{fig_lDk-PL}
\end{figure}

Before we investigate other choices for the perturbative term $P_{\rm pert}$
in the next sections, we first compare in Figs.~\ref{fig_Pk-PL} and \ref{fig_lDk-PL} 
our model with the more standard approach, where one simply uses the linear
power $P_L$ for the 2-halo contribution. Thus, we replace Eq.(\ref{Pkxq-pert2})
by the approximation $\PtH(k)=P_L(k)$, while we keep the same 1-halo contribution
(\ref{Pk-1H}), with its new counterterm that ensures a satisfactory behavior
at low $k$.
In agreement with previous works, we can see in Fig.~\ref{fig_Pk-PL}
that on the large scales associated with the baryon acoustic oscillations the
approximation $\PtH(k)=P_L(k)$ is not sufficient to obtain a good match to
the numerical simulations, contrary to the use of the one-loop perturbative term
$P_{\rm pert}(k)$ obtained by the ``steepest-descent'' scheme recalled in
Sect.~\ref{2-halo-1}. This means that on these scales the departure from the
linear power is not yet due to the non-perturbative contribution associated with
the 1-halo term, but to higher order perturbative terms. This agrees with
the results of \cite{Valageas2010a}, based on the Zeldovich dynamics,
which show that many orders of perturbation theory are relevant before
the power spectrum is dominated by the non-perturbative contribution
associated with shell crossing or halo formation (typically one can go up to
order $P_L^9$ at $z=0$ and $P_L^{66}$ at $z=3$ for a $\Lambda$CDM
cosmology, at least for this simpler case).
This motivates the use of perturbative approaches that include higher order
contributions.

On smaller scales, shown in Fig.~\ref{fig_lDk-PL}, the difference between
our model for $\PtH(k)$ and the popular approximation $\PtH(k)=P_L(k)$ is
negligible. This agrees with Fig.~\ref{fig_lDk-200}, where we could see that at
high $k$ the one-loop resummed contribution to $\PtH(k)$ is near
the linear power $P_L(k)$ and subdominant as compared with the 1-halo
contribution.

\subsection{Eulerian perturbation theories}
\label{Eulerian}

\begin{figure}[htb]
\begin{center}
\epsfxsize=9 cm \epsfysize=8 cm {\epsfbox{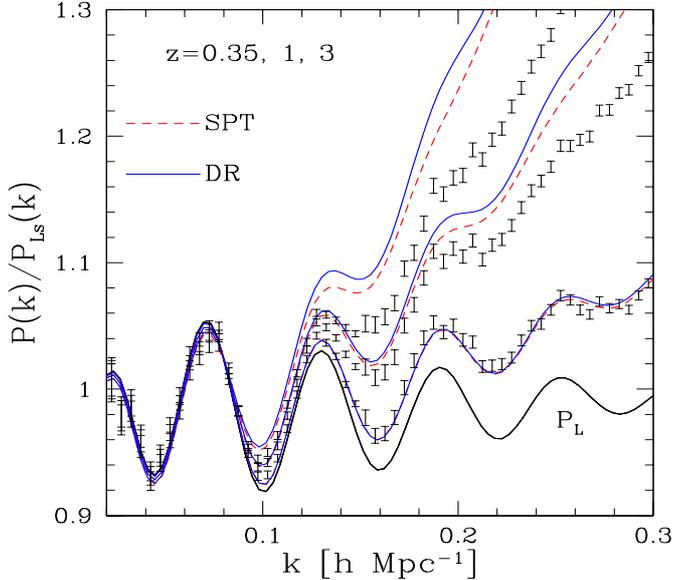}}
\end{center}
\caption{The ratio of the nonlinear power spectrum $P(k)$ to a ``no-wiggle''
linear power spectrum $P_{Ls}(k)$, as in Fig.~\ref{fig_Pk-200}, at redshifts
$z=0.35$, $1$, and $3$. The red dashed lines ``SPT'' correspond to the use
of 1-loop ``Standard Perturbation Theory'' for the term $P_{\rm pert}$ in the
2-halo contribution (\ref{Pkxq-pert2}), whereas the blue solid lines ``DR''
correspond to the use of the ``Decaying Response'' function from \citet{Crocce2006b}
in the external double lines of the two upper diagrams of Fig.~\ref{fig_Csd}.
The 1-halo term is the same as the one used in Sect.~\ref{Comparison}.}
\label{fig_Pk-st}
\end{figure}

We consider here two other choices for the perturbative term
$P_{\rm pert}$ of the 2-halo contribution (\ref{Pkxq-pert2}).
A first choice is simply to use the standard perturbation theory and to
write at one-loop order $P_{\rm pert}=P_{1\rm loop}(k)$, where $P_{1\rm loop}(k)$
was given in Eq.(\ref{P1loop-def}).
A second choice is to keep the diagrammatic expression given by the first line
in Fig.~\ref{fig_Csd}, but to replace the external double lines by the Gaussian-decay
response function (propagator) obtained by \citet{Crocce2006a,Crocce2006b} instead
of using the response function obtained by the steepest descent resummation
scheme. This is not identical to the ``RPT'' resummation used in
\citet{Crocce2008}, where the ``bubble'' in the upper right diagram of
Fig.~\ref{fig_Csd} is also replaced by resummed two-point correlations,
obtained from this same nonlinear propagator (which corresponds to the
partial resummation of further diagrams).
Here we keep linear two-point correlation functions in this ``bubble''
to keep factorizable integrals, as in Eq.(\ref{C-fact}),
since our goal is simply to estimate the dependence
on the choice of resummation scheme. Moreover, it is interesting to evaluate
various factorizable schemes of this kind as they allow faster computations.
Two other differences with the prescription used in \citet{Crocce2008} are the
presence of the prefactor $\FtH(1/k)$ in Eq.(\ref{Pkxq-pert2}), and the fact that
we do not introduce an extra tuning in the decay of the response function
by replacing the linear velocity dispersion $\sigma_v$, which governs its high-$k$
cutoff, by a nonlinear velocity dispersion obtained from fits to the nonlinear power
spectrum from numerical simulations.
Indeed, one of the main points of our approach is precisely that the term
$P_{\rm pert}(k)$ in the 2-halo contribution (\ref{Pkxq-pert2}) should be
obtained from systematic perturbation theories, to ensure a good accuracy
and robustness for a variety of cosmologies, so that we avoid introducing
fitting parameters in this factor.
Then, these two alternatives, ``standard perturbation theory'' (SPT) and
``Gaussian-like decaying response function'' (DR), and the complete 
steepest-descent resummation used in the previous sections, agree with each
other up to one-loop order.

We first show in Fig.~\ref{fig_Pk-st} the power spectra obtained on large scales
for redshifts $z=0.35, 1$, and $3$. We plot the results obtained by using
either the 1-loop ``standard perturbation theory'', or
the ``decaying response'' function from \citet{Crocce2006b}
in the external double lines of the two upper diagrams of Fig.~\ref{fig_Csd},
for the term $P_{\rm pert}$ in the 2-halo contribution (\ref{Pkxq-pert2}).
In both cases the 1-halo term is the same as the one used in
Sect.~\ref{Comparison}, so that these curves only differ from those shown in
Fig.~\ref{fig_Pk-200} through the term $P_{\rm pert}$.
As was already noticed in Fig.~\ref{fig_Pk-200}, and in agreement with
previous works \citep{Crocce2008,Taruya2008,Carlson2009},
the 1-loop standard perturbation theory overestimates the power on these scales,
especially at low redshift. The dashed curves shown in Fig.~\ref{fig_Pk-st}
are not identical to those shown in Fig.~\ref{fig_Pk-200} and in other works,
because the perturbative term $P_{\rm pert}$ is multiplied by the prefactor
$\FtH(1/k)$ in Eq.(\ref{Pkxq-pert2}). Moreover, the results also include the (small)
1-halo contribution. However, on these scales the prefactor $\FtH(1/k)$ remains
very close to unity (see Fig.~\ref{fig_F2H}) and it is not sufficient to significantly
improve the agreement with the N-body results.
The use of the  ``decaying response'' function leads to results that are very close
to those obtained with standard perturbation theory and also overestimates the
power spectrum on these large scales.
Using the full ``renormalized perturbation theory''
(RPT) described in \citet{Crocce2008} leads to a smaller prediction for $P(k)$.
This can be understood from the fact that inserting the
``decaying response'' function into the bubble of the upper right diagram
of Fig.~\ref{fig_Csd}, as in ``RPT'', clearly suppresses the nonlinear power
and explicit computations show that this yields a good match to results from
simulations \citep{Crocce2008,Carlson2009}.

\begin{figure}[htb]
\begin{center}
\epsfxsize=9 cm \epsfysize=8 cm {\epsfbox{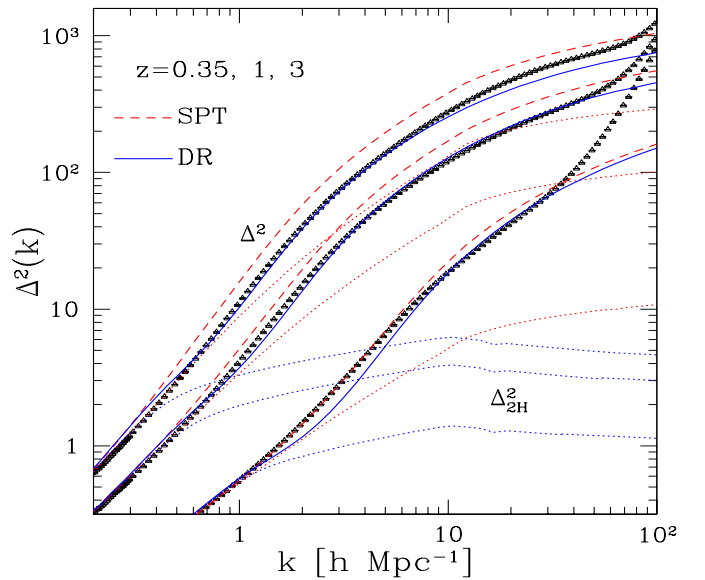}}
\end{center}
\caption{The power per logarithmic interval of $k$, as in Fig.~\ref{fig_lDk-200},
at redshifts $z=0.35$, $1$, and $3$. We plot the curves obtained using 1-loop
``standard perturbation theory'' (red dashed lines for the full nonlinear power
$\Delta^2$, and lower red dotted lines for the associated 2-halo contribution)
or the ``decaying response'' function from \citet{Crocce2006b} (blue solid lines 
for the full nonlinear power $\Delta^2$, and lower blue dotted lines for the
associated 2-halo contribution). Both the full and 2-halo power spectra obtained
with ``SPT'' are larger than their ``DR'' counterparts.}
\label{fig_lDk-st}
\end{figure}

Next, we show in Fig.~\ref{fig_lDk-st} the power per logarithmic interval of $k$,
as in Fig.~\ref{fig_lDk-200}. We plot both the full nonlinear power $\Delta^2$,
as in Fig.~\ref{fig_Pk-st}, and the 2-halo contribution, for the ``SPT'' and
``DR'' cases.
As already pointed out in Sect.~\ref{2-halo-1}, we can see that standard
perturbation theory fails at a qualitative level as the 2-halo contribution
keeps growing at high $k$ and becomes very large, whereas 
the term $P_{\rm pert}$ being associated with particle pairs that have not
collapsed within a single object it should remain near $P_L$ at most.
Moreover, this spurious growth is not suppressed by the prefactor $\FtH(1/k)$,
which does not decrease sufficiently fast at high $k$, as seen in Fig.~\ref{fig_F2H}.
Even though this 2-halo contribution remains smaller than the 1-halo
contribution on these scales, it leads to a significant extra power that worsens the
agreement with the numerical simulations. As can be seen from 
Figs. ~\ref{fig_lDk-200} and \ref{fig_lDk-50}, this extra power cannot be compensated
by modifications to the underlying halo model (such as the concentration $c(M)$
or the halo truncation radius), especially in the range $5<\Delta^2<200$
where the dependence on the details of the halo model is rather weak.
Of course, this problem would become increasingly severe as one includes
higher order terms that grow increasingly fast at high $k$
\citep{Bernardeau2002,Crocce2006a}.
This means that standard perturbation theory cannot be used to build systematic
models for the density power spectrum, such as those studied in this paper.
As pointed out in Sect.~\ref{2-halo-1}, this provides a further motivation to
investigate resummation schemes that would be better behaved.

We can see that using the ``decaying response'' function in the external double
lines of Fig.~\ref{fig_Csd} leads to a 2-halo contribution that remains small
and close to the linear power spectrum at high $k$, as for the
direct steepest-descent scheme used in Fig.~\ref{fig_lDk-200}.
This shows again the benefit of using resummation schemes instead of standard
perturbation theory to obtain a well-behaved perturbative term $P_{\rm pert}(k)$
that also agrees with standard perturbation theory up to the order of truncation.
In fact, it happens that using this ``decaying response'' function yields
a somewhat larger perturbative term than the one obtained in
Sect.~\ref{Comparison} on large scales, as seen in Fig.~\ref{fig_Pk-st},
and this extra power actually improves the agreement with the N-body results
in the range where $1<\Delta^2<100$. Indeed, as discussed in Sect.~\ref{Comparison},
the implementation used in Fig.~\ref{fig_lDk-200} underestimates the nonlinear
power spectrum in this range, and the additional power due to the use
of this ``decaying response'' function helps to bridge the gap. It actually
provides a very good quantitative match, although at $z=3$ we again underestimate
somewhat the power spectrum in the transition range.
Thus, for practical quantitative purposes this may provide a good variant of the
model developed in this paper to compute the power spectrum, if one is
interested in these transition scales.
However, because on large scales this prescription actually gives a worse match
to numerical simulations, as seen in Fig.~\ref{fig_Pk-st}, the better agreement
on the transition scales is probably not more than a lucky coincidence. Thus,
it is likely to compensate
a meaningful lack of power (due to the various reasons discussed in
Sect.~\ref{Comparison}, such as the neglect of some higher order perturbative
terms or the limitations of the description in terms of spherical virialized halos)
by an unrelated artificial overestimate.

\subsection{Lagrangian perturbation theories}
\label{Lagrangian}

The Lagrangian re-interpretation of the halo model developed in this paper,
leading to the expression (\ref{Pkxq-pert1}) for the 2-halo contribution,
would most naturally lead us to consider Lagrangian perturbation theories.
It is possible to substitute such schemes into Eq.(\ref{Pkxq-pert1}), which
allows us to integrate in a self-consistent fashion over the $q-$dependent
prefactor $\FtH(q)$. Unfortunately, this does not yield very successful results
while leading to somewhat more complex expressions than those encountered
in Eulerian frameworks. Therefore, in this section we illustrate the problems
faced by current Lagrangian schemes by using the simpler expression
(\ref{Pkxq-pert2}), where the term $\FtH(q)$ has been factorized as the prefactor
$\FtH(1/k)$. Thus, as for the Eulerian perturbation theories used in the previous
sections, we only need the prediction for the resummed power $P_{\rm pert}(k)$ 
to compute the 2-halo contribution. This simplifies matters and only makes
a negligible change on large scales where $\FtH\simeq 1$.

\begin{figure}[htb]
\begin{center}
\epsfxsize=9 cm \epsfysize=8 cm {\epsfbox{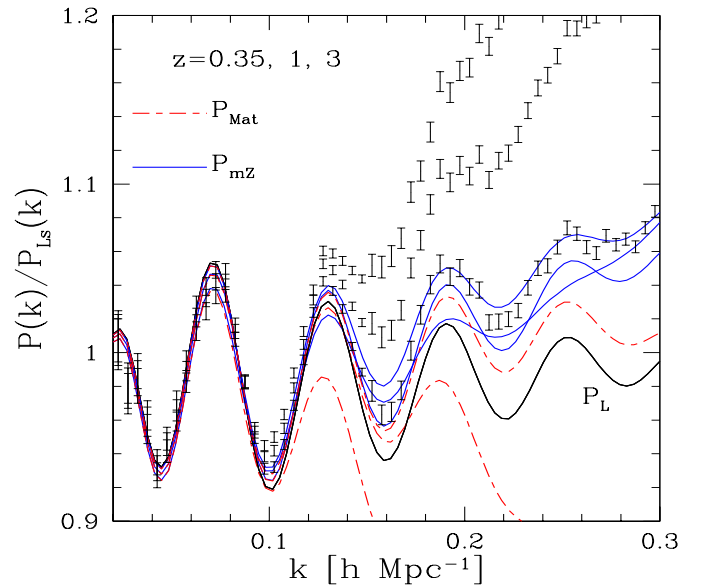}}
\end{center}
\caption{The ratio of the nonlinear power spectrum $P(k)$ to a ``no-wiggle''
linear power spectrum $P_{Ls}(k)$, as in Fig.~\ref{fig_Pk-200}, at redshifts
$z=0.35$, $1$, and $3$. The red dot-dashed lines ``$P_{\rm Mat}$'' correspond to
the use of the Lagrangian-based 1-loop expansion (\ref{P-Mat}) for the term
$P_{\rm pert}$ in the 2-halo contribution (\ref{Pkxq-pert2}), whereas the blue
solid lines ``$P_{\rm mZ}$'' correspond to the use of the ``modified Zeldovich''
expansion (\ref{PmZ-def}).
The 1-halo term is the same as the one used in Sect.~\ref{Comparison}.}
\label{fig_Pk-Z}
\end{figure}

\begin{figure*}
\begin{center}
\epsfxsize=6.1 cm \epsfysize=6 cm {\epsfbox{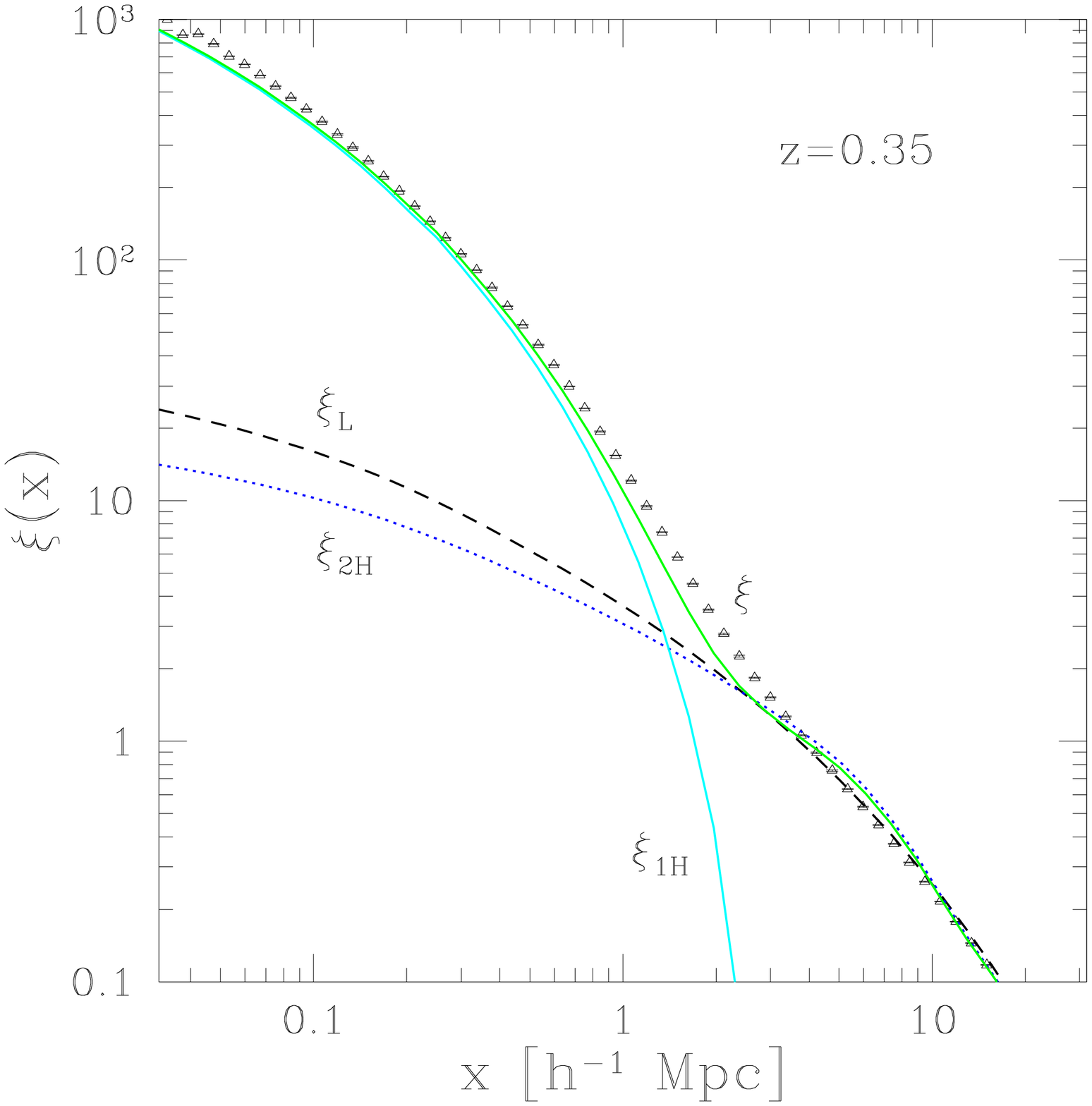}}
\epsfxsize=6.05 cm \epsfysize=6 cm {\epsfbox{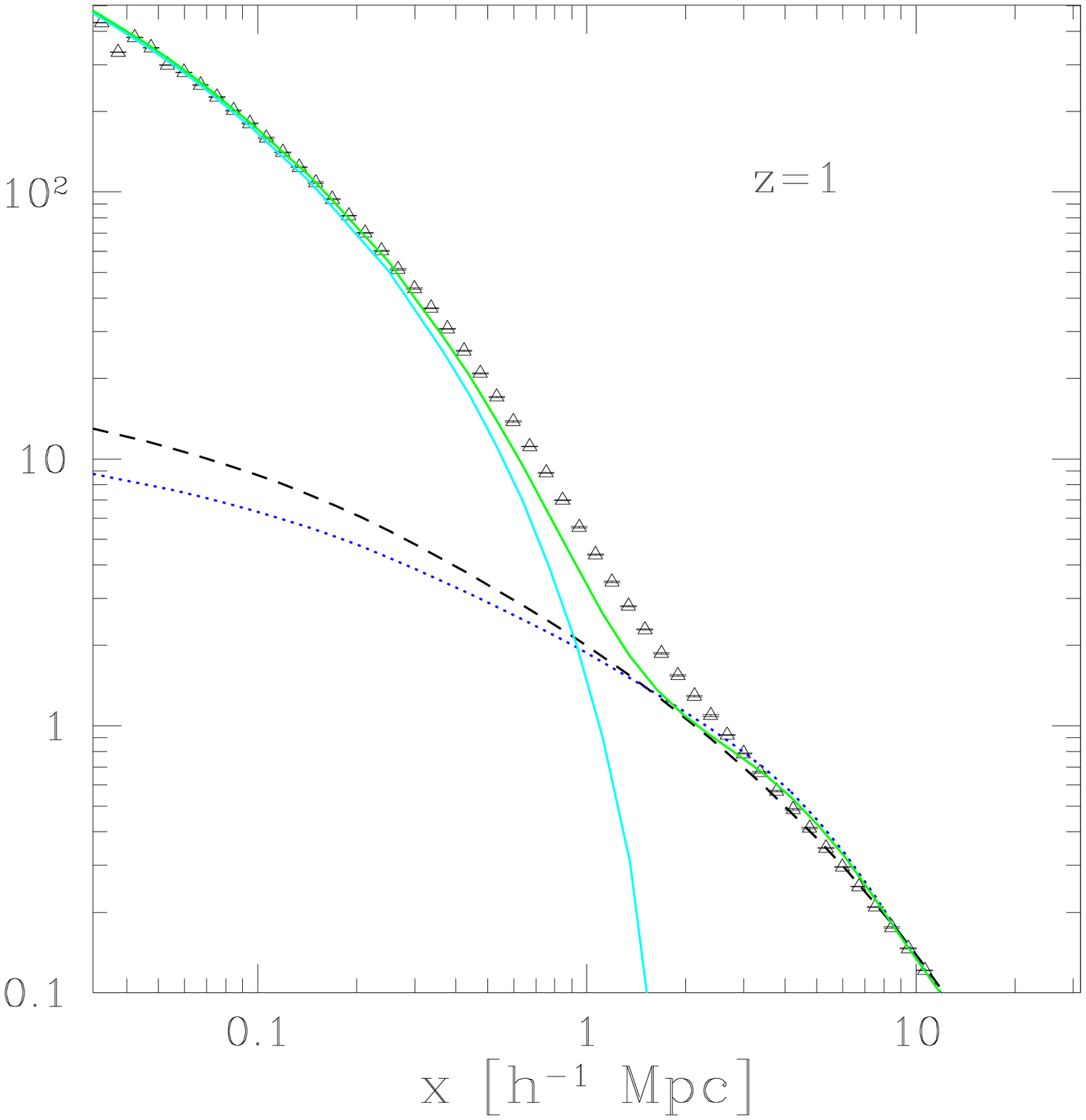}}
\epsfxsize=6.05 cm \epsfysize=6 cm {\epsfbox{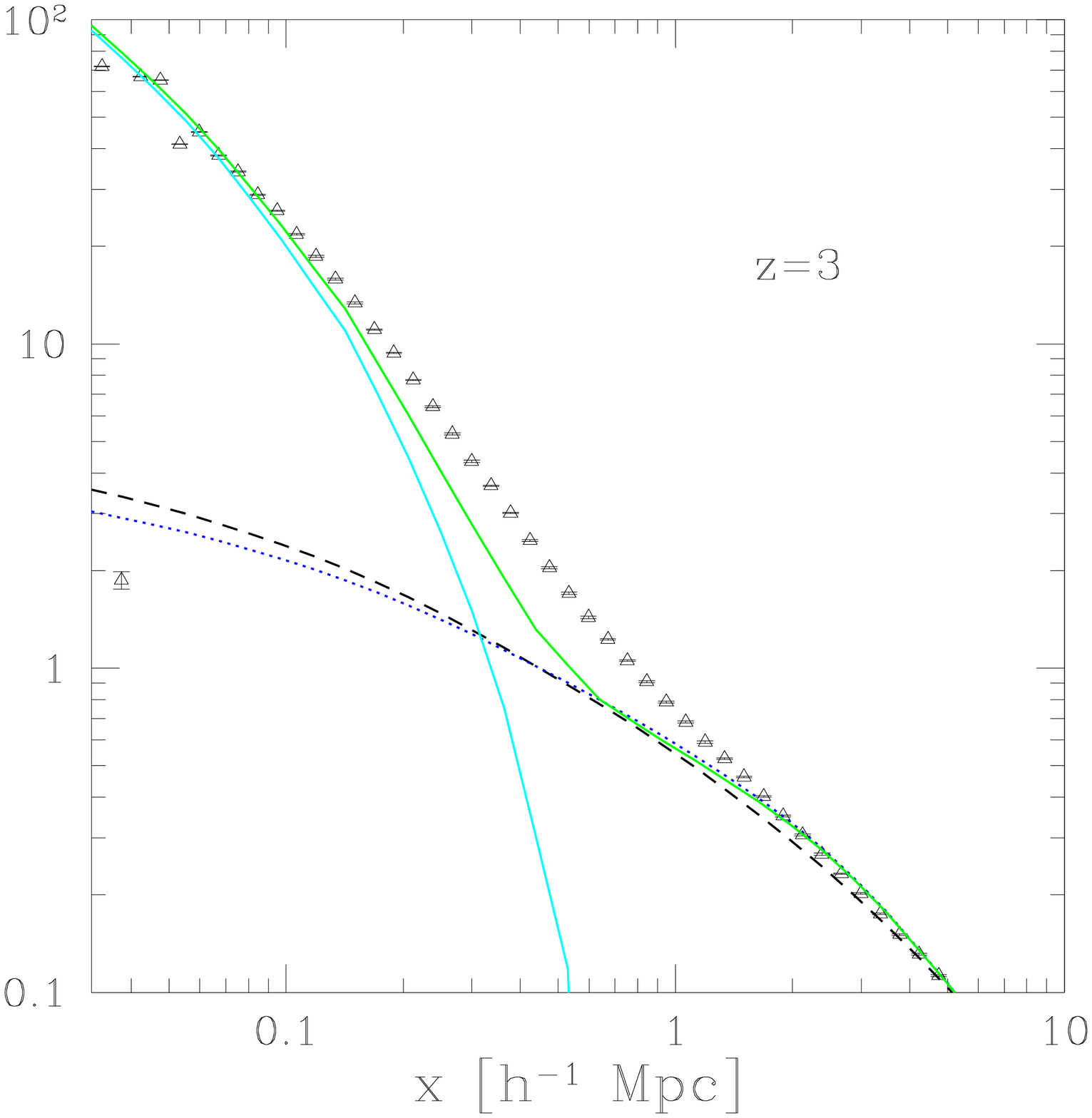}}
\end{center}
\caption{The real-space two-point correlation function $\xi(x)$, at redshifts $z=0.35$,
$1$, and $3$. The symbols are the results from the numerical simulations described
in Sect.~\ref{N-body-simulations}. The black dashed line is the linear correlation,
$\xi_L$, and the blue dotted line which is somewhat below at low $x$ is the 2-halo
contribution $\xi_{2\rm H}$. The steep solid line $xi_{1\rm H}$
is the 1-halo contribution, for the halo model described in
Sect.~\ref{1-halo-1}. The green solid line is the full nonlinear correlation function,
$\xi=\xi_{2\rm H}+\xi_{1\rm H}$.}
\label{fig_lxix-200}
\end{figure*}

We consider two schemes that are inspired by a Lagrangian framework.
A first approach is the one presented in \citet{Matsubara2008}, which for our
purposes amounts to expand $P(k)$ after factorization of a Gaussian
damping term $e^{-k^2\sigma_v^2}$ (see also \citet{Crocce2006a,Valageas2010a}),
where $\sigma_v$ is the rms one-dimensional linear displacement (or
the rms one-dimensional linear velocity, up to a time-dependent factor) given by
\beq
\sigma_v^2 = \frac{4\pi}{3} \int_0^{\infty} \dd k \, P_L(k) .
\label{sigmav-def}
\eeq
Thus, at 1-loop order the density power spectrum reads as \citep{Matsubara2008}
\beq
P_{\rm Mat}(k) = e^{-k^2\sigma_v^2} \left( P_L + P_{22} + P_{31}
+ k^2\sigma_v^2 P_L \right) ,
\label{P-Mat}
\eeq
where $P_{22}$ and $P_{31}$ are the 1-loop terms obtained in the standard
Eulerian perturbation theory, as in Eq.(\ref{P1loop-def}).
Of course, Eq.(\ref{P-Mat}) could be obtained at once by requiring consistency
with Eq.(\ref{P1loop-def}) at order $P_L^2$.
As discussed in \citet{Valageas2007b,Valageas2008,Valageas2010a},
and as can be seen from the computation of Lagrangian response functions
\citep{BernardeauVal2008,BernardeauVal2010a}, the prefactor $e^{-k^2\sigma_v^2}$
is associated with the coherent displacement of density structures by the long
wavelengths of the velocity field and should only appear in different-time
quantities, under the form $e^{-(D_1-D_2)^2 k^2\sigma^2_{v0}/2}$,
where $D_i$ is the linear growth factor at redshift $z_i$ and $\sigma_{v0}$
is the linear dispersion today, where $D(z=0)=1$.
Thus, looking for an expansion under the form (\ref{P-Mat}) leads to an artificially
strong Gaussian damping at high $k$, which has no physical meaning.
In order to remedy to this problem, which can also be seen as a consequence of the
truncation to a finite number of terms between the parenthesis in
Eq.(\ref{P-Mat}) (once we insist on looking for an expansion of this form),
we note that the nonlinear power spectrum generated by the Zeldovich dynamics
\citep{Zeldovich1970}
can also be written under this form,
\beq
P_{\rm Z}(k) = e^{-k^2\sigma_v^2} \sum_{n=1}^{\infty} P_{\rm Z}^{(n)}(k) ,
\label{PZ-def}
\eeq
where each term $P_{\rm Z}^{(n)}(k)$ scales as $(P_L)^n$.
Moreover, both the full nonlinear expression of the Zeldovich
power spectrum, $P_{\rm Z}(k)$, (i.e. the resummation of the series (\ref{PZ-def})),
and the expression of the terms $(P_L)^n$ of all orders, are explicitly known
\citep{Crocce2006a,Valageas2007b,Valageas2010a}. Then, a possible recipe is to
use for the nonlinear power spectrum (more precisely the perturbative
term $P_{\rm pert}$) the ``modified Zeldovich'' expression
\beqa
P_{\rm mZ}(k) & = & e^{-k^2\sigma_v^2} \left( P_L(k) + P^{(2)}(k) + 
\sum_{n=3}^{\infty} P_{\rm Z}^{(n)}(k) \right) \\
& = & P_{\rm Z}(k) + e^{-k^2\sigma_v^2} \left( P^{(2)}(k) - P_{\rm Z}^{(2)}(k) \right) ,
\label{PmZ-def}
\eeqa
where $P^{(n)}(k)$ is the term of order $(P_L)^n$ in the expansion of the form
(\ref{PZ-def}) for the gravitational power spectrum, whence
$P^{(2)}(k)=P_{22}(k) + P_{31}(k)+ k^2\sigma_v^2P_L(k)$ from Eq.(\ref{P-Mat}).
In other words, we complete the expansion (\ref{P-Mat}) by substituting the
terms obtained for the Zeldovich power spectrum for all higher orders, which allows
us to perform an explicit resummation as in the second line of Eq.(\ref{PmZ-def}).
This is clearly a systematic procedure, that can be pushed up to any order,
and agrees with standard perturbation theory up to the order of the substitution.
Since the resummed Zeldovich power spectrum no longer shows the
spurious Gaussian decay of Eq.(\ref{P-Mat}) but a more physical power-law decay
\citep{Taylor1996,Valageas2007b,Valageas2010a}, one may hope that this could
improve the agreement with results from numerical simulations.

We show in Fig.~\ref{fig_Pk-Z} the results we obtain using either Eq.(\ref{P-Mat})
or Eq.(\ref{PmZ-def}), at redshifts $z=0.35$, $1$, and $3$.
As pointed out above, we can see that the Gaussian damping term of Eq.(\ref{P-Mat})
leads to a fast decay for the nonlinear density power spectrum, which cannot be
compensated by the 1-halo term. The ``modified Zeldovich'' expansion
(\ref{PmZ-def}), which only implies a slower power-law decay at high $k$,
fares better as the nonlinear power spectrum keeps growing on these scales.
However, this is not sufficient to obtain a satisfactory match with the N-body
results. Thus, the comparison with Fig.~\ref{fig_Pk-200} shows that these
Lagrangian-based expansions are not competitive with Eulerian schemes, and
we do not investigate further these approaches in this paper.
We note however that this is rather unfortunate, since in principles Lagrangian
perturbation theories would seem better suited to compute the 2-halo
term (\ref{Pkxq-pert1}), and more importantly they should be better suited to
computations in redshift space \citep{Matsubara2008}.
Therefore, it remains of interest to investigate whether other resummation
schemes could be developed within the Lagrangian framework, but
Fig.~\ref{fig_Pk-Z} suggests that some important new ingredients are required
to make significant progress.

\section{Real-space two-point correlation}
\label{Real-space}

\begin{figure*}
\begin{center}
\epsfxsize=6.1 cm \epsfysize=6 cm {\epsfbox{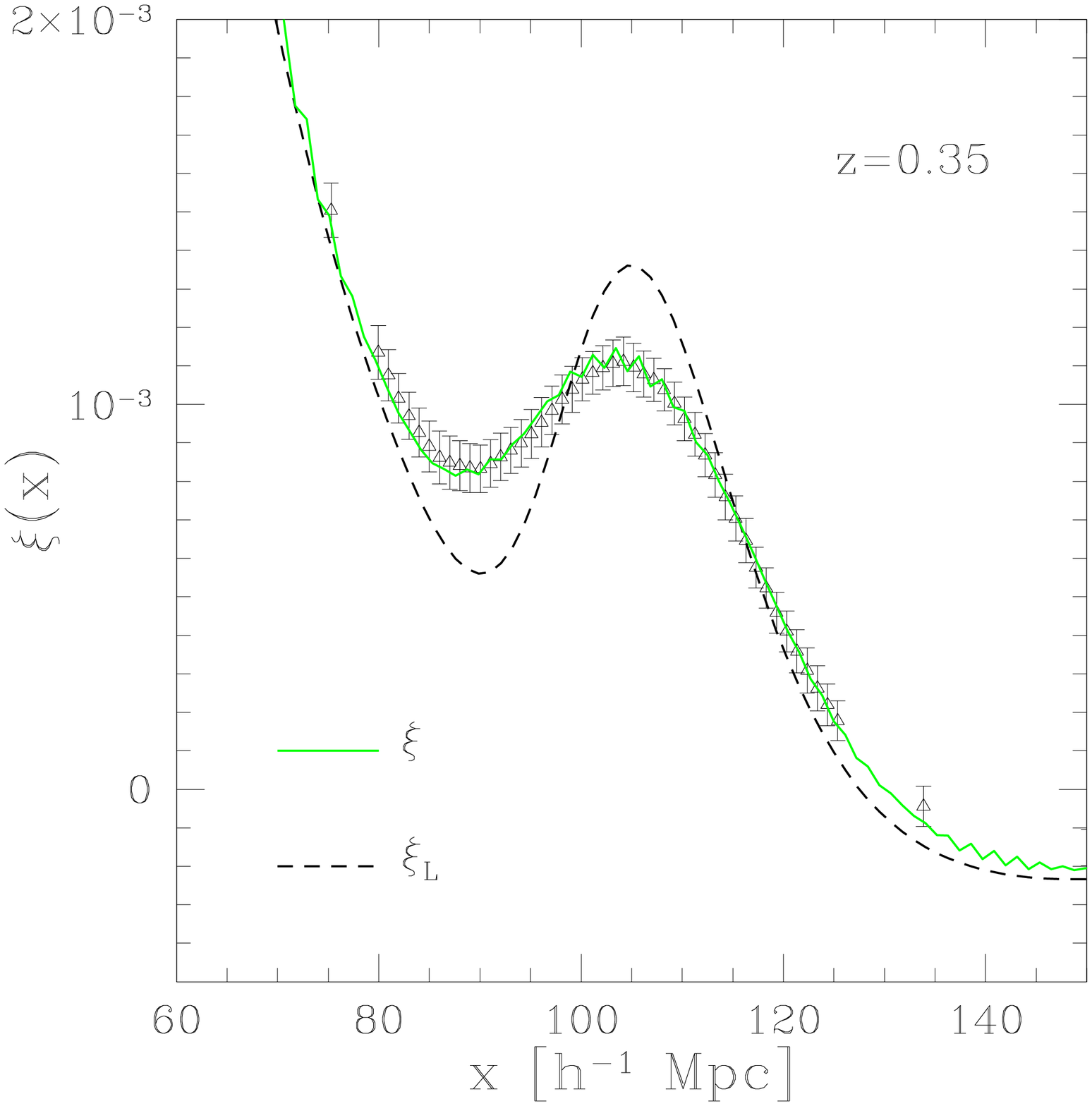}}
\epsfxsize=6.05 cm \epsfysize=6 cm {\epsfbox{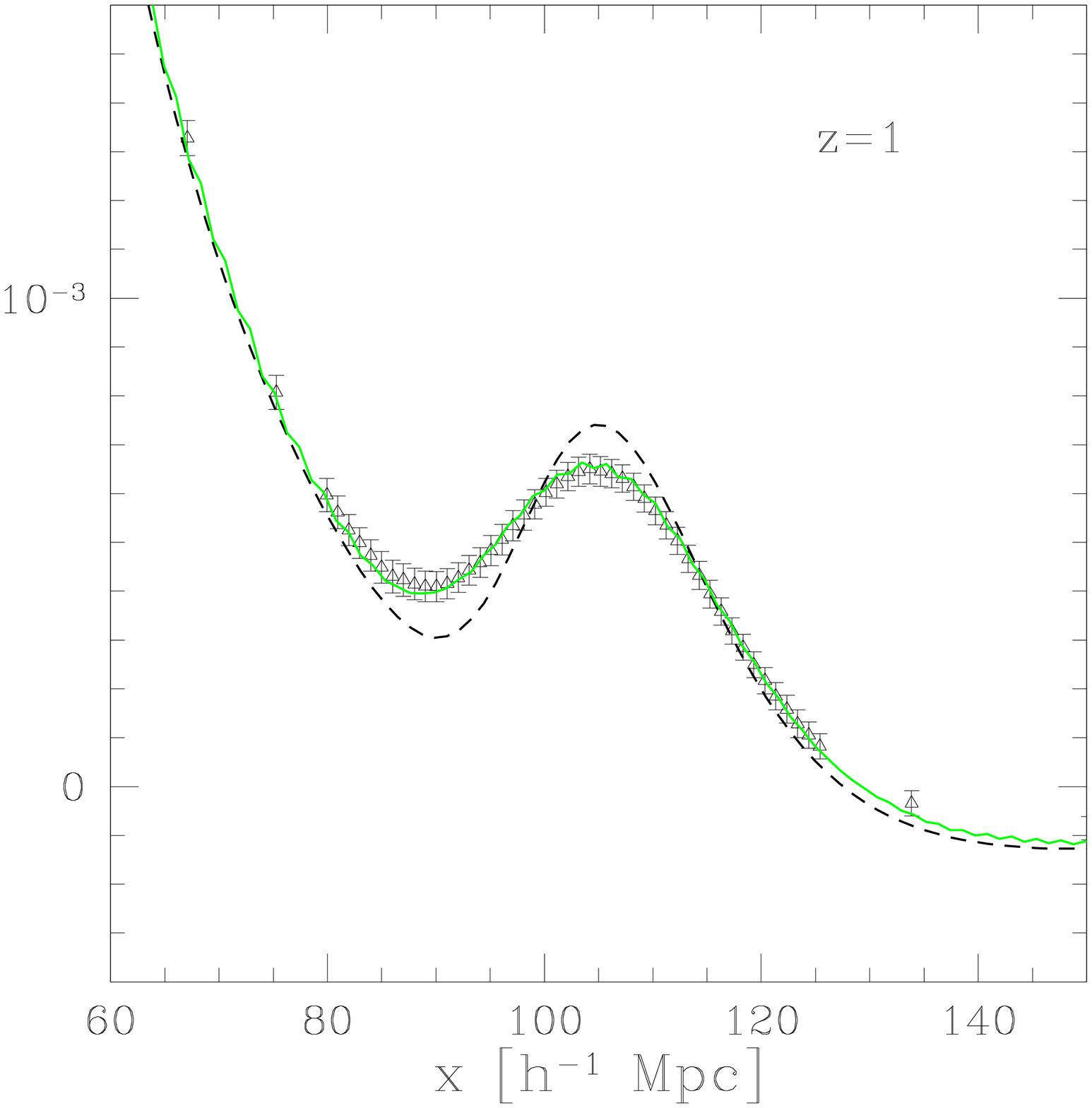}}
\epsfxsize=6.05 cm \epsfysize=6 cm {\epsfbox{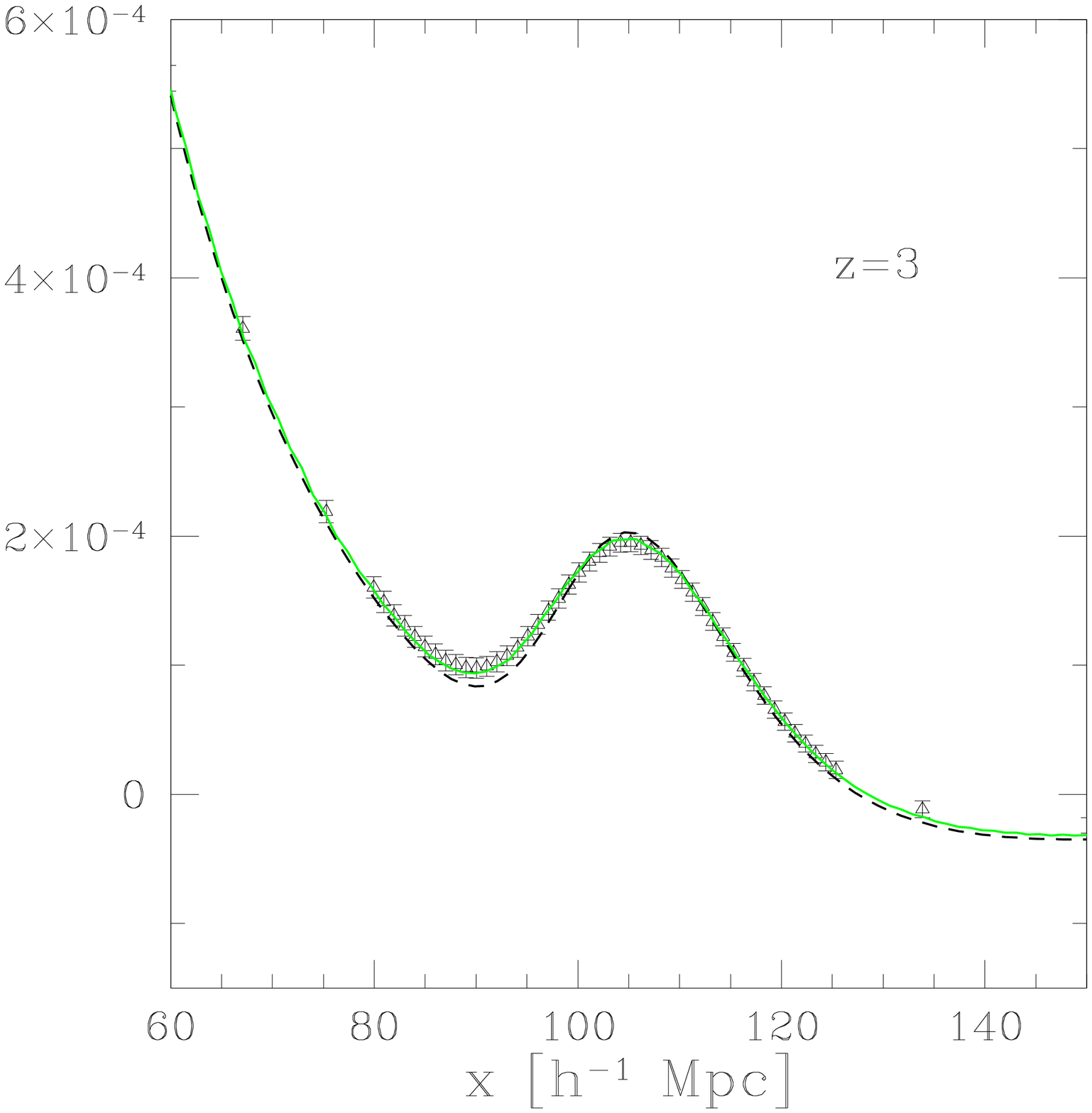}}
\end{center}
\caption{The real-space two-point correlation function $\xi(x)$, at redshifts $z=0.35$,
$1$, and $3$, as in Fig.~\ref{fig_lxix-200} but on larger scales.
Since the 1-halo contribution is negligible on these scales we only show the
linear correlation (black dashed line) and the full nonlinear correlation (green solid
line).}
\label{fig_xix-200}
\end{figure*}

Finally, we consider in this section the real-space two-point correlation function,
$\xi(|\vx_2-\vx_1|)=\lag\delta(\vx_1)\delta(\vx_2)\rag$, that is predicted by our
model. It is related to the Fourier-space power spectrum through
\beqa
\xi(x) & = & 4\pi \int_0^{\infty} \dd k \, k^2 P(k) \, \frac{\sin(kx)}{kx} \\
& = & \int_0^{\infty} \frac{\dd k}{k} \, \Delta^2(k) \, \frac{\sin(kx)}{kx} ,
\label{xi-def}
\eeqa
so that we obtain $\xi(x)$ by integration over $k$ of the nonlinear power
spectrum (\ref{Pk-halos}).
In particular, from the decomposition (\ref{Pk-halos}) we can write the nonlinear
correlation as the sum of 1-halo and 2-halo contributions,
\beq
\xi(x) = \xi_{1\rm H}(x) + \xi_{2\rm H}(x) ,
\label{xi-halos}
\eeq
which are the Fourier transforms of the power spectra (\ref{Pkxq-1H}) and
(\ref{Pkxq-2H}).

We show our results at redshifts $z=0.35$, $1$, and $3$, in Figs.~\ref{fig_lxix-200}
and \ref{fig_xix-200}.
Here we consider our fiducial model presented in Sects.~\ref{implementation}
and \ref{Comparison},
with the 2-halo perturbative term $P_{\rm pert}(k)$ given by the
direct steepest-descent scheme detailed in App.~\ref {direct-steepest-descent},
and the 1-halo contribution obtained for halos defined by a density contrast
$\delta=200$ and the mass-concentration relation (\ref{cM-1}).

In agreement with our results for the power spectrum shown in Fig.~\ref{fig_lDk-200},
we can see in Fig.~\ref{fig_lxix-200} that our model provides a good match to
numerical simulations on both small and large scales. As is well known,
on large scales the two-point correlation is governed by the 2-halo contribution
$\xi_{2\rm H}$, whereas on small scales it is governed by the
1-halo contribution. 
As for the power spectrum, the transition scales where the 2-halo and 1-halo
contributions are of the same order are more difficult to reproduce, and we again
underestimate the power (here the real-space correlation) in this range.
This is a direct consequence of the same feature observed in Fourier space
in Fig.~\ref{fig_lDk-200}. In particular, we also find that the range and amplitude
of this discrepancy is larger at $z=3$ than at $z=0.35$.
Again, improving theoretical predictions on this domain would require pushing
perturbation schemes to higher orders and/or improving the description of
halo outer shells.

We focus in Fig.~\ref{fig_xix-200} on the large scales around the baryon acoustic
peak. We can check that our model agrees very well with the N-body results and
reproduces the well-known damping of the BAO peak by the nonlinearities of
the dynamics (mode-coupling effect). This shows that for the purpose of
BAO studies, the accuracy of our theoretical model is certainly sufficient.
In particular, on these large scales, which are far from the transition region
seen in Fig.~\ref{fig_lxix-200}, the 1-halo contribution is completely negligible
and the two-point correlation is set by the perturbative term $P_{\rm pert}$ of the
2-halo contribution. Thus, it is given by a systematic perturbation theory,
and we can check that the simple 1-loop direct steepest-descent scheme detailed
in App.~\ref {direct-steepest-descent} already provides quite satisfactory results.
For practical purposes, the main improvement that remains to be achieved
is to extend these results to redshift space, but this is beyond the scope of
this paper and we leave this point for future studies.

\begin{figure}[htb]
\begin{center}
\epsfxsize=9 cm \epsfysize=8 cm {\epsfbox{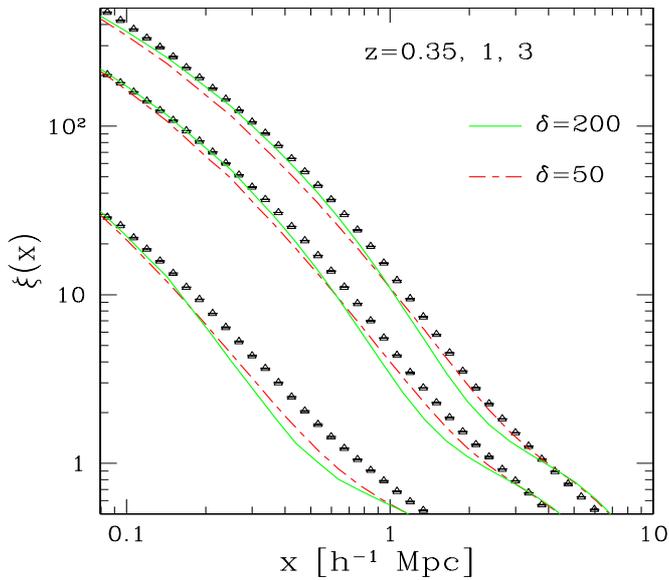}}
\end{center}
\caption{The two-point correlation $\xi(x)$, as in Fig.~\ref{fig_lxix-200}, but for
halos that are defined either by a density contrast $\delta=200$
(solid curves, identical to those of Fig.~\ref{fig_lxix-200}), or by $\delta=50$
(dot-dashed curves).}
\label{fig_lxix-50}
\end{figure}

We compare in Fig.~\ref{fig_lxix-50} the correlation functions obtained with halos
defined by a density threshold $\delta=200$ or $\delta=50$. This is the real-space
counterpart of Fig.~\ref{fig_lDk-50}. In agreement with Fig.~\ref{fig_lDk-50}, we can
see that extending the truncation radius of the halo profiles increases somewhat
the power in the transition range between the 2-halo and 1-halo
contributions. However, this is not sufficient to obtain a perfect match to the
numerical simulations, especially at $z=3$. Again, this suggests that part of
the discrepancy is due to the incomplete resummation of higher order terms
of the perturbative contribution $P_{\rm pert}(k)$.

We do not plot the counterpart of Fig.~\ref{fig_xix-200}, because on these large
scales the 1-halo contribution is very small and we have checked that
the curves obtained with halos defined by $\delta=200$ or $\delta=50$
cannot be distinguished in the figure: the two curves coincide within the thickness
of the curves of Fig.~\ref{fig_xix-200}. This also shows that the predictions on these
large scales are quite robust, since they are independent of the underlying halo
model and are given by systematic perturbation theory.
The effect of the 1-halo counterterm is also very small so that we do not
plot a specific figure to compare the curves obtained with and without this
counterterm. Note that the difference between various curves in
Figs.~\ref{fig_Pk-50} and \ref{fig_Pk-noWth} was amplified by the fact that we
divided by a ``no-wiggle'' linear power spectrum $P_{Ls}(k)$.

\section{Typical accuracy of combined models}
\label{accuracy}

\begin{figure*}
\begin{center}
\epsfxsize=6.1 cm \epsfysize=5 cm {\epsfbox{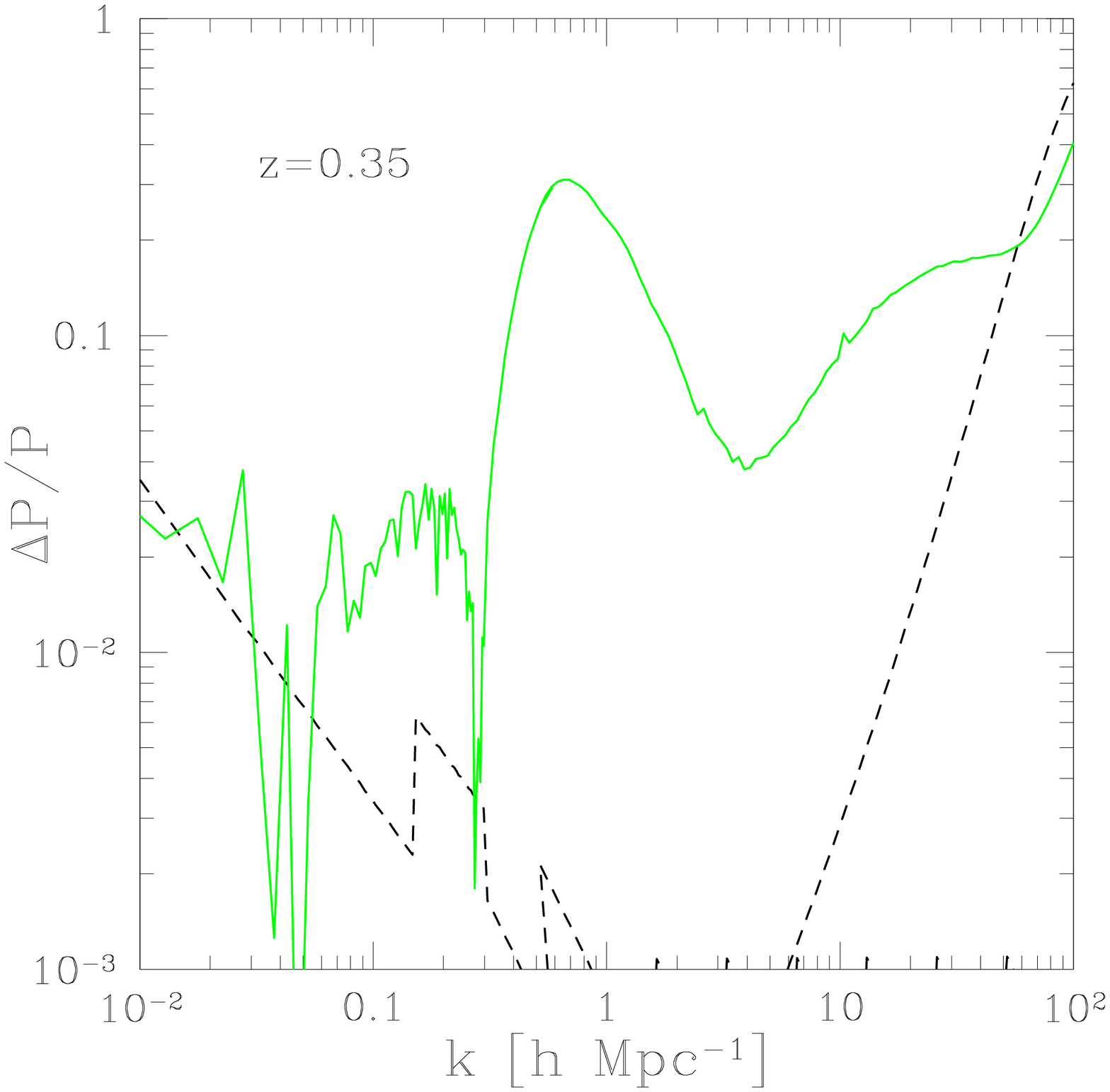}}
\epsfxsize=6.05 cm \epsfysize=5 cm {\epsfbox{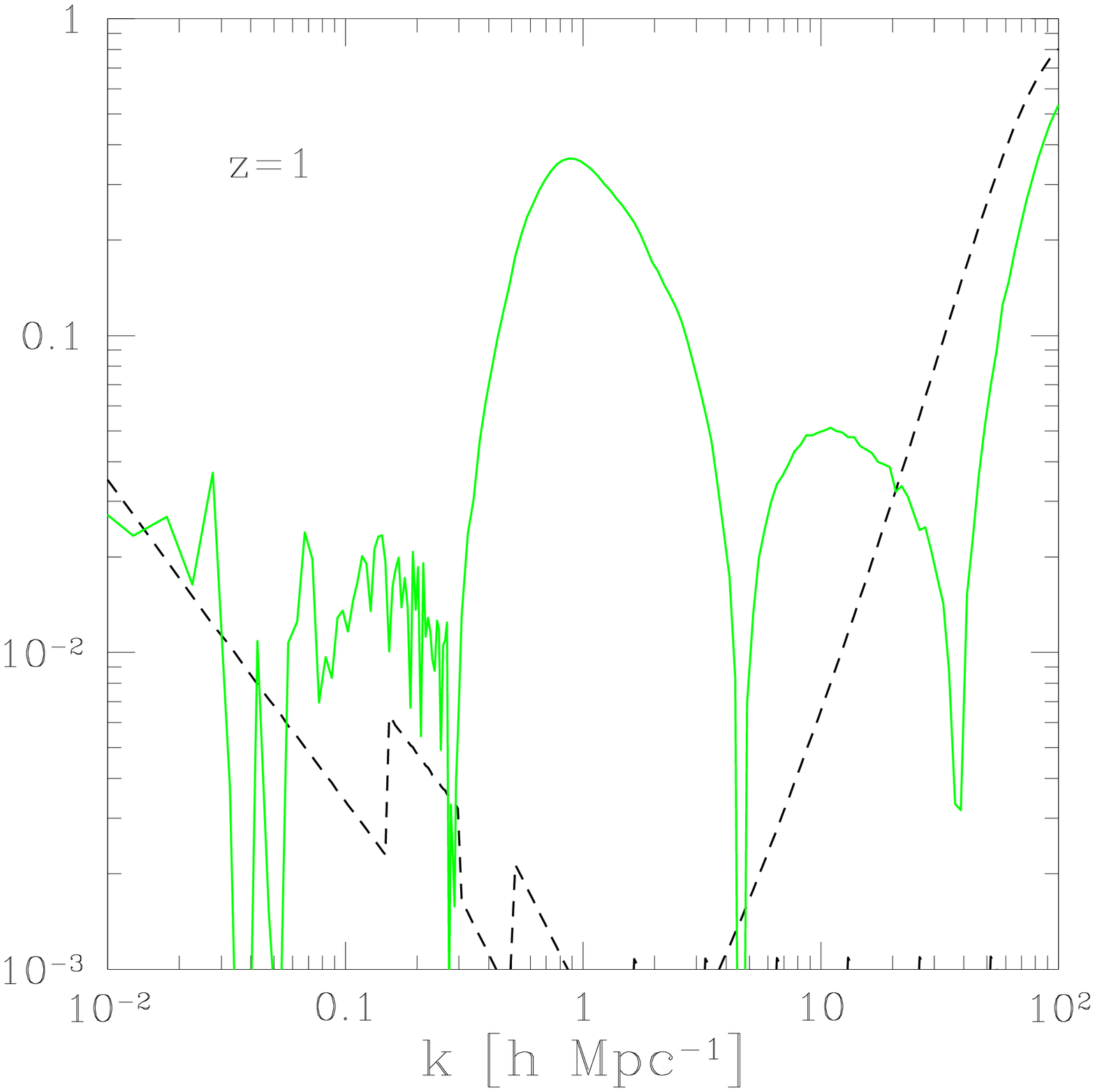}}
\epsfxsize=6.05 cm \epsfysize=5 cm {\epsfbox{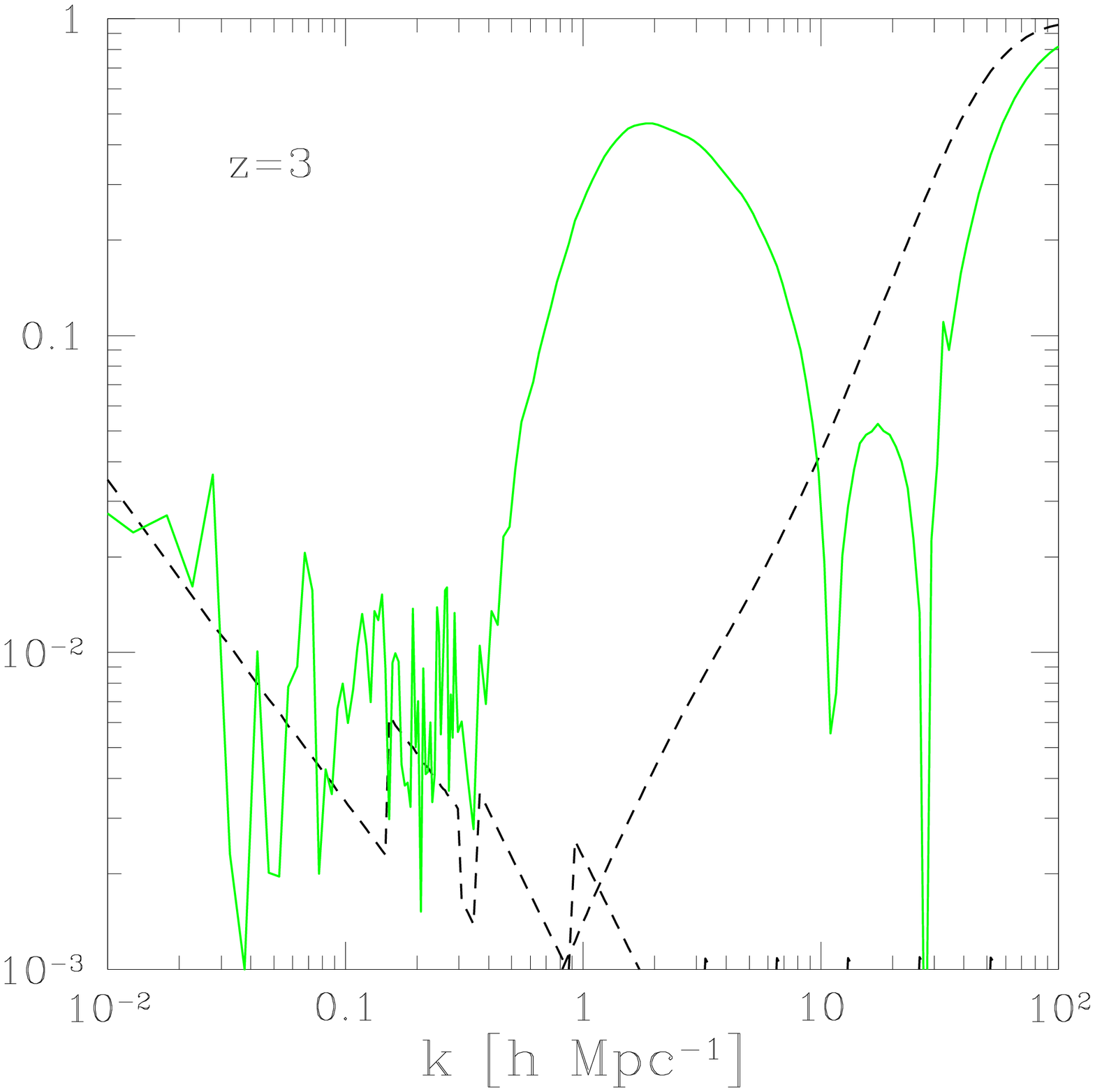}}
\end{center}
\caption{The accuracy of our model and our simulations at redshifts $z=0.35, 1$,
and $3$, for the density power spectrum. The green solid line is the relative
difference between the model and the simulations, from Eq.(\ref{dP-def}).
The dashed line that grows at low $k$ is the relative statistical error (\ref{eq:error})
of the simulations, while the dashed line that grows at high $k$ is the relative
shot-noise error.}
\label{fig_dP-200}
\end{figure*}

In this last section before conclusion, as a summary of what can be achieved
from simple analytical models that combine perturbation theories with halo
models, we estimate the accuracy reached by our fiducial model.
Thus, we plot in Fig.~\ref{fig_dP-200} the relative difference between our model,
described in Sects.~\ref{implementation} and \ref{Comparison},
and the results from the N-body simulations presented in
Sect.~\ref{N-body-simulations},
\beq
\frac{\Delta P}{P}(k) = \frac{|P_{\rm model}(k) - P_{\rm N-body}(k)|}{P_{\rm N-body}(k)} .
\label{dP-def}
\eeq
We also show the statistical error due to the finite number of modes in the
simulation box, given by Eq.(\ref{eq:error}), and the shot-noise error given by
$\Delta P_{\rm shot-noise} = L_{\rm box}^3/N$.
Since the number of modes within a bin of fixed size $\Delta k$ scales as
$k^2\Delta k$, the relative statistical error decreases at higher $k$ as $1/k$, see
Eq.(\ref{eq:error}). The sudden jumps are due to the folding procedure,
see Sect.~\ref{subsec:measure} and Fig.~\ref{fig:statistical}.
On the other hand, the relative shot noise grows as $1/P(k)$ and dominates at
high $k$. It seems that the simple approximation
$\Delta P_{\rm shot-noise} = L_{\rm box}^3/N$
overestimates somewhat the inaccuracy of the simulations, however we
do not look for a better estimate here, as this is already sufficient to understand
the high-$k$ part of Fig.~\ref{fig_dP-200} and to mark the wavenumber
where shot noise becomes dominant.

We can see that on the large scales associated with the BAO oscillations,
$0.03<k<0.3h$Mpc$^{-1}$, the analytical approach is competitive with the  
numerical simulations. It typically gives an accuracy of $1\%$, which
however worsens somewhat at lower redshift to reach $2\%$ at $z=0.35$.
However, a part of $\Delta P/P$ shown in Fig.~\ref{fig_dP-200} arises
from the inaccuracies of the simulations themselves, so that the true
value of relative error between the analytical model and the exact nonlinear
power spectrum may be slightly smaller. 
Of course, on even larger scales, $k<0.03h$Mpc$^{-1}$, the analytical model becomes
exact as it agrees with linear theory and the relative difference $\Delta P/P$ is
solely due to the inaccuracies of the simulation results
(as can be checked by comparison with the relative statistical error of the
simulations).
In the 1-halo region, $3<k<100h$Mpc$^{-1}$, we reach an accuracy on the
order of $10\%$, and even better ($4\%$) in the region where
$10<\Delta^2(k)<100$.
Moreover, as we have seen in Sect.~\ref{ingredients}, the predictions of our model
are rather robust in this range, since the dependence on the details of the
underlying halo model is quite weak.
At very high $k$ the numerical simulations are strongly affected by shot noise,
so that the curve $\Delta P/P$ mostly measures this source of error and the
accuracy of the model may be better than what Fig.~\ref{fig_dP-200} suggests.

In agreement with the discussion in Sect.~\ref{Comparison}, a salient
feature in Fig.~\ref{fig_dP-200} is the peak in the transition range between
the 2-halo and 1-halo contributions, at $k \sim 1h$Mpc$^{-1}$.
This is due to the underestimation noticed in Fig.~\ref{fig_lDk-200},
which can reach $30\%$ around the peak. This feature shifts to higher
$k$ and somewhat broadens at higher redshift, in agreement with
the discussion in Sect.~\ref{Comparison}.

\begin{figure*}
\begin{center}
\epsfxsize=6.1 cm \epsfysize=5 cm {\epsfbox{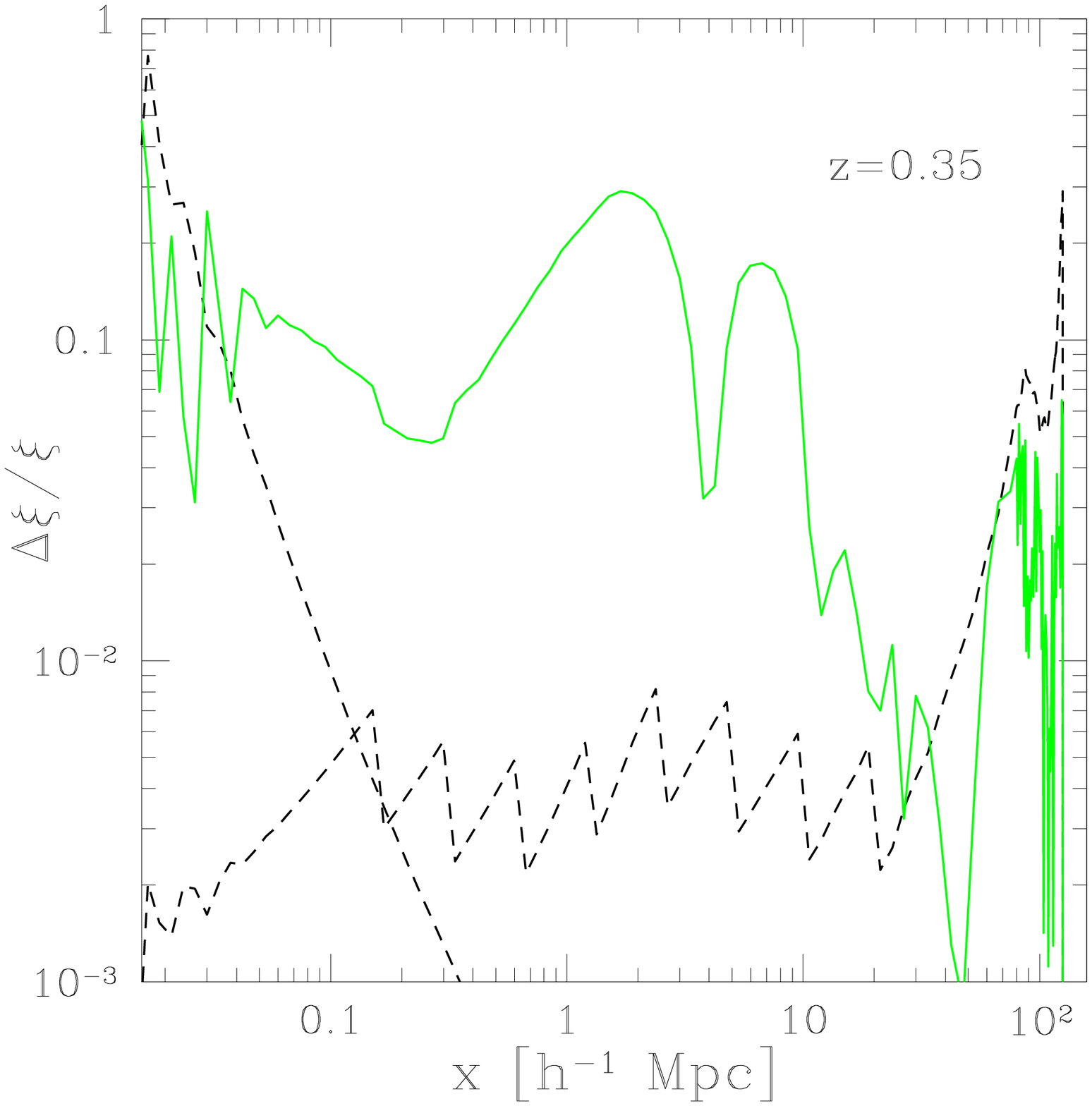}}
\epsfxsize=6.05 cm \epsfysize=5 cm {\epsfbox{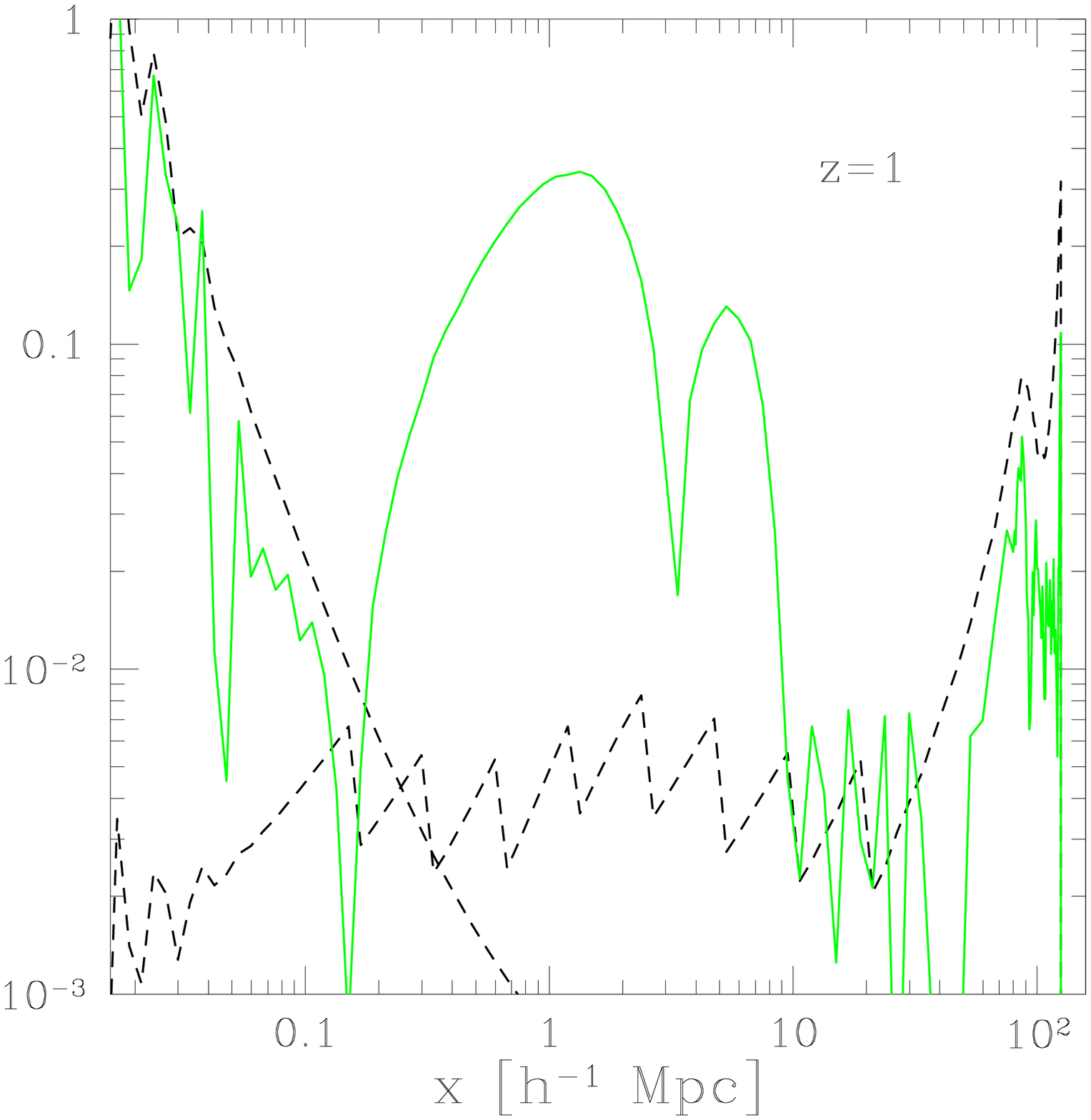}}
\epsfxsize=6.05 cm \epsfysize=5 cm {\epsfbox{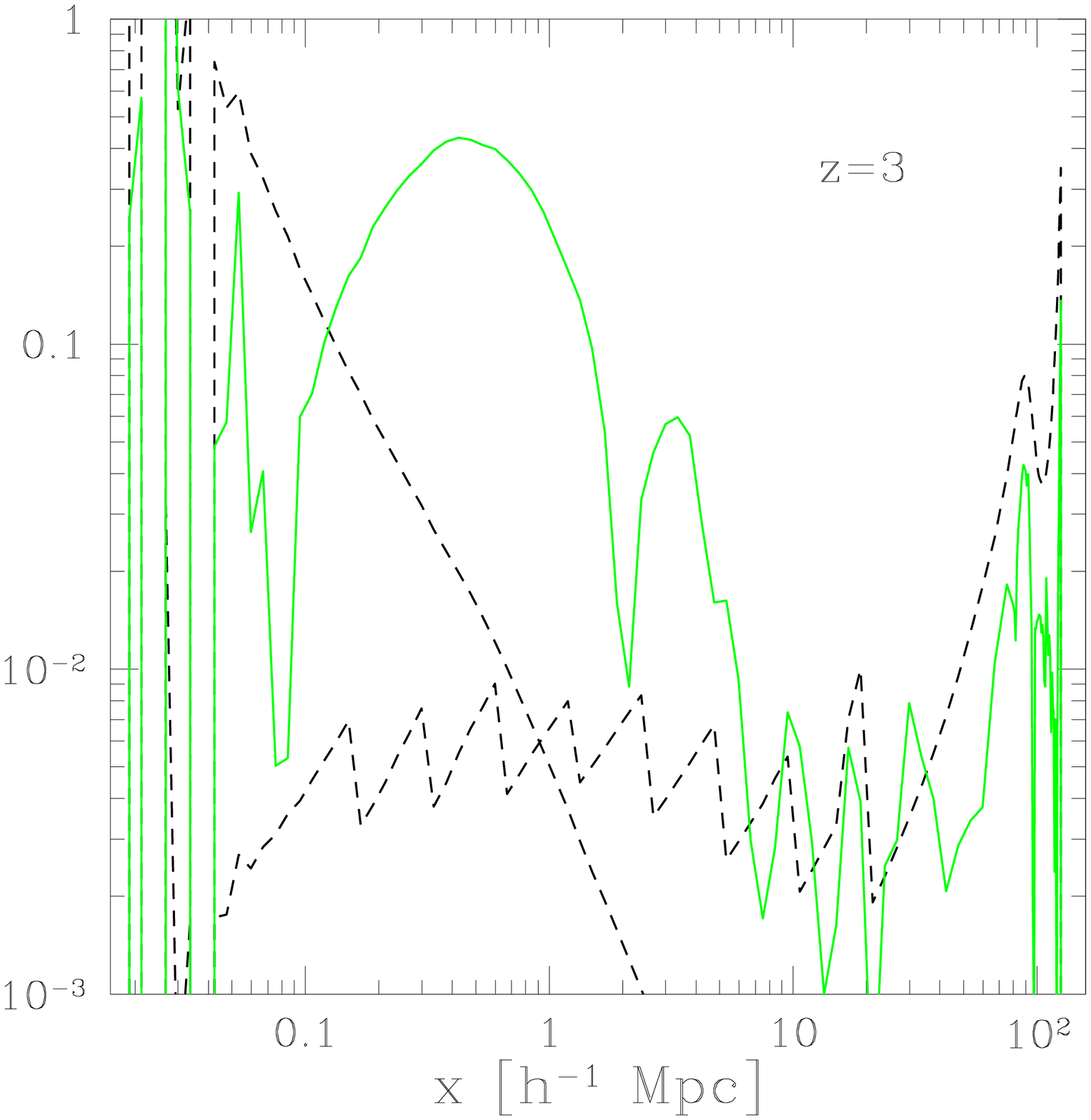}}
\end{center}
\caption{The accuracy of our model and our simulations at redshifts $z=0.35, 1$,
and $3$, for the density two-point correlation. The green solid line is the relative
difference between the model and the simulations, from Eq.(\ref{dxi-def}).
The dashed line that grows at large $x$ is the relative statistical error (\ref{eq:xi_error})
of the simulations, while the dashed line that grows at small $x$ is the relative
shot-noise error.}
\label{fig_dxi-200}
\end{figure*}

Next, we show in Fig.~\ref{fig_dxi-200} the relative accuracy of our model and
our simulations for the real-space two-point correlation function.
As in the Fourier-space figure ~\ref{fig_dP-200}, we plot the relative difference
between our model, presented in Sect.~\ref{Real-space} (for halos defined by
$\delta=200$), and our simulations,
\beq
\frac{\Delta \xi}{\xi}(x) = \frac{|\xi_{\rm model}(x) - \xi_{\rm N-body}(x)|}
{\xi_{\rm N-body}(x)} .
\label{dxi-def}
\eeq
We also show the statistical error, given by Eq.(\ref{eq:xi_error}), and the shot-noise
error given by
$\Delta \xi_{\rm shot-noise} = [L_{\rm box}^3/(4\pi x^3/3)]/N$.
As for the power spectrum, the relative statistical error grows on large scales,
as $\propto x$, because of the smaller number of modes, and the sudden jumps
are due to the folding procedure, see Sect.~\ref{subsec:measure} and
Fig.~\ref{fig:statistical}.
On the other hand, the relative shot noise grows as $1/(x^3\xi(x))$ on small scales
and dominates at low $x$.

We can check that Fig.~\ref{fig_dxi-200} is consistent with Fig.~\ref{fig_dP-200}.
On large scales, $x>10h^{-1}$Mpc, the analytical approach is again competitive
with the numerical simulations and it reaches an accuracy of $1\%$.
On very large scales the analytical model converges to the linear two-point
correlation and becomes exact, and the relative difference $\Delta\xi/\xi$
is solely due to the inaccuracies of the simulations.
On small scales dominated by the 1-halo contribution, $x<0.1h^{-1}$Mpc, 
we reach an accuracy of $10\%$, and even better at $z\geq 1$.
On very small scales the difference $\Delta\xi/\xi$ mostly measures the level of
shot noise in the simulations, so that the accuracy of the model could be better
than shown in the figure.

Again, the most difficult region to reproduce is the transition between the
2-halo and 1-halo contributions, $x\sim 1h^{-1}$Mpc.
The underestimation seen in Fig.~\ref{fig_lxix-200} leads to a peak for the
relative error of the analytical model that can reach about $30\%$ and is
somewhat greater at higher redshift.
Of course, the boundaries of these various domains shift to smaller scales
at higher redshifts.

Thus, it appears that analytical models such as the one studied in this paper
can already provide reasonably good estimates for the density power spectrum
and two-point correlation function
over a large range of scales. We have not scanned all possible models in this paper
(although we have investigated several variants in Sect.~\ref{ingredients}),
but we expect similar approaches, based on a combination of perturbation
theory and halo model, to yield similar levels of accuracy, for current resummation
schemes and halo models.
At low and high $k$, and at small and large $x$,
the level of accuracy is already rather satisfactory, and may
be improved in a straightforward and systematic manner by pushing
perturbation theories to higher orders and measuring halo profiles down to
smaller scales in simulations (and possibly taking into account additional effects
such as substructures and baryon impact on gravitational collapse).
An advantage of such models, especially on large scales, it to provide robust
predictions that can be applied to a variety of initial conditions or cosmological
parameters. This is obvious on the large scales described by systematic
perturbation theories, but this remains true to some extent at high $k$ and small
$x$ as the power spectrum is expressed in terms of more elementary quantities
that satisfy physical constraints (such as the normalization or scaling of the
mass function, and reasonable halo profiles). This ensures at the very least
that the power spectra and two-point correlations obtained in this fashion make
sense and scale in the appropriate fashion, even for (reasonable) cases where they
have not been tested.

What may prove to be the most difficult regime for systematic improvement
is the transition range. Although we have not investigated such an approach here,
it is certainly possible to build fitting formulas that improve the agreement with
simulations on this range. Thus, one may choose some interpolation formula
between the low-$k$ and high-$k$ domains and obtain a good match by tuning
some free parameters that govern the curvature of the interpolation curve in this
regime.
However, one must take some care not to spoil the good agreement on BAO scales
and to keep a percent accuracy there.
On the other hand, if we insist on avoiding such a direct interpolation procedure, so
as to derive the power spectrum from more elementary quantities (such as
perturbative expansions and halo profiles) without further modifications, it is not
clear how much progress can be made on this range.
Indeed, the description of the density field in terms of halos is by itself
an approximate picture, especially for the moderate-density regions of space
associated with outer shells or filaments that have already experienced  
shell crossing but have not fully virialized and are not enclosed within
well-identified objects.
This may prevent a perfect matching on the transition scales, unless it is
somehow enforced a posteriori.

\section{Conclusion}
\label{Conclusion}

In this article we have explained how to build unified models to predict the matter
density power spectrum, and the real-space two-point correlation function, by
combining perturbation theories with halo models.
First, we have shown how a Lagrangian point of view allows us to re-interpret the
decomposition of the power spectrum into 2-halo and 1-halo contributions.
This provides a convenient route to this splitting, which automatically ensures
the conservation of mass and gives an explicit relationship with the dynamics
of the system, through the direct Lagrangian map $\vq\mapsto\vx$.
We have also emphasized the relationship between this decomposition and
the separation into perturbative and non-perturbative terms.
This explicitly shows how one can achieve consistency with perturbation theory
through the 2-halo term, as the 1-halo term is fully non-perturbative
and does not contribute to any order of perturbation theory.

Next, we have shown that the 1-halo contribution contains a counterterm,
which was missed in previous studies, that gives the expected low-$k$
tail $P_{1\rm H}(k) \propto k^2$ associated with the conservation of matter.
This ensures that the 1-halo term becomes subdominant on large scales,
as linear CDM power spectra typically behave as $P_L(k) \propto k$ at low $k$.
On the other hand, we have explained why the $k^4$-tail associated with
momentum conservation is not recovered because of the approximations
that we use (such as instantaneous virialization and loss of memory), but this
is not a problem for practical purposes.
If needed, in future works aiming at greater accuracy that include higher orders
of perturbation theory, it may be possible to enforce the $k^4$-tail by
modifying this counterterm.
Then, we have pointed out that the 2-halo contribution, and more precisely
its term $P_{\rm pert}(k)$ that should be consistent with perturbation theory,
cannot be computed by truncating standard perturbation theory at a finite order.
Indeed, higher order terms of this perturbative expansion grow increasingly fast
at high $k$ (with large cancellations between various orders), so that an abrupt
truncation at a given order leads to a 2-halo contribution that is non-negligible
at high $k$ (and would even be dominant on small scales if we go to two loops or
higher orders), which is not physical and prevents a good agreement with
numerical simulations. The remedy is to use resummation schemes that agree
with standard perturbation theory, up to the required order, while remaining
well-behaved at high $k$ (typically close to or smaller than the linear power
spectrum). 
In addition to the superior accuracy of such schemes on large scales, this is a
second important motivation to develop such perturbative approaches.
We have described in particular a simple implementation, based on the
direct steepest-descent resummation at 1-loop order, which satisfies this
property while being fast to compute, thanks to its factorized form.
In this fashion, both the 1-halo and 2-halo contributions remain well-behaved
beyond the domain where they dominate. This allows us to build a meaningful model
from their combination, which can describe all scales and regimes.

We have compared a simple implementation of this model to N-body simulations.
We can reach an accuracy of $1\%$ on quasi-linear scales, associated with
baryon acoustic oscillations, and $10\%$ on highly nonlinear scales dominated
by the 1-halo contribution. The agreement on large scale is actually better than
the one obtained with standard 1-loop perturbation theory. Since it is based
on systematic perturbation theory and contains no free parameter
(except through the prefactor $F_{2\rm H}$, which however is very close to
unity on these scales), this gives a robust and reliable prediction that also improves
over simple fitting formulas. On the other hand, the good agreement on highly
nonlinear scales, $\Delta^2(k)>200$, depends on the properties of the halos
included in the underlying halo model, and more specifically on the
mass-concentration relation $c(M)$. We have given a simple model that provides
a reasonable match to the N-body results, but this regime is clearly less reliable
than the quasi-linear regime. In particular, to apply this framework to very high
redshifts one should use a better constrained prescription $c(M)$ or a
physically motivated model.
Nevertheless, in this high-$k$ regime where $200<\Delta^2(k)<1000$, we find that
using various formulas for $c(M)$ that have been proposed in the literature
already improves over simple fitting formulas for $P(k)$, in spite of the
scatter of their predictions. Indeed, on these scales the dependence on the details
of the model (the low-mass tail of $c(M)$) is not too large yet.
Moreover, on somewhat less nonlinear scales where $10<\Delta^2<200$,
the predictions appear even more robust and also provide a good match to
simulations.

The transition range where the 2-halo and 1-halo contributions are of the
same order, roughly where $1<\Delta^2<10$, proves to be the most difficult to
reproduce. In agreement with previous works \citep{Giocoli2010}, we find that
the model underestimates the power spectrum on these scales, and the
inaccuracy is somewhat larger at higher $z$.
Some of the discrepancy may arise from the truncation of halo profiles at a
density contrast $\delta=200$, and we have shown that going to larger radii,
defined by $\delta=50$, gives a slightly better match to the simulations, but this is
not sufficient.
Another route to improve the model would be to truncate the resummation
scheme (involved in the 2-halo contribution) at a higher order, that is to include
exactly all perturbative terms up to a higher loop order, while higher orders are
only partially resummed.
As seen in previous works \citep{Valageas2010a}, the potential for improvement
is greatest at higher $z$ for CDM power spectra, since the range where perturbative
expansions are relevant is somewhat broader and one can push to higher orders
before non-perturbative terms become dominant. This would explain why
the underestimate of the power spectrum appears to be more significant
at $z=3$ than at $z=0.35$.

However, it is likely that even a complete resummation of the perturbative
term would not provide a perfect match, because of the approximations involved
by the splitting into 2-halo and 1-halo contributions.
Indeed, even though the decomposition (\ref{Pk-halos}), with
Eqs.(\ref {Pkxq-1H})-(\ref{Pkxq-2H}), can be made exact if we have a well-defined
criterion to define halos, the computation of the 2-halo term involves
some approximation as we replace the conditional average $\lag .. \rag_{2\rm H}$
by the perturbative average $\lag .. \rag_{\rm pert}$.
Going beyond this approximation requires going beyond perturbation theory
(which does not know about halos).
On the other hand, the 1-halo contribution also involves approximations that
are difficult to improve, such as the instantaneous virialization, that is,
the fact that particles are located at random in a halo, independently of their
initial location.
Nevertheless, this is a range of scales where there is still room for improvement.

Next, we have investigated the dependence of such predictions on the details of the
model. We have shown that it is important to take into account the 1-halo
counterterm in order to get a good accuracy on large scales, $k \sim 0.1$ to
$0.3 h$Mpc$^{-1}$. Otherwise, the 1-halo contribution does not decrease fast
enough at low $k$ and spoils the good match with simulations (and would even 
become dominant on very large scales).
On the perturbative side, we have shown that standard perturbation theory
cannot be used to build unified model, because it yields higher order terms
that grow increasingly fast at high $k$ and prevent a good match to numerical
simulations in the nonlinear regime. This cannot be compensated by details
of the underlying model and would require at best the addition of some ad-hoc
damping prefactor. Another perturbative scheme than the 1-loop
direct steepest-descent resummation we have focused on in this paper,
based on a more realistic ``decaying response'' function, gives similar results to
the former, as it also provides a well-behaved 2-halo contribution at high $k$.
It even improves somewhat the match with simulations on transition scales,
but this is likely to be a coincidence as this comes at the price of an overestimate
on the larger scales probed by BAO.
Next, we have shown that current Lagrangian resummation schemes, which
a priori could have been expected to be best suited to our purposes,
actually fare much worse than Eulerian resummation schemes. Indeed,
although they yield a well-behaved perturbative term at high $k$,
the associated damping is so large that it leads to a significant underestimate
for the power spectrum already on BAO scales, $k \sim 0.1$ to
$0.3 h$Mpc$^{-1}$.
However, if it is possible to devise new Lagrangian schemes that do not suffer
from this effect they may provide a promising route to improve unified models
for the power spectrum.

Finally, we have shown that our model also provides
reliable predictions for the density two-point correlation function.
As for the power spectrum, it describes rather well both small and large scales,
but underestimates somewhat the correlation in the transition range between
the 2-halo and 1-halo contributions. On very large scales, around
the BAO peak, we obtain a very good agreement with numerical simulations,
so that the simple 1-loop resummation scheme used in this work is probably
sufficient to describe this feature.

Thus, we have shown that it is possible to build successful unified models for the
density power spectrum and two-point correlation, that combine systematic and
reliable perturbation
theories on large scales with phenomenological halo models on small scales.
Two issues where there is still room for improvement are the underestimate
of the power spectrum and correlation function on transition scales and the building
of accurate Lagrangian
resummation schemes (but this is a more theoretical goal since for practical
purposes Eulerian schemes are sufficient).
A third issue is the modeling of the underlying virialized halos, which plays a key
role at very high $k$. There, many points can be improved, such as taking into
account substructures, deviations from spherical profiles, or the effect of baryons on
the halo density profile. In particular, one should be cautious when using such models
beyond the range where they have been tested ($\Delta^2(k)<1000$ and $z\leq 3$).
There, the main uncertainty comes from the low-mass tail of the mass-concentration
relation $c(M)$, and it is probably safer to use physically motivated models for $c(M)$
(even though they may not be very accurate), or at least to check that
the formula used for $c(M)$ is still reasonable.

The model presented in this paper could be extended to higher order statistics,
such as the bispectrum or three-point correlation function. It should also be possible
to handle the case of non-Gaussian initial conditions, at least as long as the
halo density profiles are not strongly modified (or this would require a separate
study to evaluate this effect).
Finally, to allow a direct comparison with galaxy surveys, it would be desirable
to extend the model to redshift space. This is likely to be a more involved task,
that we leave for future works.

\begin{acknowledgements}

T. N. is supported by a Grant-in-Aid for Japan Society for the Promotion of Science (JSPS) Fellows and
by World Premier International Research Center Initiative (WPI Initiative), MEXT, Japan.
Numerical computations for the present work have been carried out in part on Cray XT4 at Center for
Computational Astrophysics, CfCA, of National Astronomical Observatory of Japan, and in part under the
Interdisciplinary Computational Science Program in Center for Computational Sciences, University of Tsukuba.

\end{acknowledgements}

 \appendix

\section{Computation of $P_{\rm pert}(k)$ with the ``direct steepest-descent method''}
\label{direct-steepest-descent}

We briefly recall here how the perturbative term $P_{\rm pert}(k)$ of the
2-halo contribution (\ref{Pkxq-pert2}) is computed within the ``direct
steepest-descent method'', see also \citet{Valageas2007a,Valageas2008} for
details.
First, to simultaneously describe the density and
velocity fields it is convenient to define the two-component vector $\psi(\vk,\eta)$
as \citep{Crocce2006a},
\beq
\psi(\vk,\eta) = \left( \bea{c} \delta(\vk,\eta) \\ -\theta(\vk,\eta)/(\dot{a} f)
\ea \right) ,
\eeq
where $\theta$ is the divergence of the velocity field, $\theta=\nabla\cdot\vv$,
$\eta=\ln D(z)$ where $D(z)$ is the linear growth factor (normalized by
$D(0)=1$ today), $\dot{a}=\dd a/\dd t$ where $a(t)$ is the scale factor,
and $f= \dd\ln D/\dd\ln a$. Then, introducing the coordinate $x=(\vk,\eta,i)$,
where $i=1$ or $2$ is the discrete index of the two-component vectors, the
equations of motion write as \citep{Valageas2007a}
\beq
\cO(x,x') \cdot \psi(x') =K_s(x;x_1,x_2) \cdot \psi(x_1) \psi(x_2) , 
\label{OKsdef}
\eeq
where the matrix $\cO$ reads as
\beqa
\cO(x,x') & = & \delta_D(\vk-\vk') \, \delta_D(\eta-\eta') \nonumber \\ 
&& \times \, \left( \bea{cc} \frac{\pl}{\pl\eta} & -1 \\ -\frac{3\Om}{2f^2} \;
& \; \frac{\pl}{\pl\eta} + \frac{3\Om}{2f^2} -1 \ea \right) ,
\label{Odef}
\eeqa
whereas the symmetric vertex $K_s(x;x_1,x_2)=K_s(x;x_2,x_1)$ writes as
\beqa
K_s(x;x_1,x_2) & = & \delta_D(\vk_1+\vk_2-\vk) \, \delta_D(\eta_1-\eta) 
\, \delta_D(\eta_2-\eta) \nonumber \\ 
&& \times \, \gamma^s_{i;i_1,i_2}(\vk_1,\vk_2) ,
\label{Ksdef}
\eeqa
with
\beqa
\gamma^s_{1;1,2}(\vk_1,\vk_2) & = & \frac{(\vk_1+\vk_2)\cdot\vk_2}{2 k_2^2} , \\
\gamma^s_{1;2,1}(\vk_1,\vk_2) & = & \frac{(\vk_1+\vk_2)\cdot\vk_1}{2 k_1^2} , \\
\gamma^s_{2;2,2}(\vk_1,\vk_2) & = & \frac{|\vk_1+\vk_2|^2 (\vk_1\cdot\vk_2)}
{2 k_1^2 k_2^2}  ,
\label{gamma-def}
\eeqa
and zero otherwise \citep{Crocce2006a}.
In addition to the usual two-point correlation function,
\beq
C(x_1,x_2) = \lag \psi(x_1) \psi(x_2) \rag ,
\label{C-def}
\eeq
resummation schemes often involve the response function (or propagator),
defined as the functional derivative of the nonlinear field $\psi$ with respect
to an infinitesimal ``noise'' $\zeta$ that would be added to the right hand side
of the equation of motion (\ref {OKsdef}),
\beq
R(x_1,x_2) = \lag \frac{\cD \psi(x_1)}{\cD \zeta(x_2)} \rag_{\zeta=0} .
\label{R-def}
\eeq

As described in \citet{Valageas2007a,Valageas2008},
the nonlinear correlation and response functions obey the exact Schwinger-Dyson
equations
\beqa
\cO(x,z)\cdot C(z,y) & = & \Sigma(x,z) \cdot C(z,y) + \Pi(x,z) \cdot R^T(z,y) , 
\nonumber \\
&& \label{O-C} \\
\cO(x,z)\cdot R(z,y) & = & \delta_D(x-y) + \Sigma(x,z) \cdot R(z,y) ,
\label{O-R}
\eeqa
where $\Sigma(x,y)$ and $\Pi(x,y)$ are ``self-energy'' terms, that can be
written as diagrammatic expansions over the linear two-point functions
$C_L$ and $R_L$. Equation (\ref{O-C}) can be integrated as
\beq
C = R \times C_L(\eta_{\rm I}) \times R^T + R \cdot \Pi \cdot R^T,
\label{C-int}
\eeq
where $C_L(\eta_{\rm I})$ is the linear correlation function at an early time
$\eta_{\rm I}$, with $\eta_{\rm I}\rightarrow -\infty$, and the first product
``$\times C_L \times$'' does not contain any integration over time.

Then, at 1-loop order the direct steepest-descent scheme amounts to use
for $\Sigma$ and $\Pi$ their lowest order terms, over powers of $C_L$ and $R_L$.
This means that Eq.(\ref{C-int}) for the correlation $C$ reads as the first line
in Fig.~\ref{fig_Csd}, while a similar diagrammatic expression holds for
the response $R$ (see Fig.8 in \citet{Valageas2008}).
In particular, solving Eq.(\ref{O-R}) as a perturbative series over powers of
$\Sigma$ gives a diagrammatic expansion for the response $R$ in terms of
``bubble'' diagrams, and substituting into Eq.(\ref{C-int}) this gives the
series of diagrams shown in the second equality of Fig.~\ref{fig_Csd}.
In practice, one does not compute the two-point function $C(x,y)$, whence
the density power spectrum, from these diagrammatic series, but from the
integro-differential equation (\ref{O-R}) and the explicit expression (\ref{C-int}).

First, using in the following the approximation $\Om/f^2\simeq 1$,
the linear correlation and response functions read as
\beq
C_L(x_1,x_2) = e^{\eta_1+\eta_2} \, P_{L0}(k_1) \, \delta_D(\vk_1+\vk_2) 
\left(\bea{cc} 1 & 1 \\ 1 & 1 \ea\right) ,
\label{CL}
\eeq
where $P_{L0}(k)$ is the linear density power spectrum today (when $\eta=0$),
and
\beqa
R_L(x_1,x_2) & = & \theta(\eta_1-\eta_2) \, \delta_D(\vk_1-\vk_2) \nonumber \\
&& \times \left[ e^{\eta_1-\eta_2} R_0^+ + e^{-3(\eta_1-\eta_2)/2} 
R_0^- \right] ,
\label{R0R0pR0m}
\eeqa
with
\beq
R_0^+= \left(\bea{cc} 3/5 & 2/5 \\ 3/5 & 2/5 \ea\right) \;\; \mbox{and} \;\; 
R_0^-= \left(\bea{cc} 2/5 & -2/5 \\ -3/5 & 3/5 \ea\right) .
\label{R0pR0m}
\eeq
In Eq.(\ref{R0R0pR0m}) the Heaviside factor $\theta(\eta_1-\eta_2)$ expresses
the causality of the response function.
Then, the ``self-energy'' terms read at lowest order as \citep{Valageas2007a}
\beqa
\Sigma_0(x_1,x_2) & = & \theta(\eta_1-\eta_2) \, \delta_D(\vk_1-\vk_2) 
\nonumber \\
&& \hspace{-0.5cm} \times \left[ e^{2\eta_1} \Sigma_0^+(k_1)
+ e^{-\eta_1/2+5\eta_2/2} \Sigma_0^-(k_1) \right] ,
\label{S0S0pS0m}
\eeqa
with
\beqa
\Sigma_{0;i_1 i_2}^{\pm}(k) & = & 4 \sum_{j_1 j_2 l_1 l_2}\int\dd\vq \; 
\gamma^s_{i_1;j_1,j_2}(\vk-\vq,\vq) \nonumber \\
&& \times \gamma^s_{l_1;i_2,l_2}(\vk,-\vq) \, P_{L0}(q) \, R_{0;j_1 l_1}^{\pm} ,
\label{S0pS0m}
\eeqa
and
\beq
\Pi_0(x_1,x_2) = \delta_D(\vk_1+\vk_2) \, e^{2\eta_1+2\eta_2} \, \Pi_0(k_1) ,
\label{Pi0}
\eeq
with
\beqa
\Pi_{0;i_1 i_2}(k) & = & 2 \sum_{j_1 j_2 l_1 l_2} \int\dd\vq \; 
\gamma^s_{i_1;j_1,j_2}(\vq,\vk-\vq) \nonumber \\
&& \hspace{-1cm} \times \gamma^s_{i_2;l_1,l_2}(-\vq,-\vk+\vq)
\, P_{L0}(q) \, P_{L0}(|\vk-\vq|) .
\label{Pi0k}
\eeqa
These ``self-energy'' terms also satisfy the symmetries
\beq
\Sigma_0^-(k) = \left(\bea{cc} \Sigma_{0;22}^+ & -\Sigma_{0;12}^+  
\\ -\Sigma_{0;21}^+ & \Sigma_{0;11}^+ \ea\right) , \;\;\;
\Pi_{0;ij}(k) = \Pi_{0;ji}(k) . 
\label{S0mS0p}
\eeq
Thanks to the simple time dependence of $\Sigma_0(x_1,x_2)$, which arises from
the fact that we have expanded $\Sigma$ over the linear two-point functions
$C_L$ and $R_L$ and only kept the lowest order term, it is possible to
eliminate the integral terms in the Schwinger-Dyson equation (\ref{O-R}) by
taking successive derivatives with respect to $\eta_1$. This yields
\beqa
\lefteqn{\frac{\pl^3 R_1}{\pl\eta_1^3} - \frac{3}{2} \frac{\pl^2 R_1}{\pl\eta_1^2} 
-\frac{\pl^2 R_2}{\pl\eta_1^2} - \frac{\pl R_1}{\pl\eta_1} 
+ \frac{3}{2} \frac{\pl R_2}{\pl\eta_1} + R_2 = } \nonumber \\
&& e^{2\eta_1} \left[ (\Sigma_{0;11}^+ + \Sigma_{0;22}^+) \frac{\pl R_1}{\pl\eta_1}
+ \frac{5}{2} \Sigma_{0;11}^+ R_1 + \frac{5}{2} \Sigma_{0;12}^+ R_2 \right]
\nonumber \\
&& \label{R01diff}
\eeqa
\beqa
\lefteqn{\frac{\pl^3 R_2}{\pl\eta_1^3} - \frac{3}{2} \frac{\pl^2 R_1}{\pl\eta_1^2} 
-\frac{\pl^2 R_2}{\pl\eta_1^2} + \frac{9}{4} \frac{\pl R_1}{\pl\eta_1} 
- \frac{7}{4} \frac{\pl R_2}{\pl\eta_1} + \frac{3}{2} R_1 - \frac{1}{2} R_2 = } 
\nonumber \\
&& e^{2\eta_1} \left[ (\Sigma_{0;11}^+ + \Sigma_{0;22}^+) \frac{\pl R_2}{\pl\eta_1}
+ \frac{5}{2} \Sigma_{0;21}^+ R_1 + \frac{5}{2} \Sigma_{0;22}^+ R_2 \right] 
\nonumber \\
&& \label{R02diff}
\eeqa
where the pair $(R_1,R_2)$ corresponds either to $(R_{11},R_{21})$ or
$(R_{12},R_{22})$. These equations apply to $\eta_1>\eta_2$ and the
initial conditions at $\eta_1=\eta_2$ are
\beqa
X^T & = & \left( \bea{cccccc} R_1 & R_2 & \frac{\pl R_1}{\pl\eta_1} & 
\frac{\pl R_2}{\pl\eta_1} & \frac{\pl^2 R_1}{\pl\eta_1^2} & 
\frac{\pl^2 R_2}{\pl\eta_1^2} \ea \right)
\label{Xdef} \\
& = & \left( \bea{cccccc} 1 & 0 & 0 & \frac{3}{2} & 
\frac{3}{2}+e^{2\eta_1}(\Sigma_{0;11}^+ + \Sigma_{0;22}^+) & - \frac{3}{4} \ea \right)
\label{X1}
\eeqa
for the pair $(R_{11},R_{21})$, and
\beq
X^T = \left( \bea{cccccc} 0 & 1 & 1 & -\frac{1}{2} & -\frac{1}{2} & 
\frac{7}{4}+e^{2\eta_1}(\Sigma_{0;11}^+ + \Sigma_{0;22}^+) \ea \right)
\label{X2}
\eeq
for the pair $(R_{12},R_{22})$.
Thus, we obtain the response function by solving the differential equations
(\ref{R01diff})-(\ref{R02diff}), with the initial conditions (\ref{X1})-(\ref{X2}),
using an ordinary Runge-Kutta algorithm. A great simplification is that there
are no longer integral terms over past times in the right hand side of 
Eqs.(\ref{R01diff})-(\ref{R02diff}), so that numerical computations are very fast.

Finally, it is possible to take advantage of the factorized time dependence
of $\Pi_0$ in Eq.(\ref{Pi0}) to factorize Eq.(\ref{C-int}).
To compute the first product in Eq.(\ref{C-int}) we define the two-component
vector
\beq
\varphi^{\rm I}_i(k,\eta) = \sum_{j=1}^2 R_{i,j}(k;\eta,\eta_{\rm I})
\, e^{\eta_{\rm I}} \, \sqrt{P_{L0}(k)} ,
\label{phi-I}
\eeq
while for the second product we diagonalize the symmetric matrix $\Pi_0$
as
\beq
\Pi_0(k) = p^+ \; \pi^+ \cdot \pi^{+T} + p^- \; \pi^- \cdot \pi^{-T} ,
\label{pi_+-}
\eeq
where $p^{\pm}$ are the eigenvalues of $\Pi_0$ and $\pi^{\pm}$ its
normalized eigenvectors, and we define the two-component vectors,
\beq
\varphi^{\pm}(k,\eta) =  \int_{-\infty}^{\eta} \dd \eta' \, e^{2\eta'} \,
R(k;\eta,\eta') \cdot \pi^{\pm}(k) .
\label{phi-pm}
\eeq
Then, Eq.(\ref{C-int}) writes as
\beqa
C(k;\eta_1,\eta_2) & = & \varphi^{\rm I}(\eta_1) \cdot \varphi^{\rm I}(\eta_2)^T
\nonumber \\
&& \hspace{-1.5cm} + p^+ \,  \varphi^+(\eta_1) \cdot \varphi^+(\eta_2)^T 
+  p^- \,  \varphi^-(\eta_1) \cdot \varphi^-(\eta_2)^T .
\label{C-fact}
\eeqa
In particular, the nonlinear power spectrum is given by the equal-time upper-left
component of the correlation matrix $C$,
\beq
P(k,\eta) = C_{11}(k;\eta,\eta) .
\label{Pk-C}
\eeq
The factorized form (\ref{C-fact}), which avoids to compute two-dimensional
integrals over past times as in generic expressions of Eq.(\ref{C-int}),
thanks to the factorized time dependence of $\Pi_0$ in Eq.(\ref{Pi0}),
again allows fast numerical computations.

\bibliographystyle{aa} 
\bibliography{15685}

\end{document}